\documentclass[12pt]{article}
\usepackage[dvips]{graphicx}
\begin{document}
 \newcommand{\intl}{\int\limits}
\newcommand{\suml}{\sum\limits}
\setcounter{page}{1}
\tolerance=10000
\newcommand{\nat}{I\!\!N}
\newcommand{\NN}{I\!\!N}
\newcommand{\ka}{I\!\!K}
\newcommand{\rgr}{{\rm grad}}
\newcommand{\ce}{I\!\!\!\!C}
\newcommand{\CC}{I\!\!\!\!C}
\newcommand{\re}{I\!\!R}
\newcommand{\RR}{I\!\!R}
\setcounter{page}{1}
\thispagestyle{empty}
\def\eg{{\it e.g.}\ } \def\ie{{\it i.e.}\ }
\def\sg{\hbox{sign}\,}
\def\sgn{\hbox{sign}\,}
\def\sign{\hbox{sign}\,}
\def\e{\hbox{e}}
\def\exp{\hbox{exp}}
\def\ds{\displaystyle}
\def\dis{\displaystyle}
\def\q{\quad}	 \def\qq{\qquad}
\def\lan{\langle}\def\ran{\rangle}
\def\l{\left} \def\r{\right}
\def\lra{\Longleftrightarrow}
\def\arg{\hbox{\rm arg}}
\def\d{\partial}
 \def\dr{\partial r}  \def\dt{\partial t}
\def\dx{\partial x}   \def\dy{\partial y}  \def\dz{\partial z}
\def\rec#1{{1\over{#1}}}
\def\log{\hbox{\rm log}\,}
\def\erf{\hbox{\rm erf}\,}     \def\erfc{\hbox{\rm erfc}\,}
\def\G{{G_{\alpha,\beta}^\theta}}
\def\K{K_{\alpha,\beta}^\theta}
\def\Gxt{\G (x,t)}
\def\Gkt{{\widehat{\G}}  (\kappa,t)}
\def\Gxs{{\widetilde{\G}}  (x,s)}
\def\Gks{{\widehat{\widetilde {\G}}} (\kappa,s)}
\def\FT{{\cal F}\,} 
\def\LT{{\cal L}\,}  
\def\L{{\cal L}} 
\def\F{{\cal F}} 
\def\M{{\cal M}}  
\def\I{{\cal I}}  
\def\pni{\par \noindent}
\def\vsh{\smallskip}
\def\vs{\medskip}
\def\vvs{\bigskip}
\def\vvvs{\bigskip\medskip} 
\def\vsp{\par}
\def\vsn{\vsh\pni}
\def\cen{\centerline}
\def\ra{\item{a)\ }} \def\rb{\item{b)\ }}   \def\rc{\item{c)\ }}
\def\alphak{{\alpha \choose k}}
\def\alphazero{{\alpha \choose 0}}
\def\alphaone{{\alpha \choose 1}}
\def\alphatwo{{\alpha \choose 2}}
\def\alphakk{{\alpha \choose k+1}}
 \cen{{\bf FRACALMO PRE-PRINT: \    www.fracalmo.org}}
\vsh
\cen{{\bf Fractional Calculus and Applied Analysis,
  Vol. 4 No 2 (2001) 153-192.}}
\vsh
\cen{An International Journal for Theory and Applications \ ISSN 1311-0454}
\vsh
\cen{{\bf www.diogenes.bg/fcaa/}}
\vs
\hrule
   \vskip 0.50truecm

\font\title=cmbx12 scaled\magstep2
\font\bfs=cmbx12 scaled\magstep1
\font\little=cmr10
\begin{center}

{\title The fundamental solution of}
\vs

{\title the space-time fractional diffusion equation}

\vvs
{Francesco MAINARDI} $^{(1)}$
{Yuri LUCHKO}$^{(2)}$, and
{Gianni PAGNINI}$^{(3)}$


\vs

$\null^{(1)}$
 {\little Department of Physics, University of Bologna, and INFN,} \\
{\little Via Irnerio 46, I-40126 Bologna, Italy} \\
{\little Corresponding Author. E-mail: {\tt francesco.mainardi@unibo.it}} 
\\ [0.25 truecm]
$\null^{(2)}$
 {\little Department of Mathematics (II),}\\
{\little Technical University of Applied Sciences Berlin,} \\
{\little Luxemburger Str. 10, D-13353 Berlin, Germany} \\
{\little E-mail: {\tt luchko@tfh-berlin.de}}
\\ [0.25 truecm]
$\null^{(3)}$
{\little ENEA: Italian Agency for  New Technologies,
  Energy and the Environment} 
\\ {\little Via Martiri di Monte Sole 4, I-40129 Bologna, Italy}\\
{\little E-mail: {\tt gianni.pagnini@bologna.enea.it}}

\end{center}

\noindent
{\it Dedicated to Rudolf Gorenflo, Professor Emeritus of the Free University of Berlin,
on the occasion of his 70-th birthday (July 31, 2000).}

\vskip 0.5truecm
\cen{{\bf Abstract}} 

\noindent
We deal with the Cauchy problem for the space-time fractional diffusion
equation, which is obtained from the standard diffusion equation
by replacing the second-order space derivative
with a Riesz-Feller derivative of order $\alpha \in (0,2]$
and skewness $\theta$ ($|\theta|\le\hbox{min}\,\{\alpha ,2-\alpha \}$),
and the first-order time derivative with a Caputo derivative
of order $\beta \in (0,2]\,.$
The fundamental solution (Green function) for the Cauchy problem is
investigated with respect to its scaling and similarity properties,
starting from its Fourier-Laplace representation. We
review the particular cases of
space-fractional diffusion $\,\{0<\alpha\le 2\,,\,\beta =1\}\,,$
time-fractional diffusion $\,\{\alpha=2\,,\,0<\beta \le 2 \}\,,$
and  neutral-fractional diffusion
$\,\{0< \alpha = \beta \le 2 \}\,,$  for which the fundamental solution can
be interpreted
as a {\it spatial  probability density function evolving in time}.
Then,
 by using the Mellin transform, we provide a general representation
of the	Green functions in terms of
Mellin-Barnes integrals in the complex plane, which
allows us
 {\it to extend the  probability interpretation
to the ranges} $\{0<\alpha \le 2\} \cap \{0<\beta \le 1 \}\,$
{\it and} $\{1<\beta \le \alpha \le 2\} $.
Furthermore,
from this representation we derive
explicit formulae (convergent series and asymptotic expansions),
which  enable us to plot the spatial probability densities
for different values of the relevant
parameters $\alpha, \theta, \beta\,.$

\section*{1. Introduction}

A space-time fractional diffusion equation, obtained
from the standard diffusion equation
by replacing the second order space-derivative
by a fractional  Riesz derivative
order $\alpha >0$
and the first order time-derivative  by a fractional
derivative of order $\beta >0$ (in Caputo or Riemann-Liouville sense),
has been recently treated  by a number of authors,
see for example
Saichev and Zaslavsky \cite{SaichevZaslavsky 97},
Uchaikin and Zolotarev \cite{UchaikinZolotarev 99},
Gorenflo, Iskenderov and Luchko \cite{GoIsLu 00},
Scalas, Gorenflo and Mainardi \cite{ScalasGorMai 00},
Metzler and Klafter \cite{MetzlerKlafter 00}.
For other treatments of the space fractional and/or time fractional
diffusion equations we refer the reader to
the references cited therein. See below for the restrictions on $\alpha $
and $\beta \,.$
In this paper  we intend to complement the results
obtained in  \cite{GoIsLu 00} by allowing
asymmetry in the space fractional derivative.
We thus consider   the
space-time fractional diffusion equation
$$
_xD_\theta^\alpha \,u(x,t) \, = \,
_tD_*^\beta \, u(x,t) \,, \q x \in \re\,, \q t \in {\re}^+\,,
\eqno(1.1) $$
where  the $\alpha \,,\,\theta\,,\, \beta $ are real parameters
{\it always} restricted as follows
$$ 0<\alpha\le 2\,,\q |\theta| \le \hbox{min} \{\alpha, 2-\alpha\}\,,
  \quad 0<\beta \le 2\,.\eqno(1.2)$$
In (1.1) $u=u(x,t)$ is the (real) field variable,
$\,_xD_\theta^\alpha \,$ is
the {\it Riesz-Feller}	 space-fractional
 derivative of order $\alpha $ and skewness $\theta\,,$
and  $\,_tD_*^\beta\,$	is
 the Caputo time-fractional derivative of order $\beta \,.$
These fractional derivatives are integro-differential operators
to be defined later.
\newpage
The paper is divided as follows.
In  Section 2 we  provide the reader with
the essential notions and notations
concerning  the Fourier, Laplace and Mellin transforms,
which are necessary in the following.

In Section 3 we introduce the Cauchy problem
for the equation (1.1) and find
the corresponding fundamental solution $\G(x,t)$
(the {\it Green function})  in terms of its
Fourier-Laplace transform from which we derive
its general scaling properties and
the similarity variable $x/t^{\beta/\alpha}\,.$
We shall get the fundamental formula
$$ \G(x,t)  =
    t^{-\gamma}\,\K (x/t^\gamma)\,,\q \gamma =\beta /\alpha
\,,   \eqno(1.3)     $$
where $\K (x)$ is referred to as the {\it reduced Green function}.

In Section 4,
we consider the particular cases
$\,\{0<\alpha\le 2\,,\,\beta =1\}\,$ ({\it space fractional diffusion}),
$\,\{\alpha=2\,,\,0<\beta \le 2 \}\,$ ({\it time fractional diffusion}),
and $\,\{0< \alpha = \beta \le 2 \}\,$ ({\it neutral fractional
diffusion}),  for which the fundamental solution can be interpreted
as   a spatial {\it probability density function} ($pdf$), evolving
in time.

In Section 5
we show a composition rule for the  Green function,
valid for
$$\{0<\alpha\le 2 \} \cap \{0<\beta \le 1\}$$
which ensures its probability interpretation in this range.

In Section 6,
we provide   a general representation
for  the (reduced)  Green function
in terms of a Mellin-Barnes integral in the complex plane,
which enables us
 to extend the	probability interpretation
to the range  
$$\{1<\beta \le \alpha \le 2\}\,. $$

In Section 7,
 we derive    for the Green function
explicit formulae (convergent series and asymptotic expansions),
whose form depends on the relation between the parameters
$\alpha$ and $\beta\,.$
By means of a suitable matching between
the convergent and the asymptotic expansion
we shall be able
to compute the Green function 
in all the cases in which it is
interpretable as a probability density.

 Finally, Section 8 is devoted to  concluding discussions,
and  a summary of the results in which we present  plots
of the Green function for a number of	cases.

\newpage

\section*{2. Notions and Notations}

For the sake of the reader's convenience  here we present an
introduction to the {\it Riesz-Feller} and {\it Caputo} fractional
derivatives
starting from their representation in the Fourier and Laplace
transform domain, respectively.
So doing we avoid
the subtleties lying in the inversion of fractional integrals,
see \eg \cite{SKM 93}, \cite{GorMai CISM}.
We also recall the main properties of the Mellin transform
that will be used later.

Since in what follows we shall meet only real or
complex-valued	functions of a real variable that
are defined and  continuous in a given open interval
$\I= (a,b)\,,$	$\,-\infty \le a < b \le +\infty\,,$
except, possibly, at isolated points where these
functions can be infinite,
we restrict our presentation
of the integral transforms to the class of functions
for which the Riemann improper integral on $\I$
absolutely converges.
In so doing we follow Marichev \cite{Marichev 83} and
we denote this class by $L^c(\I)$
or $L^c(a,b)\,.$

\vskip 0.5truecm
\noindent
{\it The Fourier transform and the Riesz-Feller space-fractional
derivative}
\vskip 0.25truecm \noindent
Let
$$  \widehat f(\kappa)	=
{\cal F} \l\{ f(x);\kappa \r\}
  = \int_{-\infty}^{+\infty} \e^{\,\ds +i\kappa x}\,f(x)\, dx\,,
  \q \kappa \in \re\,, \eqno(2.1a)$$
be the	Fourier  transform  of a 
function  $f(x) \in L^c (\RR)$, and let
$$ f(x) =
{\cal F}^{-1} \l\{ \widehat f(\kappa );x \r\}
  = \rec{2 \pi}\, \int_{-\infty}^{+\infty} \e^{\,\ds -i\kappa x}\,
 \widehat f(\kappa )\, d \kappa\,,	\q x \in \re\,, \eqno(2.1b)$$
be the inverse Fourier transform\footnote{
If $f(x)$ is piecewise differentiable, then the formula (2.1b)
holds true at all points where $f(x)$ is continuous and
the integral in it must be understood in the sense of the Cauchy
principal value.}.
For a sufficiently well-behaved function $f(x)$ we
define the {\it Riesz-Feller} space-fractional derivative
of order $\alpha  $ and skewness $\theta$ as 
$$ {\cal F} \l\{\, _xD_\theta^\alpha\, f(x);\kappa \r\} =
  - \psi_\alpha ^\theta(\kappa ) \,
  \, \widehat f(\kappa) \,, \eqno(2.2)$$
 $$
   \psi_\alpha ^\theta(\kappa ) =
|\kappa|^\alpha \, \e^{\ds  i (\sgn \kappa)\theta\pi/2}
\,,\q
 0<\alpha  \le 2\,, \q
 |\theta| \le  \,\hbox{min}\, \{\alpha ,2-\alpha \}\,.
\eqno(2.3)$$
 We note that the allowed region for the 
parameters $\alpha $ and $\theta$
turns out to be
 a {diamond} in the plane $\{\alpha \,,\, \theta\}$
with vertices in the points
$(0,0)\,, \,(1,1)\,, \, (1,-1) \,,\,(2,0)\,,$
that we call the {\it Feller-Takayasu diamond},
see Fig. 1.
\begin{figure}
\centering
\includegraphics[width=0.49\textwidth]{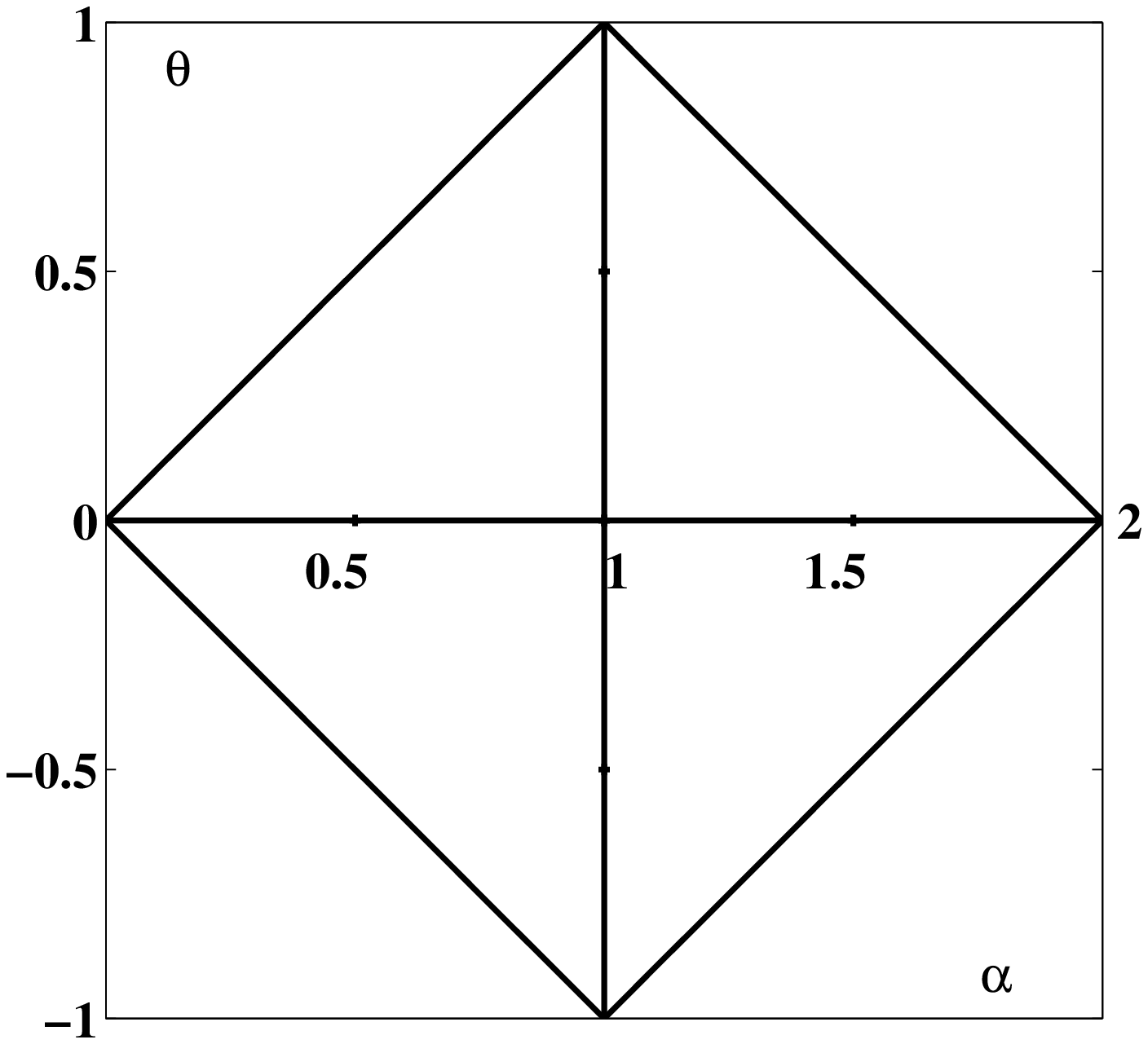}
\vskip 0.25truecm
\cen{{\bf Fig. 1}The Feller-Takayasu diamond}
\end{figure}
Thus, we recognize that the {\it Riesz-Feller}
derivative is required to be
the pseudo-differential operator\footnote{
Let us recall that
a generic pseudo-differential operator $A$,
acting with respect to the variable $x \in \RR\,,$
is defined through its Fourier representation, namely
  $    \int_{-\infty}^{+\infty}
  \e ^{\, i\kappa x} \,  A \,[ f(x)] \, dx =
 \widehat A(\kappa )\, \widehat f (\kappa )\,,	$
  where
$\widehat A(\kappa)\,$ is referred to as  symbol of $A\,,$
   given as
 $ \widehat A (\kappa ) = \l( A\, \e^{\, -i\kappa x}\r)\,
  \e^{\, +i\kappa x}\,. $}
 whose symbol
$- \psi_\alpha ^\theta(\kappa ) $
is the logarithm of the characteristic function of a
general {\it L\'evy strictly stable} probability density
with {\it index of stability} $\alpha $ and asymmetry parameter
$\theta$ (improperly called {\it skewness})
according to  Feller's parameterization \cite{Feller 52,Feller 2},
as revisited
by Gorenflo and Mainardi,
see \cite{GorMai FCAA98,GorMai ZAA99,GorMai CHEMNITZ01}.

The operator defined by (2.2)-(2.3) has been referred to as
the {\it Riesz-Feller} fractional derivative since it is obtained
as the left inverse  of a fractional integral originally
introduced (for $\theta =0$ and $\alpha \ne 1$) by Marcel Riesz in the late
1940's, known as the {\it Riesz potential}, and  then
generalized (for $\theta \ne 0$) by
William Feller in 1952,
see \cite{Feller 52}, \cite{SKM 93}.

For more details on L\'evy stable
densities we refer the reader to  specialistic treatises,
as Feller \cite{Feller 2},  Zolotarev \cite{Zolotarev 86},
   Samorodnitsky and Taqqu \cite{Taqqu 94},
 Janicki and  Weron   \cite{Janicki 94},
 Sato\cite{Sato 99}, Uchaikin and Zolotarev\cite{UchaikinZolotarev 99},
where different notations are adopted.
We like to refer also to the  
1986 paper by Schneider \cite{Schneider LNP86},
where he first provided  the   Fox $H$-function
representation of the  stable distributions (with $\alpha  \ne 1$)
and	to the 1990 book by Takayasu \cite{Takayasu FRACTALS}
where he first gave the diamond  representation in the plane
$\{\alpha ,\theta\}$.


For $\theta=0$	we have a symmetric operator
with respect to $x\,,$ which can be interpreted as
$$
  _xD_0^\alpha = - \l ( - {d^2\over dx^2}  \r) ^{\alpha /2}\,,
\eqno(2.4) $$
as can be formally deduced by writing
$- |\kappa| ^\alpha = - (\kappa^2)^{\alpha /2} \,.   $
 We thus recognize that the  operator $D_0^\alpha $ is related
to a power of the positive definitive operator
  $\, - _xD^2= -{d^2\over dx^2}$
and  must not be confused with a
power  of the first order differential operator
  $\,_xD= {d\over dx}$
for which the symbol is $-i\kappa \,. $
An alternative illuminating  notation for the symmetric
fractional derivative	is due to Zaslavsky, see
\eg \cite{SaichevZaslavsky 97}, and reads
$$ _xD_0^\alpha =     {d^\alpha  \over d |x|^\alpha}\,.
\eqno(2.5) $$

In its regularized form valid for $0<\alpha <2$
the {\it Riesz} space-fractional derivative
admits the explicit representation\footnote{
The representation (2.6), based on a suitable regularization of a
hyper-singular integral,
can be	found in  Samko, Kilbas \& Marichev \cite{SKM 93}
as formula (12.1') and
is more explicit and convenient than other ones available in the
literature,   see \eg
 Saichev \& Zaslavsky \cite{SaichevZaslavsky 97},
 Uchaikin \& Zolotarev \cite{UchaikinZolotarev 99},
in that it is valid in the whole range $ 0<\alpha <2\,. $
Gorenflo and Mainardi  have used it
in \cite{GorMai CHEMNITZ01},   where they have shown that it
holds also in the singular case $\alpha =1\,. $}
$$ _xD_0^\alpha \,f(x)
 = {\Gamma(1+\alpha) \over \pi } \,
 \sin \,\l({\alpha \pi \over 2}\r) \,
 \int_0^\infty
 {f(x+\xi)- 2f(x) + f(x-\xi) \over {\xi}^{1+\alpha}}\, d \xi
 \,.  \eqno(2.6)$$
For $\alpha =1$ the Riesz derivative  is related to the
Hilbert transform
as first noted by Feller in 1952
in his pioneering paper \cite{Feller 52}, resulting in the formula
$$_xD_0^1 f(x)
 =  -{1 \over \pi }\,{d \over dx}\, \int_{-\infty}^{+\infty}
   {f(\xi)\over x-\xi} \,d \xi \,.\eqno(2.7) $$
 For $0<\alpha <2$ and
$|\theta| \le \hbox{min} \,\{\alpha, 2-\alpha \} $
the {\it Riesz-Feller} derivative  reads
$$\q\q \q\q _xD_\theta^\alpha \,f(x)
 = {\Gamma(1+\alpha) \over \pi } \,
 \l\{\sin \,[(\alpha+\theta) \pi/2] \,
 \int_0^\infty
 {f(x+\xi)- f(x)  \over {\xi}^{1+\alpha}}\, d \xi \r.
\q\q\q\q \eqno(2.8)$$
$$ + \l.\sin \,[(\alpha-\theta) \pi/2] \,
 \int_0^\infty
 {f(x-\xi)- f(x)  \over {\xi}^{1+\alpha}}\, d \xi \r\}\,.$$
For $\alpha =1$ we obtain  the composite formula
$$   _xD_\theta ^1 \, f (x) =
\l[ \cos (\theta \pi/2) \,_xD_0^1 + \,\sin (\theta \pi/2) \, _xD \r]\,
    f(x)  \,,\eqno(2.9)$$
 that in the extremal cases $\theta = \pm 1$ reduces to
$$ _xD_{\pm 1}^1 \, =\, \pm \, _xD = \,\pm \, {d\over dx}\,.
 \eqno(2.10)$$

\noindent
{\it The Laplace transform and the Caputo fractional derivative}
\vskip 0.25truecm
\noindent
Now we present an
introduction to the {\it Caputo} fractional derivative starting from its
representation in the Laplace transform domain and
contrasting it to  
the standard {\it Riemann-Liouville} fractional derivative.

Let
$$ \widetilde f(s) =
{\cal L} \l\{ f(t);s\r\}
 = \int_0^{\infty} \e^{\ds \, -st}\, f(t)\, dt\,, \q
\Re\,(s) > a_f\,,\eqno(2.11a) $$
be the Laplace transform of   a 
function $f(t) \in \L^c(0,T)\,,\, \forall T>0\, $ and let
$$ f(t) =
  {\cal L}^{-1} \l\{ \widetilde f(s);t\r\}
 = \rec{2\pi i}\, \int_{\gamma -i\infty} ^{\gamma + i\infty}
\e^{\ds \, st}\, \widetilde f(s) \, ds\,, \q
\Re\,(s) = \gamma > a_f\,,\eqno(2.11b) $$
with $t>0\,,$
be the inverse Laplace transform\footnote{
A sufficient condition of the existence of the Laplace transform
is that the original function is of exponential
order as $t \to \infty\,. $  This means that some constant
$a_f $	exists such that the product
$ \e^{-a _f t}\,  |f(t)|$ is bounded for all $t$ greater
than some $T\,. $
Then $\widetilde f(s)$	exists and
is analytic in the half plane $\Re (s) > a_f\,. $
If $f(t)$ is piecewise differentiable, then the formula (2.11b)
holds true at all points where $f(t)$ is continuous and
the (complex) integral in it must be understood in the sense of the Cauchy
principal value.}
For a sufficiently well-behaved function $f(t)$ we
 define the {\it Caputo} time-fractional derivative  of
order $\beta  $ ($m-1 <\beta \le m\,,$ $\, m\in \NN$)
through
$$ {\cal L} \l\{ _tD_*^\beta \,f(t) ;s\r\} =
      s^\beta \,  \widetilde f(s)
   -\sum_{k=0}^{m-1}	s^{\beta  -1-k}\, f^{(k)}(0^+) \,,
  \q m-1<\beta	\le m \,. \eqno(2.12)$$
This leads to define, see \eg \cite{Caputo 69}, \cite{GorMai CISM},
$$
    _tD_*^\beta \,f(t) :=
\cases{
    {\ds \rec{\Gamma(m-\beta )}}\,{\ds\int_0^t
 {\ds {f^{(m)}(\tau)\, d\tau \over (t-\tau )^{\beta  +1-m}}}} \,,
  & $\; m-1<\beta  <m\,, $\cr\cr
     {\ds {d^m\over dt^m}} f(t)\,,
    & $\; \beta  =m\,. $\cr\cr }
   \eqno(2.13) $$
The operator defined by (2.12)-(2.13) has been referred to as
the {\it Caputo} fractional derivative since it
was introduced by Caputo in the late 1960's
for modelling the energy dissipation in some anelastic materials
with memory, see  \cite{Caputo 67,Caputo 69}.
A former review
of the theoretical aspects of this derivative
with applications in visco\-elasticity
was given  in 1971 by {Caputo and Mainardi}
\cite{CaputoMaina 71},
with special emphasis to the
long-memory  effects.

The reader should observe that the {\it Caputo} fractional derivative
differs from the usual {\it Riemann-Liouville} fractional derivative
which, defined as the left inverse of the Riemann-Liouville fractional
integral, is  here denoted as
$\, _tD^\beta \,f(t)\,. $
We have,  see \eg \cite{SKM 93},
$$
 _tD^\beta  f(t) :=
\! \cases{
  {\ds {d^m\over dt^m}}\left[
  {\ds \rec{\Gamma(m-\beta )}\,\int_0^t
    {f(\tau)\,d\tau  \over (t-\tau )^{\beta  +1-m}} }\right],
 & $ m-1 <\beta  < m,$ \cr\cr
     {\ds {d^m\over dt^m}} f(t)\,,
    & $ \beta  =m\,. $}
   \eqno(2.14) $$
When the order is not integer,	Gorenflo and Mainardi
have shown the following relationships between the two fractional
derivatives  (when both of them exist), see \eg \cite{GorMai CISM},
$$ _tD_*^\beta	\,f(t)	\, = \, _tD^\beta  \,\l[ f(t) -
  \sum_{k=0}^{m-1} f^{(k)}(0^+)\,{t^k\over k!} \r]\,,\q
    m-1 <\beta <m\,,\eqno(2.15)$$
or
$$ _tD_*^\beta	\,f(t)	\, = \, _tD^\beta  \, f(t) -
    \sum_{k=0}^{m-1}	f^{(k)}(0^+) \,
{t^{k-\beta }\over \Gamma(k-\beta+1)}\,,
\q	 m-1 <\beta <m\,.  \eqno(2.16)$$
The {\it Caputo} fractional derivative is of course more restrictive than
the {\it Riemann-Liouville} fractional derivative
in that   the derivative of order $m$  is required to exist
and be absolutely Laplace transformable.
Whenever we use the operator   $\,_tD_*^\beta\,,$   we (tacitly) assume
that	  this condition is met.

The {\it Caputo} fractional derivative\footnote{
According to Samko, Kilbas \& Marichev\cite{SKM 93} and
 Butzer \& Westphal\cite{ButzerWestphal 00}
 the "regularized" fractional  derivative was
 considered by Liouville  himself (but then disregarded).},
 practically ignored in the  mathematical treatises,
represents a sort of regularization in the time origin for the
{\it Riemann-Liouville} fractional derivative.
Recently, it has been extensively investigated by Gorenflo \& Mainardi
\cite{GorMai CISM} and by Podlubny \cite{Podlubny 99}
in view of its major utility in treating physical and engineering problems
which require standard initial conditions.
Several applications have been treated by Caputo himself
up to nowadays, see \eg
\cite{Caputo 00,Caputo FCAA} and references therein.
 
We point out that the  {\it Caputo} fractional derivative
  satisfies the  relevant property
of being zero when applied to a constant, and, in general,
to any power function  of non-negative integer degree less than $m\,,$
if its order $\beta $ is such that $m-1<\beta <m\,. $
Furthermore,  since
$$ _t D^{\beta }\, t^{\gamma}=
   {\Gamma(\gamma +1)\over\Gamma(\gamma +1-\beta )}\,
     t^{\gamma-\beta }\,,
 \q \beta  >0\,,
  \q \gamma >-1\,, \q t>0\,,
\eqno (2.17)
$$
we note that
$$   _t D^\beta  \, f(t) \,=\, _tD^\beta   \, g(t)
   \,  \Longleftrightarrow  \,
  f(t) = g(t) + \sum_{j=1}^m c_j\, t^{\beta -j} \,,
    \eqno(2.18) $$
whereas, using also  (2.15) or (2.16),
$$    _tD_* ^\beta  \, f(t) \,=\,  _tD_*^\beta	 \, g(t)
   \,  \Longleftrightarrow  \,
  f(t) = g(t) +  \sum_{j=1}^m c_j\, t^{m-j} \,.
    \eqno(2.19) $$
In these formulae the coefficients $c_j$ are arbitrary constants.
We also note the different behaviour  of $\, _tD_*^{\beta}$ at the
end points of the interval $(m-1,m)\,,$
$$ \lim_{\beta \to (m-1)^+}\,_tD_*^{\beta} \,f(t)=
  f^{(m-1)}(t) - f^{(m-1)} (0^+)\,, \;
  \lim_{\beta \to m^-}\,
 _tD_*^\beta  \,f(t)=  f^{(m)}(t)\,.
\eqno(2.20) $$
The last limit can be formally obtained
by recalling the formal representation of the $m$-th derivative of the
Dirac  function,
$\delta^{(m)} (t) = t^{-m-1}/\Gamma(-m)\,,$ $\, t \ge 0\,,$
see  \cite{Gel'fand 64}. 
As a consequence of (2.20),
with respect to the order,
the {\it Caputo} derivative is an operator left-continuous
 at any positive integer.


\vskip 0.5truecm
\noindent
{\it The Mellin transform}
\vskip 0.5truecm

Let
$$
   {\cal M} \, \{ f(r ); s\} = f^*(s)=
   \int_0^{+\infty} f(r)\,
 r^{s-1}\,  dr,  \q  \gamma_1< \Re\, (s) <\gamma_2 \eqno(2.21a)
$$
be the Mellin transform of a sufficiently well-behaved function
$f(r)\,,$ and let
$$
 {\cal M}^{-1}\, \{  f^*(s ); r \} =f(r)=
{1\over 2\pi i}\int_{\gamma -i
\infty}^{\gamma +i\infty} f^*(s)\, r^{-s} \,ds\eqno(2.21b)
$$
 be the inverse Mellin transform\footnote{
For the existence of the Mellin transform and the validity of the
inversion formula  we need to
recall the  following theorems, $^{(6a)}$ and $^{(6b)}$,
adapted from Marichev's treatise \cite{Marichev 83}, see THEOREMS 11, 12,
at page 39.    \\
\noindent
$^{(6a)}$ Let $f(r) \in L^c(\epsilon,E)\,,\,0<\epsilon <E<\infty\,,$ be
continuous   in the intervals $(0,\epsilon]\,,\,[E,\infty)\,,$
and let $\,|f(r) | \le M\, r^{-{\gamma_1}}$ for $0<r<\epsilon \,,$
$\,|f(r) | \le M\, r^{-{\gamma_2}}$ for $r>E\,,$ where $M$ is a constant.
Then for the existence of a strip in the $s$-plane in which
$f(r)\, r^{s-1}$ belongs to $L^c(0,\infty)$ it is
sufficient that $\gamma_1<\gamma _2\,. $
When this condition holds, the Mellin transform $f^*(s)$ exists and
is analytic in the vertical strip
$\gamma _1 <\gamma =\Re (s) <\gamma _2\,.   $
\\  \noindent
$^{(6b)}$ If $f(t)$ is piecewise differentiable,
and $f(r)\, r^{\gamma -1}   \in L^c(0, \infty)\,,$
 then the formula (2.21b)
holds true at all points where $f(r)$ is continuous and
the (complex) integral in it must be understood in the sense of the
Cauchy principal value.},
where $\ r>0\,,$ $\, \gamma = \Re\,(s) \,,$
$\, \gamma_1< \gamma <\gamma_2\,.$

We refer to  specialized treatises and/or
handbooks, see \eg \cite{Erdelyi TIT}, \cite{Marichev 83},
\cite{Prudnikov IS3}, for more details and tables on the Mellin transform.
Here, for our convenience we  recall the main rules
that are  useful to adapt the formulae from the handbooks
and, meantime,	are relevant in the following.

Denoting by   $\stackrel{{\cal M}} {\leftrightarrow}$
the juxtaposition of a function $f(r)$ with its
Mellin transform $f^*(s)\,,$ the main rules are:
$$ f(a r )
\,\stackrel{{\cal M}}{\leftrightarrow}\,
a^{-s}\, f^*(s)\,, \q a>0\,, \eqno(2.22)$$
$$ r ^a\, f(r )
\,\stackrel{{\cal M}}{\leftrightarrow}\,
 f^*(s+a)\,, \eqno(2.23)$$
$$ f(r ^p)
\,\stackrel{{\cal M}}{\leftrightarrow}\,
{1 \over |p|}\, f^*(s/p)\,,  \q p \ne 0\, ,
\eqno(2.24)$$
$$ h(r ) = \int\limits_0^\infty
 \rec{\rho }\, f(\rho)\,g(r/\rho  )\,{d\rho  }
\,\stackrel{{\cal M}}{\leftrightarrow}\,
 h^*(s) = f^*(s)\,g^*(s) \,.
\eqno(2.25)$$
The  Mellin convolution formula (2.25)
is useful in treating  integrals of Fourier type
for $ x= |x| >0\,:$
$$I_c(x)=\rec{\pi}\,\int_0^\infty f(\kappa)\,\cos\,({\kappa \,x})\,d\kappa
\,,\eqno(2.26)$$
$$I_s(x)=\rec{\pi}\,\int_0^\infty f(\kappa)\,\sin\,({\kappa \,x})\,d\kappa
\,, \eqno(2.27)$$
when the Mellin transform $f^*(s)$ of	$f(\kappa )$ is known.
In fact we recognize that the  integrals $I_c(x)$
and $I_s(x)$ can
be interpreted as Mellin convolutions (2.25) between  $f(\kappa )$ and the
functions $g_c(\kappa )\,,\, g_s(\kappa )\,,$ respectively,
with $r = 1/|x|\,,\, \rho =\kappa \,, $
where  
$$ g_c(\kappa ):= \rec{\pi\, |x|\, \kappa } \cos \l( \rec{\kappa}\r)
       \,\stackrel{{\cal M}}{\leftrightarrow} \,
{\Gamma(1-s)\over \pi \, |x|}
  \sin \l( {\pi s\over 2}\r)
 :=g_c^*(s)   ,
\; 0< \Re (s) <1 \, ,\eqno(2.28) $$
$$ g_s(\kappa ):= \rec{\pi\, |x|\,\kappa}\, \sin \l(\rec{\kappa}\r)
       \,\stackrel{{\cal M}}{\leftrightarrow} \,
{\Gamma(1-s) \over \pi\, |x|}
  \cos \l( {\pi s\over 2}\r)
 :=g_s^*(s)  ,
\; 0< \Re (s) <2 \,.\eqno(2.29)  $$
The Mellin transform pairs  (2.28)-(2.29) have been adapted from the tables
in \cite{Marichev 83} by using (2.22)-(2.24) and the duplication
and reflection formulae for the Gamma function.
Finally, the inverse Mellin transform representation (2.21b) provides the
required integrals as
$$I_c(x)=\rec{{\pi}\, x}\, \rec{2\pi i}\,
  \int_{\gamma -i\infty}^{\gamma +i\infty}
   \!\! f^*(s)\,
\Gamma(1-s)\,\sin\l({\pi \,s\over 2}\r) \, x^{s}\, ds\,,
 \; x>0\,,\; 0<\gamma <1\,,
\eqno(2.30)
$$
$$I_s(x)=\rec{{\pi}\, x}\, \rec{2\pi i}\,
  \int_{\gamma -i\infty}^{\gamma +i\infty}
  \!\! f^*(s)\,
\Gamma(1-s)\,\cos\l({\pi \,s\over 2}\r) x^{s}\, ds\,,
 \; x>0\,,\; 0<\gamma <2\,.
  \eqno(2.31)
$$

\section*{3. Scaling and similarity properties of the Green function}

For the equation (1.1) we consider the
Cauchy problem
$$
u(x,0)=   \varphi(x)\,,\q x\in \re\,,\qq
 u(\pm\infty,t)=   0\,,\q t>0\,, \eqno(3.1) $$
where $\varphi(x)\in L^c(\RR)$ is a sufficiently well-behaved function.
If $1 < \beta \le 2$ we add to (3.1)  the condition
$u_t(x,0)=   0$, where $u_t(x,t) =    {\partial\over \partial t} u(x,t)$.

By solution
of the Cauchy	problem for the
equation (1.1) we mean a function $u_{\alpha, \beta}^\theta(x,t)$
 which satisfies the conditions (3.1).
By the Green function (or fundamental solution)
of the Cauchy problem we mean the (generalized)
function $G_{\alpha, \beta}^\theta(x,t)$
which, being the formal solution of (1.1) corresponding
to $\varphi(x) = \delta (x)\, $   (the Dirac delta function),
allows us to represent the solution of the Cauchy problem by
the integral formula
$$ u_{\alpha, \beta}^\theta(x,t) =
\int_{-\infty}^{+\infty} G_{\alpha, \beta}^\theta(\xi ,t)\,
 \varphi (x-\xi) \, d\xi \,. \eqno(3.2)$$
It is straightforward to derive  from (1.1) and (3.1) the
Fourier-Laplace transform of the Green function by taking
into account the Fourier transform for the {\it Riesz-Feller}
space-fractional derivative, see (2.2),
and the Laplace transform for the {\it Caputo}
time-fractional derivative,
see (2.12).
We have (in an obvious notation):
 $$- \psi_\alpha^\theta(\kappa ) \,\Gks  \,= \,
 s^\beta \,\Gks - s^{\beta -1}	  \,,
\eqno(3.3)  $$
where
$$ \psi_\alpha^\theta (\kappa) :=
 |\kappa|^\alpha \, \e^{\ds  i (\sgn \kappa)\theta\pi/2}
  = \overline{ \psi_\alpha^\theta (-\kappa)}
 =  \psi_\alpha^{-\theta} (-\kappa)
  \,. \eqno(3.4) $$
We therefore obtain
$$
  \Gks = {s^{\beta -1} \over
 s^\beta + \psi_\alpha^\theta (\kappa)} \,. \eqno(3.5)$$
By using the known scaling rules
for the Fourier  and Laplace  transforms,
$$   f(ax) \,\stackrel{{\cal F}} {\leftrightarrow} \,
  a^{-1}\, \widehat   f(\kappa/a)\,,\q	 a> 0\,,
 \qq   f(bt) \,\stackrel{{\cal L}} {\leftrightarrow}\,
 b^{-1}\, \widetilde f(s/b)\,, \q b> 0\,,\eqno(3.6)$$
we  infer directly from (3.5) (thus without inverting the
two transforms)   the  following {\it scaling property}
of the Green function,
$$    \G(ax\,,\,bt)
= b^{-\gamma} \G (ax/b^\gamma\,,\,t)\,,\q \gamma=\beta/\alpha\,.
  \eqno(3.7)  $$
Consequently, introducing the {\it similarity variable} $x/t^\gamma\,,$
we  can write

$$ \G(x,t)  =
    t^{-\gamma}\,\K (x/t^\gamma)\,,\q \gamma =\beta /\alpha
\,,   \eqno(3.8)     $$
where the one-variable function $\K$ is to be determined
as indicated below.


Let us first invert the Laplace transform in (3.5).
To this purpose we recall
the Laplace transform pair,
$$ E_\beta (ct^\beta)\,
  \,\stackrel{{\cal L}} {\leftrightarrow}\,
 {s^{\beta-1} \over  s^\beta - c}\,, \q \Re \,(s) > |c|^{1/\beta}\,,
\eqno(3.9)$$
with $ c\in \ce\,, \; 0<\beta \le 2\,, $
where $E_{\beta}$ denotes the entire transcendental function, known as
the Mittag-Leffler function of order $\beta\,,$
defined in the complex plane by the power series
$$ E_\beta (z) :=
    \sum_{n=0}^{\infty}\,
   {z^{n}\over\Gamma(\beta\,n+1)}\,, \q \beta >0\,, \q z \in \CC\,.
 \eqno	(3.10)$$
Then, by comparing (3.5) with (3.9), we  obtain
the Fourier transform of the Green function as
  $$   \Gkt =
     E_\beta \l[- \psi_\alpha ^\theta(\kappa ) \,t^\beta \r]\,,
 \q \kappa \in \re\,, \q t\ge 0\,.
  \eqno(3.11)$$
For detailed information on the Mittag-Leffler-type functions
and their Laplace transforms the reader  may  consult \eg
\cite{Erdelyi HTF}, \cite{Djrbashian 66},
\cite{SKM 93}, \cite{Kiryakova 94},  \cite{GoLuRo 97},
\cite{GorMai CISM}, \cite{Podlubny 99}.
We note for later use that the Mittag-Leffler function (3.10) admits
a Mellin transform type representation, originally due to
Barnes \cite{Barnes 06}
(see  also \cite{Marichev 83} p. 118 (7.79)), as
$$   E_\beta (z)
= \rec{2 \pi i} \,
\int_{L_{-\infty}}  \!\!
{ \Gamma(s)\, \Gamma (1-s) \over \Gamma (1-\beta  s)} \,
 (-z) ^{-s} \, ds  \,,
 \eqno(3.12)$$
where	the integration is over a left-hand loop
$L_{-\infty}$ drawn round all the left-hand poles
$s =0, -1, -2, \dots$ of the integrand in a
positive direction. Indeed, by the residue theorem
it is not difficult to write the integral in (3.12)
as the power  series in (3.10).
 As a matter of fact,  from (3.12) we can deduce  the following Mellin
transform pair (see also \cite{Marichev 83}, p. 300)
which will be used to get the Fourier antitransform of (3.11),
$$	   E_\beta (-r)
     \,\stackrel{{\cal M}}{\leftrightarrow} \,
      {\Gamma (s)\, \Gamma(1-s) \over \Gamma (1-\beta  s)} \,,
\q r>0\,,\q 0 <\beta \le 2\,,\q 0< \Re (s) <1\,. \eqno(3.13)$$

In view of the symmetry properties of
$\psi_\alpha ^\theta(\kappa )\,,$ see	 (3.4),
and of the Mittag-Leffler function
\ie $E_\beta (z) = \overline{E_\beta (\bar z)}\,,$ $\; z \in \ce\,,$
we have
$$ \Gkt = \overline{\widehat{\G}(-\kappa,t)}
 = \widehat{G_{\alpha,\beta }^{-\theta}} (-\kappa ,t)
 \,.\eqno(3.14)$$
Furthermore,
we easily recognize  from (3.10)-(3.11),
$$ \widehat{\G}(0,t) = E_\beta(0) =1\,,\q t\ge 0\,.  \eqno(3.15)   $$
Provided that $\Gxt$ does exist
as inverse Fourier transform of (3.11),
equations (3.14)-(3.15) ensure	that $\Gxt$ is real and
normalized, \ie
 $\int_{-\infty} ^{+\infty}  \!\! \Gxt \,dx =1\,. $

The inversion of the Fourier transform (at most as a improper integral)
requires (in view of the Riemann-Lebesgue lemma) that for $ t\ge 0\,,$
$$\l\vert E_\beta \l[- \psi_\alpha ^\theta(\kappa ) \,t^\beta \r]\r\vert
 \to 0 \q \hbox{as} \q |\kappa |\to \infty\,. \eqno(3.16)$$
Taking into account the growth properties
of the Mittag-Leffler function, see \eg Gorenflo, Luchko and
Rogosin \cite{GoLuRo 97},
this means that the LHS of (3.16) is required to be bounded
for $ \kappa \in \re\,$ and $ t\ge 0\,.$
It turns out that for this we must require
$$ |\theta| \le 2-\beta\,.  \eqno(3.17)$$
This means that for $1<\beta < 2$ the allowed $\{\alpha \,,\, \theta\}$
region could be no longer the {\it Feller-Takayasu diamond}
$|\theta| \le  \,\hbox{min}\, \{\alpha\, ,\,2-\alpha\}\,;$ in fact,
in the cases $0<\alpha <1<\beta <2$ with $\alpha +\beta >2$ and
$ 1<\alpha <\beta <2$ we must cut-off the upper and
lower corners of the diamond with the lines $\theta = \pm (2-\beta )\,. $


As far as the determination of the one-variable function $\K(x)$
is concerned, see (3.8),  we note the {\it symmetry relation} 
$$ K_{\alpha ,\beta}^\theta(-x) = K_{\alpha ,\beta}^{-\theta}(x)
\,. \eqno(3.18)$$
As a consequence, we can restrict our attention to $x>0\,,$ and obtain
$$ \K (x) =	 \rec{2\pi}\,
\int_{-\infty}^{+\infty} \e^{\,\ds -i \kappa x} \,
  E_\beta \l[- \psi_\alpha ^\theta(\kappa ) \r] \, d\kappa
 \,= \, _c\K (x) + _s\K (x)\,,
\eqno(3.19) $$
$$  _c\K (x) = \rec{\pi}\,\int_{0}^{\infty}
\cos\,(\kappa  x)  \, \Re \Big[
  E_\beta \Big(-\kappa^\alpha\,\e^{\,\ds i\theta \pi/2}\Big) \Big]\,
d\kappa  \,,\eqno(3.20) $$
$$  _s\K (x) = \rec{\pi}\,\int_{0}^{\infty}
  \sin\,(\kappa  x)\, \Im \Big[
  E_\beta \Big(-\kappa^\alpha\,\e^{\,\ds i\theta \pi/2}\Big) \Big]
  \,d\kappa \,. \eqno(3.21)$$
From (3.19) we can easily obtain the value attained  by
 $K_{\alpha   ,\beta}^\theta(x)$
in $x=0\,.$  To this purpose we extend the argument in
\cite{GoIsLu 00}
writing
$$
\begin{array}{ll}
\K(0)  & =
{\ds {1\over \pi} \Re \l[ \int_{0}^{\infty}
\,E_{\beta}
\l(-\kappa^{\ds \,\alpha}\, \e^{\,\ds i\theta \pi/2} \r)
 \, d\kappa \r]} \\
& =	
{\ds \cos \l({\ds {\theta \pi\over 2\alpha }}\r)\,
{1 \over \pi \alpha } \,
 \int_{0}^\infty E_\beta(-r)\,r^{\ds {1/ \alpha}-1}\,dr}\,.
\end{array}
\eqno(3.22)
    $$
Thus, the last integral above is
the Mellin
transform of the Mittag-Leffler function at the point
$s=   {1/ \alpha}\,; $
it turns out to be  convergent under the conditions
$\beta=   1\,$ with $\alpha >0\,, $
and $0<\beta\le 2\,,\, \beta\ne   1$ with $\alpha >1\,. $
For the finite value  of $\K(x)$ at
 $x=0$	  we thus use (3.13)   obtaining
 $$\K(0) =   \cases{
   {\ds \rec{\pi \alpha}}
 {\ds{\Gamma({1/\alpha})
 \Gamma(1-{1/ \alpha})\over\Gamma(1-{\beta/\alpha}) } }
      \, \cos \l({\ds {\theta \pi\over 2\alpha }}\r),
   &  $1<\alpha \le 2,\; \beta \ne 1, $ \cr\cr
 {\ds \rec{\pi \alpha }}\,  {\Gamma({1/ \alpha} )}
     \, \cos \l({\ds {\theta \pi\over 2\alpha }}\r),
   &  $0<\alpha \le 2,\;\beta =1. $}
\eqno(3.23) $$
We note that $\K(0)$ is non negative except for
 $1<\alpha < \beta \le	2\,.$
 \vfill\eject 

\section*{4. Particular cases  for the Green function}

In this Section we are going to consider
the important particular cases of our space-time
fractional diffusion equation, \ie

\vskip 0.1truecm
\noindent
$\{\alpha =2\,, \,\beta =1\}$ ({\it standard diffusion}),
\par\noindent
$\{0 < \alpha \le 2\,, \,\beta =1 \}$ ({\it space fractional diffusion}),
\par\noindent
$\{ \alpha =2\,, \,0<\beta\le 2\,,\, \beta \ne 1\}$ ({\it time fractional
	  diffusion}),
\par\noindent
$\{ 0< \alpha =\beta \le 2\}$ ({\it neutral fractional
	  diffusion}),

It is well known that for the {\it standard diffusion}
$\{\alpha =2\,, \,\beta =1\}$
the Green function is the Gaussian  $pdf$
$$ G_{2,1}^0 (x,t) = t^{-1/2}\, {1\over 2\sqrt{\pi}} \,\exp [-x^2/(4t)]
   \,,
\q -\infty<x<+\infty\,, \q t\ge 0\,,
 \eqno(4.1)$$
with similarity variable $x/t^{1/2}\,,$
that evolves in time with
moments (of even order)
$$ \mu_{2n}(t) := \int_{-\infty}^{+\infty} \!\!\! x^{2n}\, G_{2,1}^0(x,t)
\,dx =	  {(2n)!\over n!}\,  t^{n}\,,
\q n=0\,, \,1\,, \,2\,, \, \dots\,,  \q t\ge 0\,.
\eqno(4.2)$$
The variance $\sigma^2 := \mu_2(t) = 2t $ is thus  proportional to the
first power of time,
according to  the Einstein diffusion law.

In Fig. 2 we report the plots of the Gaussian.
$ K_{2,1}^0 (x) = 1/(2\sqrt \pi) \, \exp (-x^2/4)$ in the interval
$-5\le x \le 5$ by adopting for the ordinate a linear scale (left) and a
logarithmic scale (right). We recognize that the logarithmic scale
is to be preferred to point out the tails. At the end of
this paper we shall exhibit a number of lin-log plots of the
fundamental solution for different values of the parameters $\alpha,
 \theta, \beta ,$   that the reader can compare with the
corresponding plot of the Gaussian, and emphasize the different
behaviour of the tails.


\begin{figure}
\centering
\includegraphics[width=0.40\textwidth]{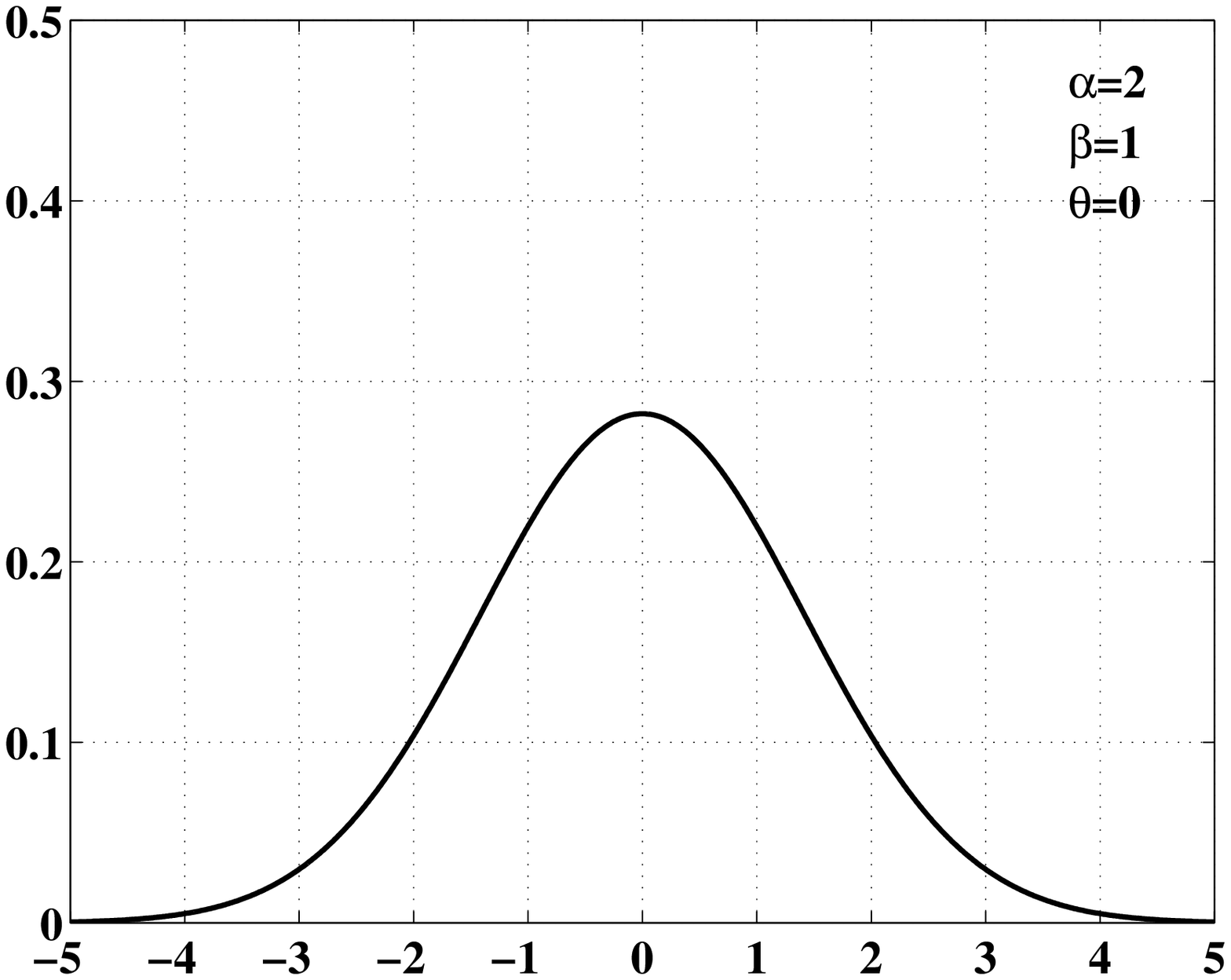}
\includegraphics[width=0.40\textwidth]{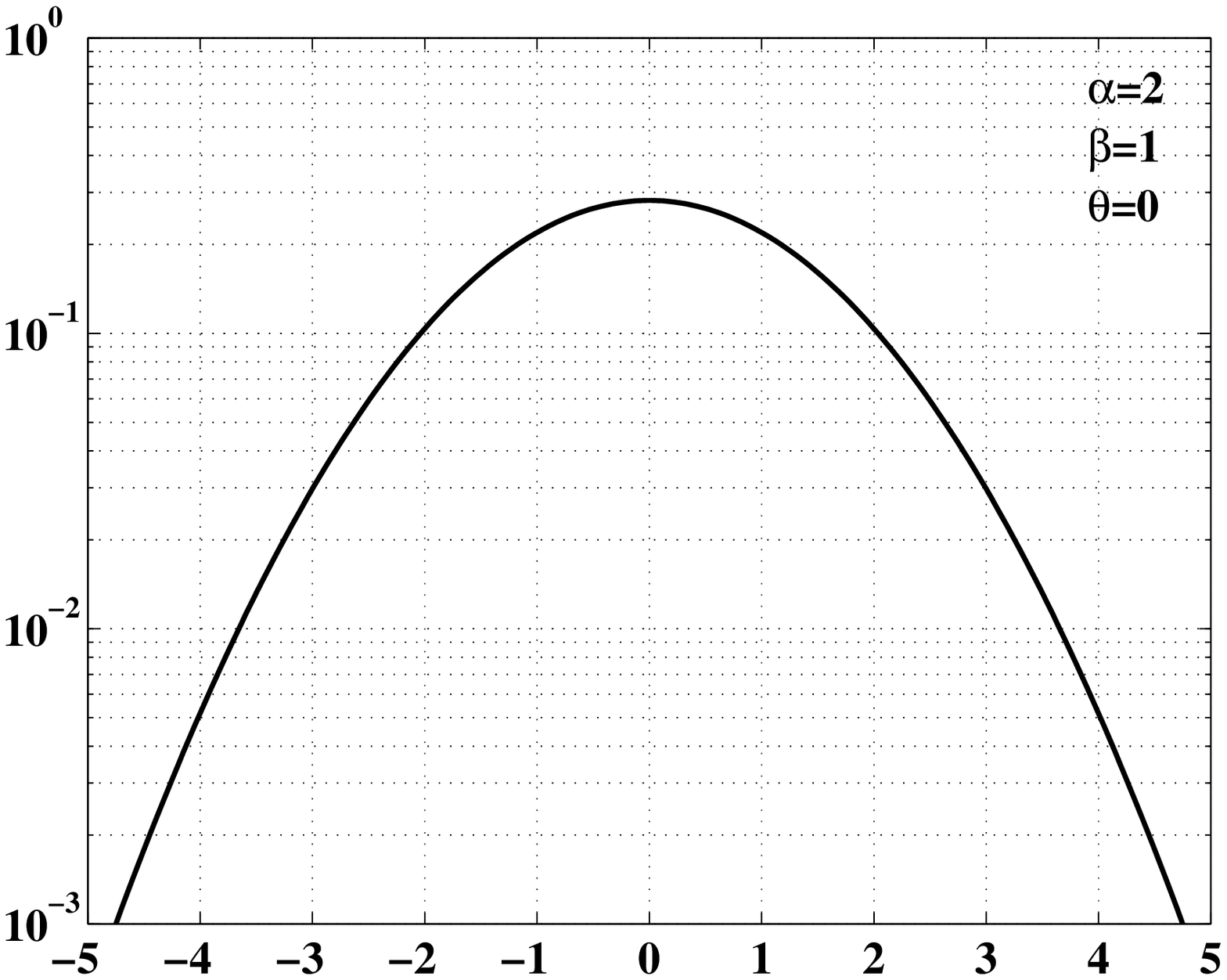}
\vskip 0.25truecm
\cen{{\bf Fig. 2} The Gaussian $pdf$.}
\end{figure}

\noindent \underbar{{\it The space-fractional diffusion}}
\vskip 0.25 truecm
\noindent
Let us now consider
$\{0<\alpha \le 2\,,\,\beta =1\} $
({\it space fractional diffusion}
including  standard diffusion for $\alpha =2$).
In this case, reducing	the Mittag-Leffler function in (3.11)
to the	exponential function,  we recover
the characteristic function
of the class of L\'evy  strictly stable densities
according to the Feller parameterization, see (2.2), (2.3)
and (3.4). In fact, denoting this class by
$\{L_\alpha ^\theta(x)\}\,,$  we have
$$ \widehat{L_\alpha ^\theta}(\kappa ) =
	 \e^{\ds  -\psi_\alpha^\theta(\kappa)}\,,
\q \hbox{and} \q
\widehat{G_{\alpha,1}^\theta}(\kappa ,t) =
    \e^{\ds\, -t \psi_\alpha^\theta(\kappa)}\,,
 \eqno(4.3)
$$
with $ \psi_\alpha ^\theta(\kappa )$ given by (3.4)
and $\alpha $ and $\theta$ restricted in the Feller-Takayasu diamond,
see (2.3).
Then the Green function of the space-fractional diffusion equation
can be interpreted as a L\'evy strictly stable $pdf$, evolving
in time, according to
$$
 G_{\alpha,1}^\theta(x,t) = t^{-1/\alpha }\,
L_\alpha ^\theta(x/t^{1/\alpha })\,,
\q -\infty<x<+\infty\,, \q t\ge 0
\,. \eqno(4.4) $$

A stable $pdf$ with extremal value of
the skewness parameter	is called {\it extremal}.
One can prove that all the extremal stable $pdf$s' with
$0<\alpha  <1$ are one-sided, the support being $\re_0^+$  if
$\theta  =-\alpha \,, $
and  $\re_0^-$	if $\theta  =+\alpha \,. $
The one-sided stable $pdf$'s with support in $\re_0^+$ can be
better characterized by their (spatial) Laplace transforms, which
turn out to be
$$ \widetilde {L_\alpha^{-\alpha}} (s) :=
   \int_0^\infty \!\! \e^{\ds\, -sx}\,L_\alpha^{-\alpha} (x) \,dx
 = \e^{\ds - s^\alpha} \,, \q \Re\, (s) >0\,, \q 0<\alpha <1\,.
\eqno(4.5)$$

The stable densities admit a representation in terms
of elementary functions only in the following particular cases
\par\noindent
$ \alpha =2 \,,\q \theta=0\,,	\q \hbox{\it Gauss}\,:$
$$  \e^{\ds\,-\kappa ^2} \,\stackrel{{\cal F}} {\leftrightarrow}\,
L_2^0 (x)  =
   {1\over 2\sqrt{\pi}} \,\e^{\,\ds -x^2/4}
   \,, \q -\infty < x <+\infty\,;
 \eqno(4.6)$$
\par\noindent
$  \alpha =1/2 \,,\q \theta=-1/2\,, \q \hbox{\it L\'evy-Smirnov}\,:$
$$  \e^{\ds \,-s^{1/2}} \,\stackrel{{\cal L}} {\leftrightarrow}\,
L_{1/2}^{-1/2} (x) =
  {\,x^{-3/2}\over 2\sqrt{\pi}}\, \,\e^{\ds\, - 1/(4x)}\,,
 \q x \ge 0	\,; \eqno(4.7)$$
\par\noindent
$   \alpha =1 \,,\q \theta=0\,, \q \hbox{\it Cauchy}\,:$
$$  \e^{\ds \,-|\kappa |} \,\stackrel{{\cal F}} {\leftrightarrow}\,
L_1^0 (x) = {1\over \pi}\,
   {1\over x^2+1}
   \,, \q -\infty < x <+\infty\,.
 \eqno(4.8)$$
We note that the case $\alpha =1$ can be easily treated
also for $\theta \ne 0$ taking into account elementary
properties  of the Fourier transform;
we have
\par\noindent
$   \alpha =1 \,,\q 0<|\theta|<1\, :$
$$ L_1^\theta (x) =
 \rec{\pi} \, {  \cos (\theta \pi/2)
 \over [x+   \sin (\theta \pi/2)]^2 +[ \cos (\theta \pi/2)]^2 }
 \,, \q -\infty < x <+\infty\,;
 \eqno(4.9)$$
\par\noindent
$   \alpha =1 \,,\q \theta = \pm 1\, :$
$$ L_1^{\pm 1} (x) =  \delta (x \pm 1)\,,
  \q -\infty < x <+\infty\,.
 \eqno(4.10)$$
We  note that in the above singular cases $\alpha =1\,,\;
\theta	=\pm 1\,,$ the corresponding Riesz-Feller derivatives
$_xD_{\pm 1}^1$ reduce to
 $ \pm\,_xD\,, $ see (2.10), so our
fractional diffusion equation (1.1)    degenerates
into   kinematic (\ie first-order) wave equations
and the corresponding Green functions  are
 $G_{1,1}^{\pm 1}(x,t) =   \delta (x \pm t)\,,$
meaning pure drift.

For $0<\alpha <2$ the stable $pdf$'s  exhibit
{\it fat tails} in such a way that their  absolute moment
of  order $\nu $ is finite only if $-1 < \nu <\alpha \,.$
In fact one can show that for non-Gaussian,
not extremal, stable densities
the asymptotic decay of the tails  is
$$ L_\alpha^\theta (x )= O\l(|x|^{-(\alpha +1)}\r)\,, \q
	       x \to \pm \infty\,. \eqno(4.11)$$
For the extremal densities with $\alpha \ne 1$
this is valid only for one
tail, the other being of exponential order.
For $0<\alpha <1$ we have one-sided $pdf$'s:
for $\theta = -\alpha $ the support is $\re^+$
and the $pdf$ tends exponentially to zero
as $x \to 0^+\,;$
for $\theta = +\alpha $ the support is $\re^-$
and the $pdf$ tends exponentially to zero
as $x \to 0^-\,.$
For $1<\alpha <2$ the extremal $pdf$'s are two-sided and exhibit
an exponential left tail  (as $x \to -\infty)$
if $\theta  =+(2-\alpha)\,,  $
or  an exponential right tail  (as $x \to +\infty $)
if $\theta  =-(2-\alpha)\,.$
Consequently, the Gaussian distribution is the unique
 stable distribution with finite variance.
Furthermore,  when $0<\alpha \le 1\,, $
the first absolute moment  is
infinite  so
we should use the  median instead of the non-existent expected value.
However, there is a fundamental property shared by all the stable
$pdf$'s that we like to point out:
for any $\alpha $ the corresponding  stable  $pdf$ is $unimodal$ and
indeed {\it bell-shaped}, \ie
its $n$-th derivative has exactly $n$ zeros,
see Gawronski \cite{Gawronski 84}.

A  general  representation of all stable $pdf$'s
in terms of  higher transcendental functions
has been  achieved only in 1986; it was  Schneider \cite{Schneider LNP86},
who first has proved that the stable $pdf$'s can be
expressed  by means of	{Fox} $H$-functions, see also
Uchaikin \& Zolotarev \cite{UchaikinZolotarev 99}.

We note that already in 1952 Feller \cite{Feller 52} had obtained, by
inverting the Fourier transform  of the characteristic function,
representations of the stable $pdf$'s in terms of convergent
and asymptotic power series. Feller's results have been
revisited (and corrected) by Schneider and
can be summarized as follows. We restrict our attention to $x>0 \,, $
since the evaluations for $x<0$  can be obtained
using the {\it symmetry relation}
$L_\alpha^\theta (-x)=	L_\alpha^{-\theta} (x)\,.$ 

The convergent expansions  are
 \par \noindent
$ 0<\alpha <1\,,\q |\theta| \le \alpha \,:$
 $$L_\alpha^\theta (x) =
{1\over \pi\,x}\,  \sum_{n=1}^{\infty}
   (-x^{-\alpha})^n \, {\Gamma (1+ n\alpha)\over n!}\,
  \sin \l[{ n\pi\over 2}(\theta -\alpha)\r]\,,
  \q x>0\,;  \eqno(4.12)  $$
\par\noindent
$ 1<\alpha \le 2\,, \q |\theta| \le 2-\alpha\, $
$$ L_\alpha^\theta (x)=
{1\over \pi\,x}\,  \sum_{n=1}^{\infty}
   (-x)^{n} \, {\Gamma (1+ n/\alpha)\over n!}\,
  \sin \l[{ n\pi\over 2\alpha }(\theta -\alpha)\r]\,,\q x>0\,.
 \eqno(4.13) $$
From the series in (4.12) and the  symmetry relation
we note that the
extremal stable distributions for $0<\alpha <1$ are
unilateral, that is vanishing for $x>0$ if $\theta =\alpha\,, $
vanishing for $x<0$ if $\theta =-\alpha\,. $
\par
The asymptotic representations are given by
\par \noindent
$ 0<\alpha <1\,,\q -\alpha <\theta \le \alpha \,:$
$$ L_\alpha^\theta (x) \sim
{1\over \pi\,x}\,  \sum_{n=1}^{\infty}
   (-x)^{n} \, {\Gamma (1+ n/\alpha)\over n!}\,
  \sin \l[{ n\pi\over 2\alpha }(\theta -\alpha)\r]\,,
 \q x \to 0^+ \,,  \eqno(4.14) $$
$ 0<\alpha <1\,,\q \theta =-\alpha \,:$
$$L_\alpha ^{-\alpha} (x) \sim
A_1\, x^{-a_1 } \, \e^{{-b_1}\,  x^{-c_1} } \,,\q x \to 0^+\,, \q
  A_1 = \l\{\l[2\pi (1-\alpha)\r]^{-1}\,
	  \alpha ^{1/(1-\alpha)}\r\} ^{1/2}\,, \q \q\q \eqno(4.15)$$
$$
    a_1 = {2-\alpha \over 2(1-\alpha)}\,\q
    b_1 = (1-\alpha)\,\alpha ^{\alpha/(1-\alpha)}\,,\q
    c_1 = {\alpha \over 1-\alpha }\,;
 $$
\par\noindent
$ 1<\alpha < 2\,,\q \alpha -2 < \theta \le 2-\alpha \,:  $
 $$L_\alpha^\theta (x) \sim
{1\over \pi\,x}\,  \sum_{n=1}^{\infty}
   (-x^{-\alpha})^n \, {\Gamma (1+ n\alpha)\over n!}\,
  \sin \l[{ n\pi\over 2}(\theta -\alpha)\r]\,, \q x\to \infty\,,
 \eqno(4.16)$$
\par\noindent
$ 1<\alpha < 2\,,\q \theta = \alpha-2 \,:  $
$$L_\alpha ^{\alpha -2 } (x) \sim
A_2\, x^{a_2} \, \e^{-{b_2} x^{c_2}} \,,\q x\to \infty \,,\q
 A_2 = \l[ 2\pi (\alpha -1)\,\alpha ^{1/(\alpha -1)}\r] ^{-1/2}\,,\q\q	\q
\eqno(4.17)$$
$$
 a_2 = {2-\alpha \over 2(\alpha -1)}\,,\q
 b_2 =(\alpha -1)\,\alpha ^{\alpha/(\alpha -1) }\,,\q
   c_2 ={\alpha \over \alpha -1}\,.
$$
From (4.12), (4.13) and (4.14), (4.16), we thus note that for non
extremal densities
the roles of convergent and asymptotic power series are interchanging
with respect to the cases $0<\alpha <1$ and $1<\alpha <2\,. $


As a consequence of the convergence of the series
in (4.12)-(4.13) and of the identity
$L_\alpha^\theta (-x)=	L_\alpha^{-\theta} (x)\,.$
we recognize that
the stable $pdf$'s with $1< \alpha \le 2$
are {\it entire} functions whereas
the stable $pdf$'s with $0< \alpha <1$
have the form
$$L_\alpha^\theta (x) = \cases{
   (1/x) \,\Phi_1(x^{-\alpha })
	&for $\; x>0\,,$ \cr
  (1/|x|) \,\Phi_2(|x|^{-\alpha })
	&for $\; x<0\,,$
	 \cr} \eqno(4.18)		 $$
where $\Phi_1(z)$ and $\Phi_2(z)$ are {\it entire} functions.
The  case $\alpha =1$ ($|\theta|< 1$) can be treated  in the limit for
$\alpha \to 1$ of  (4.12) and (4.13), where the corresponding
series reduce to geometric series
in $1/x$ and $x\,,$ respectively, with a finite radius of convergence.
The corresponding stable $pdf$'s are no longer represented by
entire functions, as can be noted directly from (4.8)-(4.9).


From (4.12) -(4.13)  a sort of {\it reciprocity relationship}	between
stable $pdf$'s with index $\alpha $ and $1/\alpha\,  $  can be derived
as noted by Feller \cite{Feller 2}.
Assuming $1/2 \le \alpha<1$ and $x>0\,, $  we obtain
$$ \rec{x^{\alpha +1}}\, L_{1/\alpha}^{\theta}(x^{-\alpha})
  =L_\alpha^{\theta^*} (x)\,,  \q
  \theta ^*=\alpha(\theta +1)-1 \,. \eqno(4.19)$$
A quick check shows that  $\theta^*$ falls within the prescribed range,
$|\theta ^*|\le\alpha \,, $ provided that $|\theta |\le 2-1/\alpha \,. $


At the end of the paper we shall exhibit some plots the fundamental
solution of the space-fractional diffusion equation,
 $$ G_{\alpha,1}^\theta (x,1) = K_{\alpha,1}^\theta (x) =
     L_{\alpha}^\theta (x)\, \q 0<\alpha <2\,, \eqno(4.20)$$
in the range $|x|\le 5\,. $
\vskip 0.5 truecm

\noindent \underbar{{\it The time-fractional diffusion}}
\vskip 0.25 truecm
\noindent
Let us now consider the   case
 $\{\alpha =2\,,\, 0< \beta <2\}$
({\it time-fractional diffusion} including  standard diffusion
for $\beta=1$) for which
(3.11) reduces to
$$   \widehat {G_{2,\beta}^0}(\kappa ,t) =
     E_\beta \l({\ds - \kappa^2  \,t^\beta }\r)
\,,\q \kappa \in \re\,,\q  t\ge 0    \,. \eqno(4.21) $$
Following Mainardi, see \eg
\cite{Mainardi CHAOS,Mainardi CISM},
the problem can be treated with the equivalent Laplace transform
$$  \widetilde{G_{2,\beta}^0}(x,s) =   \rec{2}\,
   s^{\beta /2-1}  \, \e^{\ds -|x| s^{\beta /2}} \,,
 \q -\infty < x <+\infty\,, \q \Re (s) >0\,,
\eqno(4.22) $$
with solution
$$ G_{2,\beta}^0 (x,t) =
{1\over 2} t^{-\beta /2}\, M_{\beta/2} \l(|x|/t^{\beta /2}\r)\,,
\q -\infty < x <+\infty\,,  \q t\ge 0\,,
\eqno(4.23)$$
where
$M_{\beta/2}$ denotes  the so-called $M$ function (of the Wright type)
of order $\beta/2\,,$
whose general properties are briefly discussed below.

The function $M_\nu (z)$  is
 defined for any order $\nu \in (0,1)$ and  $\forall z \in \CC$
by
$$ M_\nu (z) =
  \sum_{n=0}^{\infty}\,
  {(-z)^n\over n!\,\Gamma[-\nu n + (1-\nu)]} =
  \rec{\pi}\, \sum_{n=1}^{\infty}\,{(-z)^{n-1} \over (n-1)!}\,
  \Gamma(\nu n)  \,\sin (\pi \nu  n)
\,,	     \eqno(4.24)$$
It is a special case of the
 Wright function on which the interested reader can inform himself
in several books and articles, \eg
\cite{Erdelyi HTF},
\cite{Kiryakova 94},
\cite{GoLuMa 99,GoLuMa 00}.
It turns out that $M_\nu (z)$ is an entire function of
order $\rho =1/(1-\nu)\,,$
which provides a generalization of the Gaussian and of the Airy
function. In fact we obtain
$$ M_{1/2}(z)
 = \rec{\sqrt{\pi}}\, \exp \l(-{\,z^2/ 4}\r)\,,\q
  M_{1/3}(z) =
  3^{2/3} \, {\rm Ai} \l( {z/ 3^{1/3}}\r) \,.	 \eqno(4.25)
$$


For our purposes (time-fractional diffusion equation)
it is relevant to consider the function $M_\nu $ ($0<\nu <1$)  for
positive argument. In the following  we  shall
denote generically by $r$ a positive variable
(it can be $|x|$, $t$ or the similarity
variable $|x|/t^{\beta/2}$) and by $c$ a positive constant
which plays the role of a scaling parameter.
We then point out
the related Laplace transform pairs,
see \eg Mainardi \cite{Mainardi CISM} and  Gorenflo, Luchko
\& Mainardi \cite{GoLuMa 99},
$$\cases{
 M_\nu \l(r/c\r)
  \,\stackrel{{\cal L}} {\leftrightarrow}\,
 c\,E_\nu(-c s )\,,& $\; \Re \,(s) >0\,, \q \hbox{(a)}$\cr
 {\ds{1 \over r^{\nu }}}\, M_\nu \l(c r^{-\nu }\r)
  \,\stackrel{{\cal L}} {\leftrightarrow}\,
 s^{\nu  -1}\,\exp(-c s^\nu )\,,& $\; \Re \,(s) >0\,,\q \hbox{(b)}$ \cr
 {\ds {c \nu	\over r^{\nu +1}}}\,
 M_\nu \l(c r^{-\nu }\r)
 \,\stackrel{{\cal L}} {\leftrightarrow}\,
   \exp(-c s^\nu )\,,& $\; \Re \,(s) >0\,. \q \hbox{(c)}$\cr}
\eqno(4.26)
$$
We note that, thanks to Bernstein theorem, Eq. (4.26-a) proves the
positivity of $M_\nu (r)$ for $r >0$
since for $0<\nu<1$ the function
$E_\nu $
is known to be completely monotonic on the negative real
axis. Furthermore, in the (singular) limit for $\nu \to 1^-\,,$
Eq. (4.26-a) shows that    $ M_\nu (r)$ tends to the
Dirac function $\delta (r-1)\,. $
Mainardi, see \eg \cite{Mainardi CHAOS,Mainardi CISM},
has provided plots for $r>0\,,$   which show
that $M_\nu (r)$ is indeed positive,
and    is monotonically decreasing
for $0 < \nu \le 1/2$, while
for $ 1/2< \nu < 1$ it exhibits a maximum 
whose position tends to $ r = 1$ as $\nu \to  1^- \,,$
consistently with the limiting case $M_1(r) = \delta(r-1)\,.$
We recognize that the result in (4.23) obtained as Laplace inversion of
(4.21) is included in the Laplace transform pair (4.26-b)
by taking  $\nu =\beta /2$, $c=|x|$ and $r=t\,. $

For the function $M_\nu (r)$ it is worth to recall the asymptotic
representation $$ \q\q\q\q\q\q\q\q M_\nu (r)  \sim  A_0  \,
     Y^{\, \nu - 1/2 } \, \exp \,\l( - Y\r)  \,,
\q r \to \infty\,, \q\q\q\q\q\q\q \eqno(4.27)$$
$$A_0 = {1\over   \sqrt{2\pi}\,(1-\nu )^\nu \, \nu ^{2\nu-1}} \,,\q
Y = (1-\nu )\, (\nu^\nu\, r)^{1/(1-\nu )}\,, $$
and the integral
$$\int_0^{+\infty}  r^{\,\ds \delta}\, M_\nu(r)\, dr
  = {\Gamma(\delta +1)\over \Gamma(\nu	\delta	+1)}\,,
   \q \delta > -1\,.
       \eqno(4.28)$$

As a consequence of 
Eqs. (4.23) and (4.27)
$G_{2,\beta}^0(x,t)$
with $ 0<\beta <2$ can be
interpreted
as a symmetric spatial $pdf$ evolving in time,
with a stretched exponential decay. More precisely, we have
$$  G_{2,\beta}^0(x,1) = \rec{2} \, M_{\beta /2} (|x|) \sim
A \, x^a \, \e^{-b x^c} \,, \q
x \to +\infty\,, \eqno(4.29) $$
$$ A=\l \{ 2\pi (2-\beta)\, 2^{\beta/(2-\beta)}
\beta^{(2-2 \beta)/(2-\beta)} \r\}^{-1/2} \,,$$
$$  \eqno(4.30) $$
$$ a={2\beta-2 \over 2(2-\beta)}\,, \q
b=(2-\beta)\,
 2^{-2/(2-\beta)}\beta^{\beta/(2-\beta)}\,, \q
c={2 \over 2-\beta} \,.$$
Furthermore, using (4.28), the moments (of even order)	of
$G_{2,\beta}^0(x,t)$ are
$$ \mu _{2n}(t) := \int_{-\infty}^{+\infty} \!\!\! x^{2n}\,
   G_{2,\beta}^0(x,t)  \,dx =
 {\Gamma(2n+1)\over \Gamma(\beta   n+1)}  t^{\beta n}\,,
\q n=0,1,2,\dots, \; t\ge 0\,.
\eqno(4.31)$$
It is interesting to compare this expression with the analogue
(4.2) for the Gaussian. In particular,
the variance $\sigma^2:= \mu_2 = 2\, t^\beta /\Gamma(\beta +1)$
is  now
proportional to the $\beta$-th power of time,
consistent with    {\it anomalous slow diffusion}
for $0< \beta <1$ and {\it anomalous fast diffusion}
for $1<\beta <2\,. $


In the slow diffusion case ($0<\beta<1$)
the $pdf$ attains its maximum value at $x =0$
(where the first derivative is discontinuous) and exhibits
exponential tails fatter than the Gaussian; in the fast
diffusion case ($1<\beta<2$) the $pdf$ attains two symmetric maxima
that move apart from the origin with time and exhibits
exponential tails thinner than the Gaussian.


We also outline
some other interesting properties of the $M$ function,
which put it in relationship with the class of extremal
stable distributions.
In fact,  we note that the Laplace transform pair (4.26-c), compared with
(4.5), allows
us to relate  the one-sided, extremal, L\'evy stable $pdf$ of index $\nu $
($0<\nu <1$)  with the	 $M$ function of order $\nu \,;$
we have
$$ \rec{c^{1/\nu}}\,L_\nu^{-\nu }\l( {r\over c^{1/\nu}}\r)
 = {c\,\nu \over r^{\nu+1}}\,
    M_\nu\l( {c\over r^\nu }\r) \,,\q
  0<\nu  <1 \,,\q c>0\,,\q r>0
 \,. \eqno(4.32) $$
Incidentally, the above relation turns out to provide an
alternative proof of the non-negativity
of the $M$ function for positive argument.
Furthermore, putting in Eqs (4.28) and (4.32)
 $\nu =\alpha\,,$ $\,c=t>0\,$ and $r=x>0\,,$
we get the Laplace transform pair related to the
(spatial one-sided) extremal stable $pdf$ (evolving in time) of index
$\alpha $ ($0<\alpha <1$),
which represents the Green function of the space-fractional
diffusion equation with $\theta = -\alpha \,,$ see Eqs. (4.4)-(4.5),
and henceforth,
$$ \e^{-t\, s^\alpha} \,\stackrel{{\cal L}} {\leftrightarrow}\,
     t^{-1/\alpha } L_\alpha ^{-\alpha }(x/t^{1/\alpha })
  = G_{\alpha ,1}^{-\alpha } (x,t) =
   { \alpha}\, {t   \over x^{\alpha+1}} \,
M_\alpha (t/x^\alpha)\,,
  \eqno(4.33)$$
with $	0<\alpha <1 \,,\; x>0\,,\; t>0 \,.$

Then, in virtue of the Feller reciprocity relation (4.19)
 and using Eqs. (4.4) and (4.23), after some manipulations we can
recover the noteworthy result
$$  G_{2, \beta}^0  (x,t)
  = \rec{\beta }\, G_{2/\beta , 1}^{2/\beta -2} (x,t)\,,
    \q 1<\beta <2 \,, \q x>0\,,  \; t>0\,. \eqno(4.34) $$
\noindent
This result (already noted
by Mainardi \& Tomirotti \cite{MainaTomi 97}
by  comparing the corresponding series expansions, see (4.13) and (4.24))
states that the two symmetric
branches of the Green function of the time-fractional
diffusion equation of order $1<\beta <2$
are proportional to
the corresponding (not symmetric) branches of
the Green function of the space-fractional
diffusion equation of order $1< \alpha = 2/\beta<2 $  with
skewness $\theta = \pm (2-\beta /2)$, namely
the exponential queues of the two extremal stable distributions
of index $2/\beta\,. $
In the limit $\beta =2$ we recover
the fundamental solution of the D'Alembert wave equation, \ie
$$G_{2, 2}^0  (x,t) =  {\delta (x-t) + \delta (x+t) \over 2} =
   {G_{1, 1}^{-1}  (x,t) + G_{1, 1}^{+1}  (x,t)\over 2}
\,,\eqno(4.35)$$
where
$ -\infty <x < +\infty\,, \; t \ge 0\,. $

At the end of the paper we shall exhibit some plots the fundamental
solution of the time-fractional diffusion equation,
 $$ G_{2,\beta}^0 (x,1) = K_{2,\beta}^0 (x) =
   \rec{2}\,  M_{\beta/2} (x)\,,\q 0<\beta <2\,, \eqno(4.36)$$
in the range $|x|\le 5\,. $

\vskip 0.5truecm
\noindent \underbar{{\it The neutral-fractional diffusion}}
\vskip 0.25 truecm
\noindent
Let us now consider
$\{0<\alpha =\beta \le 2\} $
({\it neutral fractional diffusion}),
which includes the Cauchy diffusion for $\alpha =\beta =1$
($\theta=0$)  and the
limiting case of wave propagation for $\alpha=\beta  =2$.

For $1 \le \alpha <2$ and $\theta=0$ the Green function
has been derived in explicit form by Gorenflo, Iskenderov and Luchko
\cite{GoIsLu 00} by using the Mellin-Barnes integral representation.
We shall adopt this representation later. Now,
we consider it conceptually more economical to remain (as  long as
possible) in the kingdom of Fourier-Laplace transforms
and we point out the following Fourier transform
pair related to the Mittag-Leffler function
of our interest:
 $$   E_\alpha \l(-|\kappa |^\alpha\r) \,\stackrel{{\cal F}}
{\leftrightarrow}\,
{{1\over\pi}}\,{|x|^{\alpha-1} \sin \l({\alpha \pi/ 2}\r) \over
1 + 2|x|^\alpha \cos\l({\alpha \pi/2}\r)+
   |x|^{2\alpha}} \,,\q 0<\alpha < 2
   \,,	\q x \in \RR\,.
\eqno(4.37)$$
This pair can be verified as an exercise in complex analysis
following the method illustrated by Gorenflo and Mainardi,
see \eg \cite{GorMai CISM}\footnote{
 In \cite{GorMai CISM} the reader can find the proof  of
the Laplace transform pair
 $$   E_\nu  \l(-s^\nu\r) \,\stackrel{{\cal L}}
{\leftrightarrow}\,
{{1\over\pi}}\,{t^{\nu -1} \sin \l(\nu	\pi\r) \over
1 + 2t^\nu  \cos\l(\nu	\pi\r)+
   t^{2\nu }}\,,   \q 0<\nu  < 1
   \,, \q t>0\,, \q\Re (s) >0\,.
$$},
  As far as we know, this case	of fractional diffusion
seems not be treated in its greatest generality
in the literature. Now, taking into account elementary properties
of the Fourier transform and (4.37), the reduced Green
function in (3.19)	    reads for $\alpha =\beta$ and
$x>0$ as
$$ K_{\alpha ,\alpha }^\theta (x) = N_\alpha ^\theta(x) =
{{1\over\pi}}\,{x^{\alpha-1} \sin[{\pi\over 2}(\alpha -\theta )] \over
1 + 2x^\alpha \cos[{\pi\over 2}(\alpha -\theta)] + x^{2\alpha}}\,,
\,\q 0<\alpha <2\,.
\eqno(4.38)$$
This solution, that can be extended to negative $x$ by
setting  $N_\alpha^\theta (-x)= N_\alpha^{-\theta} (x)\,,$
is evidently not negative in all of $\RR$, so it can
be interpreted as  a {\it probability density}.
In other words, $N_\alpha ^\theta(x)$ may be
considered the fractional generalization with skewness
of the well-known Cauchy density (4.8).
In the limiting case $\alpha \to 2^{-}$ (with $\theta =0$)
the density tends
to the combination $[\delta (x-1) + \delta (x+1)]/2\,,$
so we recover the Green function
of the D'Alembert wave equation quoted in (4.35).


\vvs

\section*{5. Composition rule for the Green function with $0 <\beta \le 1$}

	 We now present a composition rule
which allows us to express the general Green function of the
space-time fractional diffusion equation
(with the restriction $0 <\beta \le 1$)
as an integral involving the two Green functions
corresponding to space-fractional
and time-fractional diffusion  equations.
To this purpose we note that the Fourier Laplace transform
of the Green function (3.5)  can be re-written
in integral form as in	\cite{SaichevZaslavsky 97}
$$ \q\q\q\q\q
 \widehat {\widetilde {G_{\alpha,\beta}^\theta}}(\kappa ,s) =
 {s^{\beta -1} \over
s^\beta+ \psi_\alpha ^\theta(\kappa )} =
s^{\beta -1}\,\int_0^\infty
 \e^{\ds - u [s^\beta +\psi_\alpha^\theta(\kappa)]} \, du
 \q\q\q\q\q  \eqno(5.1)$$
$$
= \int_0^\infty \e^{\ds - u \psi_\alpha ^\theta(\kappa )} \,
 \l(s^{\beta -1}\, \e^{\ds - u s^\beta}\r)
 \, du \,. $$
In view of Eqs. (4.3) and (4.26)  we can
interpret the above formula  as
$$  \widehat {\widetilde {G_{\alpha,\beta}^\theta}}(\kappa ,s)
= 2 \int_0^\infty
   \widehat {G_{\alpha,1}^\theta}(\kappa ,u)	  \,
  \widetilde {G_{2,2\beta}^0}(u,s) \, du
 \,. \eqno(5.2)$$
Then, by inversion, we obtain the required composition rule
$$  G_{\alpha,\beta}^{\theta}(x ,t) =
   2\, \int_0^\infty   {G}_{\alpha,1}^\theta(x,u  ) \,
      {G}_{2,2\beta}^0(u,t)\, du \,. \eqno (5.3)$$

\noindent
Note  the presence of ${G}_{2,2\beta}^0$ instead of
${G}_{2,\beta}^0 \,. $
Hence Eq. (5.3) is a  formula separating variables.
It states that
the  Green function for
the space-time-fractional diffusion equation of
order $\{\alpha,\beta \}\,,$
with $0<\alpha \le 2$ and $0<\beta \le 1\,,$
can be expressed in terms  of the  Green function for
the space-fractional diffusion equation of order $\alpha$
and the  Green function for
the time-fractional diffusion equation of order $2\beta\,. $

We now present alternative, equivalent	forms of
the composition rule, that directly involve the functions
$L$ and $M\,,$ limiting ourselves  to $x>0\,. $
Because of Eqs. (4.4) and (4.23), we can write Eq. (5.3) as
$$  G_{\alpha,\beta}^\theta(x ,t) =
   t^{-\beta}	 \,
  \int_0^\infty \!\!
   u^{- 1/\alpha}\,
  {L}_{\alpha}^\theta\l({x \over u^{1/\alpha}}\r) \,
      {M}_{\beta}\l( {u \over t^{\beta}}\r)\, du \,. \eqno (5.4)$$
{\sc Remark} 5.1.
The formulae (4.4) and (4.23) corresponding
to the particular cases $\{\alpha, 1\}$ ({\it space-fractional diffusion})
and  $\{2, \beta \}$ ({\it time-fractional diffusion})
are recovered from (5.4) as follows:
$$  G_{\alpha,1}^\theta(x ,t) =
   t^{- 1}
  \int_0^\infty  \!\!
 u^{-1/\alpha}\,{L}_{\alpha}^\theta \l({x\over u^{1/\alpha}}\r)
      \delta \l( {u \over t} -1 \r) \, du      =
   t^{- 1/\alpha}    \,
    {L}_{\alpha}^\theta\l({x \over t^{1/\alpha}}\r)\,,	\eqno (5.5)$$
$$  G_{2,\beta}^0(x ,t) =
   { t^{ -\beta} \over 2\sqrt{\pi}}   \,
  \int_0^\infty  \!\! \e^{\ds \,- x^2/(4u)} \,
      {M}_{\beta}\l({u \over t^{\beta}}\r)\, {du \over u^{1/2}} =
   {t^{-\beta /2} \over 2}\,{M}_{\beta/2}\l({|x|\over t^{\beta/2}}\r)\,.
 \eqno (5.6)$$
Eq. (5.6) is of high interest for the theory of the $M$ functions
since it is a sort of (integral) duplication formula with respect to the
order; it is worth to note the presence of the Gaussian with
a spreading variance in the kernel of the integral.

\vvs

Taking into account  the relation (4.32), the function $M$
in (5.4) can be expressed in terms of an $L$ function.
We have
$$  G_{\alpha,\beta}^\theta(x ,t) =
     \int_0^\infty
  \l[u^{- 1/\alpha}\,{L}_{\alpha}^\theta\l({x\over u^{1/\alpha}}\r)\r]
 \, \l[ {t\over u\,\beta} \,u^{-1/\beta} \,
      {L}_{\beta}^{-\beta}\l({t \over u^{1/\beta}}\r)\r]\, du \,.
 \eqno (5.7)$$

Putting $y  = u^{-1/\beta}$ we derive from
(5.7) the composition rule in the form recently obtained by
Uchaikin, see \cite{Uchaikin 00,Uchaikin CARRY},
(for $t=1$) :
$$  G_{\alpha,\beta}^\theta(x ,1) =  K_{\alpha,\beta}^\theta(x) =
  \int_0^\infty \!\! y^{\beta/\alpha  }\,
 {L}_{\alpha}^\theta\l(x\,y^{\beta /\alpha }\r) \,
      {L}_{\beta}^{-\beta}(y)\,  dy\,. \eqno (5.8)$$

The composition rule, in its equivalent forms  (5.3), (5.4), (5.8),
shows that the Green function
of the space-time fractional diffusion equation
is non negative for any $x \in \RR$ and $t\in \RR_0^+\,,$
when  $\{0<\alpha <2\,,$ $\, 0<\beta \le 1 \}\,.$
We also know that this is  the case
when $\{\alpha =2\,,\, 0<\beta\le 2\}$ ({\it time-fractional diffusion})
and when $\{0< \alpha = \beta\le 2\}$ ({\it neutral-fractional
diffusion}).
In the next Section, by using the Mellin convolution,
we shall derive a  new composition rule which ensures
the non-negativity of $\K(x)$ in all of $\RR$
for $\{0<\alpha \le 2\} \cap \{0<\beta/ \alpha \le 1\} \,.$
Then we can conclude that
the Green function can	be surely interpreted
as a spatial $pdf$, evolving in time,
in the	ranges $\{0<\alpha \le 2\} \cap \{0<\beta \le 1 \}\,$
and  $\{0<\beta \le \alpha \le 2\}\,.$

\section*{6. Mellin-Barnes integral representations for the Green function}

Let us now consider  the Fourier representation
of the general Green function $K_{\alpha ,\beta}^\theta(x)$
(restricting to $x>0$) as stated at the end of the Sect 3, see Eqs
(3.18)-(3.21).
We intend to use the Mellin
convolution to invert the relevant Fourier transforms according
to the scheme (2.26)-(2-31). To this purpose we also need
the Mellin transform pair deduced from (3.12)-(3.13)
$$
 E_\beta \Big(-\kappa  ^\alpha\,\e^{\,\ds i\theta \pi/2}\Big)
\,\stackrel{{\cal M}} {\leftrightarrow}\,
{1\over \alpha} \,
{\Gamma({s\over \alpha}) \, \Gamma(1-{s\over \alpha})
\over \Gamma(1-{\beta\over \alpha}s)}
\,\exp \l(- i{\theta \pi \over 2} {s \over \alpha}\r)
\,,  \eqno(6.1) $$
where $ \kappa >0\,,$ $\;|\theta| \le 2-\beta \,,$
$\; 0<\Re(s)< \alpha \,.$

Using Eq. (6.1) in Eqs (2.28) and (2.29), Eq. (3.12) yields
$$ \K (x) \,= \, _c\K (x) + \,_s\K (x)\,,\q x>0\,,\eqno(6.2) $$
with
$$  _c\K(x) =
{1\over \pi \alpha x}  \,
{1\over 2\pi i} \int_{\gamma-i\infty}^{\gamma+i\infty}
{\Gamma({s\over \alpha}) \, \Gamma(1-{s\over \alpha}) \, \Gamma(1-s)
 \over \Gamma(1-{\beta\over \alpha}s)} \,
      \sin \l( {\pi s\over 2}\r)
 \, \cos \l( {\theta \pi s\over 2\alpha} \r) \, x^{s}\,
  ds\,,  \eqno(6.3) $$
$$ _s\K(x) = -
{1\over \pi \alpha x} \,
{1\over 2\pi i} \int_{\gamma-i\infty}^{\gamma+i\infty}
{\Gamma({s\over \alpha}) \, \Gamma(1-{s\over \alpha}) \, \Gamma(1-s)
 \over \Gamma(1-{\beta\over \alpha}s)} \,
      \cos \l( {\pi s\over 2}\r)
 \,\sin \l( {\theta \pi s\over 2\alpha} \r) \, x^{s}\,
  ds\,,  $$
where
$ 0<\gamma<\hbox{min}\{\alpha, 1 \}\,, $ and $|\theta|\le 2-\beta \,. $
We thus obtain from Eqs (6.2)-(6.3)
$$ \K(x) =
{1\over \pi \alpha x}
{1\over 2\pi i} \int_{\gamma-i\infty}^{\gamma+i\infty}
{\Gamma({s\over \alpha}) \, \Gamma(1-{s\over \alpha}) \,\Gamma(1-s)
 \over \Gamma(1-{\beta\over \alpha}s)} \,
     \sin \l[{s\over \alpha } {\pi \over 2} (\alpha -\theta)\r] \,
 x^{s}\,  ds.\eqno(6.4)  $$
By setting
$$ \rho =   { \alpha -\theta \over 2\,\alpha }\,, \eqno(6.5) $$
and using the reflection formula for the gamma function,
we finally obtain
$$  \K(x) =
{1\over  \alpha x}
{1\over 2\pi i} \int_{\gamma-i\infty}^{\gamma+i\infty}
{\Gamma({s\over \alpha}) \, \Gamma(1-{s\over \alpha}) \,\Gamma(1-s)
 \over \Gamma(1-{\beta\over \alpha}s) \,
 \Gamma ( \rho \,s)\,
 \Gamma (1-\rho \,s)}
 \, x^{\,\ds s}\,  ds\,.  \eqno(6.6)  $$
By using a standard notation for the "gamma fraction",
see \eg \cite{Marichev 83}, p. 65 (4.13),
we can re-write (6.6) as
$$    \K(x) =
   {1\over  \alpha x}
{1\over 2\pi i} \int_{\gamma-i\infty}^{\gamma+i\infty}
   \Gamma \l[
{\q {s\over \alpha }\,,\; 1-{s\over \alpha }\,,\; 1-s
\atop
  1- {\beta \over \alpha}s\,,\;\rho s\;,\,1-\rho s}
\r]
  \, x^{\,\ds s}\,  ds \,. \eqno(6.6')  $$
 The  formulae (6.4) and (6.6)	are valid for
$0 <\gamma< \hbox{min}\{\alpha, 1 \}\,$ with $|\theta|\le  2-\beta\,,$
and   provide equivalent integral
representations of the Green function for the general
{\it space-time fractional diffusion} equation.
In the following we  shall
use (6.4) or (6.6) [(6.6')] according to our convenience.

The integral at the RHS of Eq. (6.6)
is a particular {\it Mellin-Barnes integral}\footnote{
The names refer to the two authors, who in the first 1910's
developed the theory of these integrals  using them
for a complete integration of the hypergeometric differential equation.
However, as pointed out by Tricomi in \cite{Erdelyi HTF}
(Vol. 1, Ch. 1, \S 1.19, p. 49), these integrals were first used
by S. Pincherle in 1888. For a revisited analysis of the pioneering work
of Pincherle (1853-1936, Professor of Mathematics at the
University of Bologna from 1880 to 1928) we refer
to the recent paper by
Mainardi and Pagnini \cite{MainardiPagnini OPSFA01}.},
according to a	usual terminology,
see \eg \cite{Erdelyi HTF}, Vol. 1, Ch. 1, \S 1.19, pp. 49-50.
The interested reader can find in \cite{Erdelyi HTF} the discussion on the
general conditions of  convergence for the typical Mellin-Barnes integral,
formerly given by Dixon \& Ferrar \cite{DixonFerrar 36},
and based on the asymptotic representation of the gamma function.
By using these results 
we can verify  once again
the  convergence condition (3.17), namely  $|\theta| \le 2-\beta\,.$

In the particular cases of  {\it standard diffusion},
{\it space-fractional diffusion},
{\it time-fractional diffusion} and
{\it neutral-fractional diffusion},
  the Mellin-Barnes
integrals  in Eq. (6.6) simplify as follows.

For the \underbar{{\it standard diffusion}},
namely for {$\,\{\alpha =2\,,\, \theta=0\,,\, \beta =1\}\,$},
we have  $\rho =1/2$  and
$$  K_{2,1}^0(x) =   L_{2}^0(x) =
{1\over  2 x}
{1\over 2\pi i} \int_{\gamma-i\infty}^{\gamma+i\infty}
{\Gamma(1-s)
 \over
 \Gamma (1-s/2)}
 \, x^{\,\ds s}\,  ds\,,\q
0 <\gamma<  1 \,.\eqno(6.7) $$

For the \underbar{{\it space-fractional diffusion}}
($\{0<\alpha<2,\,|\theta|\le\hbox{min}\{\alpha,2-\alpha\},$
$\,\beta  =1\}$)
we distinguish a number of cases.
\pni
For {$\,\{0<\alpha <1\,,\, |\theta| <\alpha \,,\,\beta =1\}\,$}
we have $ 0<\rho <1$ and
$$  K_{\alpha ,1}^\theta (x) =	 L_\alpha ^\theta(x) =
{1\over  \alpha x}
{1\over 2\pi i} \int_{\gamma-i\infty}^{\gamma+i\infty}
{\Gamma({s/ \alpha})  \,\Gamma(1-s)
 \over	\Gamma ( \rho \,s)\, \Gamma (1-\rho s)}
 \, x^{\,\ds s}\,  ds\,, \q
0 <\gamma< \alpha\,.\eqno(6.8) $$
Let us consider the extremal cases. For $\theta = -\alpha \,,$ we have
$\rho =1$ so $\Gamma(1-s)$  cancels with
$\Gamma (1-\rho s)\,.$ Then the  "gamma fraction"
in (6.8) reduces to
$\Gamma(s/\alpha )/\Gamma(s)\,. $
For $\theta = \alpha \,,$
we have $\rho =0$ so $K_{\alpha ,1}^0(x) =0$ for $x>0\,. $

\noindent
For {$ \,\{\alpha =1\,, |\theta|  <  1\,,\, \beta =1\}\,$}
we    have $0 <\rho <1$ and
$$  K_{1 ,1}^\theta (x) =  L_1^\theta(x)
 = {1\over  x}
{1\over 2\pi i} \int_{\gamma-i\infty}^{\gamma+i\infty}
{\Gamma({s})  \,\Gamma(1-s)
 \over	\Gamma ( \rho \,s)\, \Gamma (1-\rho s)}
 \, x^{\,\ds s}\,  ds\,, \q
0 <\gamma<1\,.\eqno(6.9) $$
Let us consider the extremal cases.
For $\theta = -1 \,,$ we have $\rho =1$ so $\Gamma(1-s)$ cancels
with $\Gamma (1-\rho s)$ and $\Gamma(s)$ cancels with
$\Gamma (\rho s)\,.$ Then the "gamma fraction"
in (6.9) reduces to 1
  to yield  the Mellin representation
of the Dirac function $\delta (x-1)\,. $
For $\theta = 1 \,,$
we have $\rho =0$ so $K_{1 ,1}^1(x) =0$ for $x>0\,. $

\noindent
For {$\,\{1<\alpha <2\,,\,|\theta| < 2-\alpha\,,\,\beta =1\}\,$}
we have $ 0<(\alpha -1)/\alpha < \rho < 1/\alpha< 1$ and
$$  K_{\alpha ,1}^\theta (x) = L_\alpha ^\theta(x) =
{1\over  \alpha x}
{1\over 2\pi i} \int_{\gamma-i\infty}^{\gamma+i\infty}
{\Gamma({s/ \alpha})  \,\Gamma(1-s)
 \over	\Gamma ( \rho \,s)\, \Gamma (1-\rho s)}
 \, x^{\,\ds s}\,  ds\,, \q
0 <\gamma< 1\,.\eqno(6.10) $$
Let us consider the extremal cases. For $\theta = -(2-\alpha) \,,$ we have
$\rho =(\alpha -1)/\alpha >0 $ so (6.10) is still valid.
For $\theta =2-\alpha$ we have $\rho =1/\alpha $ so  $\Gamma(s/\alpha)$
cancels with $\Gamma (\rho s)\,,$ and the
"gamma fraction" in (6.10) reduces to
$\Gamma(1-s)/\Gamma(1- s/\alpha)\,. $

For the \underbar{{\it time-fractional diffusion}}
({$\{\alpha= 2,\,\theta =0,\,0<\beta <2,\,\beta\ne 1\}$})
we put $\alpha =2$ and $\rho =1/2$ in (6.6) and  obtain
$$  K_{2 ,\beta }^0 (x) = {1\over 2} M_{\beta /2}(x) =
{1\over  2 x}
{1\over 2\pi i} \int_{\gamma-i\infty}^{\gamma+i\infty}
{\Gamma(1-s)
 \over	 \Gamma (1- \beta s/2)}
 \, x^{\,\ds s}\,  ds\,, \q
0 <\gamma< 1\,.\eqno(6.11) $$

For the \underbar{{\it neutral-fractional diffusion}}
($\{0<\alpha=\beta <2\}$)	we put $\alpha =\beta$ in (6.6)
and obtain
$$ K_{\alpha,\alpha}^\theta (x) = N_\alpha ^\theta(x) =
{1\over  \alpha x}
{1\over 2\pi i} \int_{\gamma-i\infty}^{\gamma+i\infty}
{\Gamma({s\over \alpha}) \, \Gamma(1-{s\over \alpha})
 \over
 \Gamma ( \rho \,s)\,
 \Gamma (1-\rho \,s)}
 \, x^{\,\ds s}\,  ds\,. \eqno(6.12)  $$
In the limiting case $\alpha =\beta =2$
($\theta =0\,,\, \rho =1/2$) we recover from (6.11)
or (6.12) the Mellin-Barnes representation of $\delta (x-1)/2\,, $
consistently with the D'Alembert  wave equation,
see Eq. (4.35).
We note that the Mellin transform pair (6.6) allows us to
compute the value of any convergent moment
of the Green function $K_{\alpha ,\beta}^\theta(x)\,.$
In fact,  recalling the basic formulae (2.21a)-(2.21b)
for the Mellin transformation
(2.21a)-(2.21b), from (6.6)
we write the Mellin transform of $x \, K_{\alpha ,\beta}^\theta(x)$ as
$$ \q\q\q \q\q
	     \int_0^{+\infty}	   \!\!
 K_{\alpha ,\beta}^\theta(x) \,x^{\,\ds s} \,dx =
 \rec{\alpha }\,
{\Gamma(-{s/ \alpha}) \, \Gamma(1+{s/ \alpha}) \,\Gamma(1+s)
 \over \Gamma(1+{\beta s/ \alpha}) \,
 \Gamma (-  \rho \,s)\,
 \Gamma (1+ \rho \,s)} \,,\q\q\q\q \eqno(6.13)$$
$$   -\hbox{min}\{\alpha, 1\}< \Re (s) <0\,.$$
In order to include $s =0$ in the convergence strip (so, in particular, the
integral of $\K(x)$ in $\RR_0^+$ can be evaluated)
we properly use in (6.13) the known property $\Gamma(1+z) = z\,\Gamma(z)$
to obtain
$$ \q\q\q \q\q \int_0^{+\infty}  \!\!
 K_{\alpha ,\beta}^\theta(x) \,x^{\, \ds s} \,dx =
 \rho \,
 { \Gamma(1-{s/\alpha})\,\Gamma(1+{s/\alpha}) \,\Gamma(1+s)
 \over
 \Gamma (1-\rho \,s)\, \Gamma (1+\rho \,s)\,\Gamma(1+{\beta\,s/ \alpha})}
\,,\q\q \q\q \q  \eqno(6.14) $$
$$ -\hbox{min}\{\alpha, 1\}< \Re (s) <\alpha \,. $$
In particular, setting $s=0$ we find
$  \int_0^{+\infty}   K_{\alpha ,\beta}^\theta(x)  \,dx = \rho \,$
($ \rho  = 1/2$ if $\theta =0$).
Eq. (6.14) is	consistent with a similar expression given by
Uchaikin \cite{Uchaikin CARRY}.
We note that it is strictly valid as soon as cancellations in the
"gamma fraction" at the RHS   cannot be done.
Then Eq. (6.14)  allows us to evaluate (in $\re_0^+$)
the (absolute) moments of the fundamental solution
of order $\delta  $
such that   $ -\hbox{min}\{\alpha, 1\} <\delta	<\alpha\,.  $
In other words, it states that, as $x \to +\infty,$
$\,K_{\alpha ,\beta}^\theta(x)= {\cal O} \l(x^{-(\alpha +1)}\r).$
When cancellations occur  in the "gamma fraction"  the range
of $\delta$ may change. In this respect
an interesting case
is $\,\{ \alpha =2\,, \, \theta =0\,, \, 0<\beta <2\}\,$
 ({\it time-fractional diffusion} including {\it standard diffusion}),
where Eq. (6.14)
reduces to $$ \int_0^{+\infty}	\!\!
 K_{2,\beta}^0(x) \,x^{\, \ds s} \,dx =
 \rec{2} \,
 {\Gamma(1+s)
 \over
\Gamma(1+{\beta\,s/2})}
\,,\q \Re (s) >-1\,.
 \eqno(6.15) $$
This result
is consistent with the existence of all moments of order $\delta  >-1$
for the corresponding Green function, see (4.28).
\newpage

Furthermore, by using the Mellin convolution formula (2.25),
and the Mellin-Barnes representations valid in the
particular cases, namely
(6.8), (6.10) for the space-fraction diffusion, (6.11)
for the time-fraction diffusion and
(6.12) for the neutral-fraction diffusion,
Eq. (6.6) can help us to investigate the
 non-negativity of  the Green function	$\K(x)\,.$
In fact, it is not difficult to recognize that, assuming $x>0$
and $0<\alpha <2$,
$$ \K(x) = \cases{
   \alpha   \, {\ds \int_0^\infty} \!\!
  \l[\xi ^{\alpha -1}\,  {M}_{\beta}\l(\xi ^{\alpha}\r)\r]\,
  {L}_{\alpha}^\theta\l({x/\xi }\r) \,
      {\ds{d\xi  \over \xi }} \,,&$\; 0<\beta <1\,,$ \cr\cr
  {\ds \int_0^{\infty}} \!\!
M_{\beta / \alpha}(\xi ) \,N_\alpha^\theta(x/\xi)\,
    {\ds{d\xi  \over \xi} }\,, & $\; 0<\beta/\alpha<1 \,.$ \cr}
\eqno(6.16)$$
The limiting cases $\beta =1$, $\alpha =\beta $ and $\alpha =2$
can be included if one	takes into account, in a proper way,
the  resulting expressions by the generalized function $\delta (x-1)\,.$
We note that the first Mellin convolution in (6.16) is
consistent with the composition rules  obtained in Sect 5
by using the Fourier transform\footnote{
To this purpose, let us consider the composition rule in form (5.4),
$$  G_{\alpha,\beta}^\theta(x ,t) =
   t^{-\beta}	 \,
  \int_0^\infty \!\!
   u^{- 1/\alpha}\,
  {L}_{\alpha}^\theta\l({x \over u^{1/\alpha}}\r) \,
      {M}_{\beta}\l( {u \over t^{\beta}}\r)\, du \,,  $$
Putting $\xi  = u^{1/\alpha}$ and taking into account the scaling property
$ \G(x,t)  =
    t^{-\beta/\alpha}\,\K \l(x/t^{\beta /\alpha}\r)\,,$
we rewrite the composition rule (for $t=1$) in the form
of the first Eq. in (6.16).}.

The  formulae in (6.16),
by involving two non negative functions
(related to the probability densities $L_\alpha ^\theta(x)$,
 $M_{\beta/2}(x)$ or $N_\alpha ^\theta(x)$),
allow us to state the  probability interpretation
 of the Green function
in  the ranges 
$$\{0<\alpha \le 2\} \cap \{0<\beta \le 1 \}\,$$
 and  $$\{0<\alpha \le 2\} \cap \{0<\beta /\alpha \le 1\}\,. $$
Therefore,  the probability interpretation
holds true for any $\alpha \in (0,2]\,$ if $0<\beta \le 1\,,$
whereas, if $1<\beta \le 2\,,$ only for $1<\beta \le \alpha \le 2\,. $
The  cases excluded from the probability interpretation
turn out to be	
$$\{0< \alpha <\beta \} \cap \{1<\beta <2\}\,.$$
\vfill\eject 

\section*{7. Computational representations for the Green function}

Those who are acquainted with  Fox $H$ functions can recognize
in (6.6) the representation of a certain function of this
class\footnote{
ADDED NOTE (2007) As a complement of the present analysis the reader can consult the paper by 
F. Mainardi, G. Pagnini and R.K. Saxena: Fox $H$ functions in fractional diffusion,
published in {\it Journal of Computational and Applied Mathematics} {\bf 178} (2005) 321-331,
where the Authors have provided the representation of the fundamental solutions of
the fractional diffusion equations here treated in terms of Fox $H functions$}
 on which the interested reader can inform himself
in several books,
\eg
\cite{MathaiSaxena  H},
\cite{Srivastava H},
\cite{Marichev 83},
\cite{Prudnikov IS3},
\cite{SKM 93},
\cite{Kiryakova 94}.
Unfortunately, as far as we know, computing routines for this general class
of special functions are not yet available.
Here, following and completing the approach adopted by
Gorenflo, Iskenderov \& Luchko \cite{GoIsLu 00},
we intend to  compute 
the Green function in any space domain	providing
for it a  {\it computational representation}
(usually to be	obtained by matching two distinct expressions).


To this purpose  we distinguish \underbar{three cases}
depending on the order relationship between $\alpha $ and $\beta $.
According to their  increasing difficulty we consider
$$ \hbox{(i)} \q\alpha	= \beta\,,\q
   \hbox{(ii)} \q\alpha < \beta\,, \q
   \hbox{(iii)} \q\alpha  > \beta\,.\eqno(7.1)$$
 \underbar{For the  case (i)} $\{\alpha  = \beta\}$
the integral
representation (6.4) simplifies  into
$$  K_{\alpha,\alpha}^\theta(x) =
 {1\over \pi \alpha x}
{1\over 2\pi i} \int_{\gamma-i\infty}^{\gamma+i\infty}
\!\! \Gamma\l({s\over  \alpha}\r) \, \Gamma\l(1-{s\over  \alpha}\r)
 \,\sin \l[{s\over \alpha } \,{\pi \over 2}\, (\alpha -\theta)\r]
  x^{\,\ds s}\,   ds .\eqno(7.2) $$
In this case the contour of integration can be transformed
to the loop ${L}_{-\infty}$ starting and ending at infinity
and encircling all the poles $s_n=   -\alpha n,\ n=   0,1,2,\dots$ of the
function $\Gamma(s/\alpha)$  for $0<x<1$ and
to the loop ${L}_{+\infty}$ starting and ending at infinity
and encircling all the poles $s_n= (1 +n)\alpha,\ n=0,1,2,\dots$ of the
function $\Gamma(1- s/\alpha)$	for $x>1\,. $
Applying the residue theorem 
we arrive at the series representations
$$ K_{\alpha,\alpha}^\theta(x)
= {1\over {\pi x}}
 \sum_{n=0}^{\infty}
\sin \l[{n \pi \over 2}(\theta - \alpha)\r] \,
(-x^{\alpha})^{n} \,, \q 0 < x < 1 \,;
\eqno(7.3)$$
$$ K_{\alpha,\alpha}^\theta(x) =
{1\over {\pi x}}
 \sum_{n=0}^{\infty}
\sin \l[{n \pi \over 2}(\theta - \alpha)\r] \,
(-x^{-\alpha})^{n} \,,	\q 1 < x < \infty
\,.\eqno(7.4)$$
So we have two different representations
by power series: the one in {\it positive powers},
convergent for $0< x<1$, the
other	in {\it negative powers},  convergent for $x>1\,. $
Remarkably, in this special case, we can obtain
the fundamental solution in all of $\RR$
in a closed form, namely expressed in term of
elementary  functions.
In fact, following \cite{GoIsLu 00}, we use
the formula
$$\sum_{n=1}^{\infty} r^n \sin(na) =
\Im \l(\sum_{n=1}^{\infty} r^n \,\e^{ina}\r) =
\Im \l( {r \e^{ia} \over 1 - r \e^{ia}}\r) =
 {r \sin a \over 1 -2r\cos a
+r^2}\,,\eqno(7.5) $$
with $\, a \in \re \,,\; |r| < 1\,, $
and we get
 $$K_{\alpha,\alpha}^\theta(x) =
{{1\over\pi}}\,{x^{\alpha-1} \sin[{\pi\over 2}(\alpha -\theta )] \over
1 + 2x^\alpha \cos[{\pi\over 2}(\alpha -\theta)] + x^{2\alpha}}\,,
\q 0<x<\infty\,. \eqno(7.6)$$
This result is in agreement with (4.38) and,
by taking into account the symmetry relation (3.18),
can be properly extended to $-\infty <x<0\,. $
We also note that from (7.6) we recover the expressions of the
particular
cases  $\{\alpha =\beta = 1\,,\,|\theta|< 1\}$ given in Eq. (4.9),
namely the stable density $L_1^\theta(x),$
and $\{1\le \alpha =\beta<2 \,,\, \theta=0\}$
given by Gorenflo, Iskenderov and
Luchko, see  Eq. (20) in \cite{GoIsLu 00}.
In the limit as $x \to 0\,,$ we get
$$\lim_{x \to 0} K_{\alpha,\alpha}^\theta(x)= \lim_{x \to 0}
{x^{\alpha-1} \over\pi}\, \sin\l[{\pi\over 2}(\alpha - \theta)\r]=
\cases{
{+\infty}\,, &	$   0 < \alpha	< 1\,, $ \cr
\rec{\pi}\, \cos[{\pi\over 2}\theta]\,, & $  \alpha =1\,, $ \cr
       0\,,  &	$   1 < \alpha < 2\,, $ \cr} \eqno(7.7) $$
which is consistent with (3.23).


\vs
\noindent
\underbar{For the case (ii)} $\,\{\alpha  < \beta\}\,$
the contour of integration in (6.6) can be transformed
to the loop ${L}_{-\infty}$ starting and ending at infinity
and encircling all the poles $s_n=   -\alpha n,\; n\in \NN_0
\equiv \{0,1,2,\dots\}$ of the
function $\Gamma(s/\alpha)\,.$
The residue theorem for simple poles 
gives us the
following representation by a convergent series
in {\it negative powers} of $x^\alpha \,,$
$$ \K(x) =
{1\over {\pi x}} \sum_{n=1}^\infty {\Gamma(1 + \alpha n) \over
\Gamma(1 + \beta n)}
\sin \l[{n \pi \over 2}(\theta - \alpha)\r]
(-x^{-\alpha})^{n} \,, \q 0<x<\infty\,. \eqno(7.8) $$
We note that for $\beta =1$
we recover the series at the RHS of (4.12)
giving the representation of  the stable density $L_\alpha ^\theta(x)$ with
$0<\alpha <1\,. $

\vs
\noindent
\underbar{For the case (iii)} $\,\{\alpha  > \beta\}\,$
the situation is more	 complicated  because we have to
transform the contour of integration in (6.6) to the loop
${L}_{+\infty}$ encircling all the poles $\,_1s_n = \alpha(1+n),
\; n \in \NN_0$ and $\,_2s_m =1+m,\; m \in \NN_0$ of the functions
$\Gamma(1-{s\over \alpha})$ and $\Gamma(1-s)$,
 respectively.

We note that for $\beta =1$ or $\alpha =2$ the poles   $\,_1s_n$ do not
appear because	$\Gamma(1-{s/ \alpha})$ cancels in (6.6)
as seen in Eqs (6.7)-(6.11).

Excluding the above  cases, \ie if
 $\beta \ne 1$ and $\alpha \ne 2\,,$  we have to consider
the possibility of double poles occurring when
 $ - \alpha(k+1) = -(m+1) \,,$ namely when
$ \alpha = {(m+1)/(k+1)} \,,\; m,k \in N_0 \,.$

When the poles are all simple	the residue theorem 
gives us the
following representation by two convergent series
in {\it positive powers} of $x^\alpha$ and $x$,
$$\q\q\q\q\q\q \K(x) = {1\over {\pi x}} \sum_{k=0}^\infty
{\Gamma(1 - {\alpha k}) \over \Gamma(1 - \beta k)}
\sin \l[{k \pi \over 2}(\theta - \alpha)\r]
(-x^{\alpha})^{k}  \q\q \q\q\q \eqno(7.9) $$
$$ \q\q\q + {1\over {\pi x}} \sum_{k=0}^\infty
{\Gamma(1 - {k \over \alpha }) \Gamma(1 + {k \over \alpha})
\over {k! \Gamma(1 - {\beta \over \alpha} k)}}
\sin \l[{k \pi \over 2 \alpha}(\theta - \alpha)\r]
 (-x)^{k}\,.\q\q\q $$
We note that in the limiting cases $\beta =1$ or $\alpha =2$ the first
series has vanishing coefficients and the second series yields	the
expected results, namely the  expansion in positive powers of $x$ for
the stable density  $L_\alpha^\theta(x)$ with $\alpha >1$, see (4.13),
and  for the  density $M_{\beta/2} (|x|)/2$, see  (4.24) and (4.36),
respectively.

When the poles are not all simple, we need to separate
the double poles from the simple ones. To this purpose
we denote the {\it simple} poles of
$ \Gamma(1-{s / \alpha}) $ and
$ \Gamma(1-s) $ by
$ _1s_k = \alpha (k+1), k\in K \,,$ and $ _2s_m = m+1, m\in M \,,$
 and the {\it double} poles of
$ \Gamma(1-{s \over \alpha}) \Gamma(1-s) $ by
 $ _3s_i = (m_0+1)(1+i), i\in \NN_0$
where
$$ K = N_0 \setminus K_0 ;
\, N_0 = \lbrace 0,1,2,... \rbrace,
\, K_0 = \lbrace k \in N \mid k = (k_0+1)+(k_0+1)i ,
	    i \in N_0 \rbrace \,,$$
$$ M = N_0 \setminus M_0 ;
\, N_0 = \lbrace 0,1,2,... \rbrace, \,
M_0 = \lbrace m \in N \mid m = (m_0+1)+(m_0+1)i ,
	    i \in N_0 \rbrace \,.$$
Then, applying the residues theorem
for simple and double poles we arrive at the following
representation with  three series (of which the first in
{\it positive powers} of $x^\alpha  $, the second
in {\it positive powers} of  $x$, and the
third of a peculiar character in that it contains
{\it positive powers} of $x$  that can be coupled with
$\log x$):
$$\qq \qq \qq
\K(x) = {1\over {\pi x}} \sum_{k \in K}
{\Gamma(1 - {\alpha k}) \over \Gamma(1 - \beta k)}
\sin \l[{k \pi \over 2}(\theta - \alpha)\r]
(-x^{\alpha})^{k} \q\q\q \q\q\q \eqno(7.10) $$
$$ \q\q\q + {1\over {\pi x}} \sum_{m\in M}
{\Gamma(1 - {m \over \alpha }) \Gamma(1 + {m \over \alpha})
\over {m! \Gamma(1 - {\beta \over \alpha} m)}}
\sin \l[{m \pi \over 2 \alpha}(\theta - \alpha)\r]
 (-x)^{m} + P_{\alpha ,\beta}^{\theta} (x)
\,. $$
Here  $P_{\alpha ,\beta}^{\theta} (x)$	denotes
the contribution from the double poles,
which for $\alpha ={m_0+1 \over k_0+1} $
reads
(see detailed derivation in  Pagnini's thesis
\cite{Pagnini TH00}):
$$ \q\q  P_{\alpha ,\beta}^{\theta} (x) =
 {{(-1)^{k_0+m_0+1}\, x^{m_0}}\over {\pi }} \,
 \sum_{i=0}^\infty
    [(-1)^{k_0+m_0}\, x^{m_0+1}]^i \,
{R_1 + R_2  + R_3 \over  D} \q\q\q
 \eqno(7.11)	$$
$$ \times  \,
   \sin\l[{(m_0+1)(1+i)\pi \over {2 \alpha}}(\theta - \alpha)\r] \,,$$
where
$$  R_1 = \psi\l[(m_0 + 1)(1 + i)\r]- \log {x} \,,\eqno(7.12)$$
$$ R_2 =   - {\beta \over \alpha} \,
    \psi\l[1 - \beta(k_0+1)(1+i)\r]  \,,\eqno(7.13)$$
$$ R_3 =  -{\pi \over 2 \alpha} (\theta - \alpha) \,
\cot \left[{(m_0+1)(1+i)\pi \over {2 \alpha}}
(\theta - \alpha) \right]\,,\eqno(7.14)  $$
$$ D= \Gamma\l[1-\beta(k_0+1)(1+i)\r]\,
  \Gamma\l[(m_0+1)(1+i)\r]\,,\eqno(7.15)
$$
and
 $\psi$ is the logarithmic derivative of the gamma function
$$\psi(z)=
{d\over dz}\log \Gamma(z)\, =\, {\Gamma^\prime (z)\over \Gamma(z)}
\,. \eqno(7.16) $$

{\sc Remark} 6.1. 
For some rational values of the parameter $\beta$,
the relation  $1-\beta(k_0 +1)(2i +1) =-l,\ l=0,1,2,\dots$ can take place
for some values of the index $i,\ i=0,1,2\dots$. In this case,
 the term $R_2/D$ in (7.11)
is an indeterminate
expression of the form ${\infty\over\infty}$. Due to the
formulae $\lim_{s \to -l}1/\Gamma(s) = 0,\; l=0,1,2\dots,$
and $\lim_{s\to -l} \psi(s)/\Gamma(s) =
(-1)^{l+1}\, l!,\; l=0,1,2,\dots,$ we can rewrite
 $P_{\alpha ,\beta}^{\theta} (x)$
 in the form (see detailed derivation in  Pagnini's thesis
\cite{Pagnini TH00}):
$$
\q P_{\alpha ,\beta}^{\theta} (x)  =
  {{(-1)^{k_0+m_0+1} \,x^{m_0}}\over {\pi }} \,
{\beta \over \alpha} \,
  \sum_{i=0}^\infty (-1)^{\beta(k_0+1)(1+i)+1} \,
  {\Gamma\l[\beta(k_0+1)(1+i)\r]\over \Gamma\l[(m_0+1)(1+i)\r]}
 \qq \qq \qq \eqno(7.17) $$
$$ \q\q \times\,
\sin \l[{(m_0+1)(1+i)\pi \over {2 \alpha}}(\theta - \alpha)\r]
\l[(-1)^{k_0+m_0}\, x^{m_0+1}\r]^i
\,, \qq \qq$$
which gives us a finite value.

\noindent
{\sc Remark} 6.2.
From the above analysis  we recognize that all series,
that are involved \underbar{in the  cases (i), (ii), (iii)},
are vanishing if
$\theta = \alpha \,.$ Since this extremal value of $\theta$
can be allowed only if $0<\alpha \le 1\,, $ we can conclude
that  the Green functions $K_{\alpha,\beta}^{\alpha} (x) \equiv 0$
for $x>0$  if $\alpha \in (0,1] $
and $ \beta \in (0,2)\,, $ subjected to the condition
$\theta= \alpha \le 2-\beta\,.$

\vvs
We must note that the representation of the
Green function	expressed  for the cases (ii)
and (iii) in terms  of convergent expansions in negative and positive
power series, respectively,
are correspondingly not suitable to
numerically evaluate the  solution
as soon as $x$ is sufficiently	small or sufficiently large.
In other words,  the convergent expansions cannot reproduce the
expected 
behaviour of their sum for $x \to 0^+$ or
$ x\to	  \infty$  when the corresponding series
 are in negative or positive
powers of $x\,, $ respectively.
So, to complete our analysis we need to give the asymptotic
representations of $\K(x)$ in the  case (ii)  $\{\alpha <\beta \}$
as $x \to 0^+$
and in the case  (iii) $\{\alpha >\beta \}$ as
 $x \to +\infty\,.$
To this purpose we shall adapt to our analysis
the results  contained in the
fundamental paper by Braaksma \cite{Braaksma 62}
on the asymptotic expansions for certain Mellin-Barnes integrals.
We shall limit ourselves to
those cases (not already treated in Sect. 4)
 where the Green function
is expected to be non-negative. As a consequence,
for  the case (ii) we consider
  $0<\alpha <\beta <1\,, $
whereas for the case (iii)
we consider
$0< \beta <\alpha < 2\, $   with $\beta \ne 1\,. $

\vs
\noindent
\underbar{For the case (ii)}
$\,\{0<\alpha <\beta <1 \,, \ |\theta|< \alpha\}\, $
the   asymptotic representation as $x \to 0^+$ turns out to be	given by
the convergent representation of the case (iii), namely
by the two series in (7.9) or by three series in (7.10)-(7.17).
 We agree to write  (for economy)
 $$ \K(x) \sim	\cases{
 (7.9), 		     & $\; \alpha \neq {m+1\over k+1}\,,$ \cr
 (7.10)\; \hbox{-} \; (7.17), & $\; \alpha  = {m+1 \over k+1}\,,$ \cr}
  \q x \to 0^+\,. \eqno(7.18)  $$
In the extremal case $\theta = -\alpha$ only the first series
in (7.18) [namely in (7.9) or (7.10)]  survives, so
we obtain the asymptotic representation
$$  \K(x) \sim
    {1\over x} \sum_{k=1}^\infty
{(-x^\alpha)^{k-1}  \over
\Gamma(\alpha k) \, \Gamma(1 - \beta k)} \,,\q x \to 0^+\,.
\eqno(7.19) $$
We note that in the limiting case $0< \alpha <\beta=1$
all the coefficients
of the powers series in (7.19) are vanishing in view of
the exponential decay just expected for the unilateral
stable distribution $L_\alpha^{-\alpha} (x)$ as $x \to 0^+\,.$

\vs
\noindent
\underbar{For the case (iii)}
$\,\{0<\beta  <\alpha\}$
we distinguish the sub-cases
$0<\alpha \le 1\,,$ where $|\theta|\le \alpha \,,$ and
$1<\alpha <2\,,$ where $|\theta| \le 2-\alpha \,.$
In both sub-cases,
when $\theta$ does not assume extremal values,
the asymptotic representation as $x \to +\infty $ turns out to be  given by
the convergent representation of the case (ii), namely
$$ \K(x) \sim
{1\over {\pi x}} \sum_{n=1}^\infty {\Gamma(1 + \alpha n) \over
\Gamma(1 + \beta n)}
\sin \l[{n \pi \over 2}(\theta - \alpha)\r]
(-x^{-\alpha})^{n} \,, \q x \to +\infty\,. \eqno(7.20) $$

\noindent
As for the extremal cases, we have as follows:

\noindent
If  $0<\alpha  \le 1\,,$ the representation (7.20) is still valid
for $\theta = -\alpha\,, $
whereas is trivially zero for $\theta=+\alpha\,, $
as expected.
If  $1<\alpha  <2\,,$ the representation (7.20) is still valid
for $\theta = 2-\alpha \,,$  whereas for $\theta=\alpha -2$
we have:
$$ \K(x) \sim A \, x^a \, \e^{-b x^c} \,, \q
x \to +\infty\,, \eqno(7.21) $$
where
$$ A=\l \{ 2\pi (\alpha-\beta) \alpha^{\beta/(\alpha-\beta)}
\beta^{(\alpha-2 \beta)/(\alpha-\beta)} \r\}^{-1/2} \,,$$
$$  \eqno(7.22) $$
$$ a={2\beta-\alpha \over 2(\alpha-\beta)}\,, \q
b=(\alpha-\beta)
\alpha^{-\alpha/(\alpha-\beta)}\beta^{\beta/(\alpha-\beta)}\,, \q
c={\alpha \over \alpha-\beta} \,.$$
We note that the above asymptotic representation  (7.21)-(7.22)
is still valid for $\beta =1\,,$  when it reduces to that
for the density $L_\alpha ^{\alpha -2}(x)\,,$  see (4.17),
and for  $\alpha =2$ (so $\theta =0$),
when it  reduces to that
for the density $K_{2,\beta }^0(x) = \rec{2} \, M_{\beta/2}(|x|)\,,$
see (4.29)-(4.30).


\vvs
\section*{8. Conclusive discussion and plots}

We can conclude with a discussion about  some general features
occurring in the Cauchy problem of our space-time
fractional diffusion equation (1.1)-(1.2).
A first general feature concerns the scaling property of the Green function
which allows us to express it in terms of a function of a single variable,
the {\it reduced Green function} $\K (x)$, see (1.3) or (3.8).
In this paper we have focused our attention to derive a computational
form for 
$\K(x)$  in all of $\re\,,$
taking into account  the symmetry relation (3.18).
The relevant particular cases of space-fractional
($\{0<\alpha \le 2\,,\, \beta =1\}$),
time-fractional  ($\{\alpha =2\,,\, 0<\beta \le 2\}$) and
neutral-fractional ($\{0<\alpha=\beta \le 2\}$)
diffusion  have been summarized in Sect. 4
where the  interpretation of the corresponding Green function
as a probability density has been pointed out.

For computational aims the representation of
$\K(x)$ through the
Mellin-Barnes integral, see (6.6), was found useful.
Incidentally, this representation
has enabled us
 to extend the	probability interpretation
of the reduced Green function
to the ranges $\{0<\alpha \le 2\} \cap \{0<\beta \le 1 \}\,$
and $\{1<\beta \le \alpha \le 2\} $.

More precisely, to compute the function $\K(x)$  we used the series
expansions (7.8)-(7.17) and asymptotic expansions (7.18)-(7.22)
which were derived from the representation (6.6).

At the end of the section we shall exhibit some plots of the
reduced Green function 
for some "characteristic" values of the parameters
$\alpha,\ \beta,$ and $\theta$. All the
plots were drown by using the MATLAB system
for the values of the independent
variable $x$ in the range $|x|\le 5$.
To give the reader a better impression about
the behaviour of the tails the logarithmic scale was adopted.

After long discussions
we chose from the uncountable set of possible plots of the
function $\K(x)$ the plots having, in our opinion, the greatest interest,
in the framework of the probability interpretation.
\vs \pni
I. For the Green function
 of the {\it space-fractional diffusion} equation ($\beta =1$),
we present plots for the cases:
\pni
Fig. 3 : $\alpha =0.50$ with a) $\theta =0$ and b) $\theta=-0.50\,;$
\pni
Fig. 4 : $\alpha =1\phantom{.50}
	  $ with a) $\theta =0$ and b) $\theta=-0.99\,;$
\pni
Fig. 5 : $\alpha =1.50$ with a) $\theta =0$ and b) $\theta=-0.50\,.$
\pni
Note that for $\alpha =1$ we have  chosen a nearly extremal
case to show a density approaching to the delta
function $\delta (x-1)$, see (4.10).
\vs \pni
II. The fundamental
solution of the {\it time-fractional diffusion} equation ($\alpha =2$)
was plotted in the following cases:
\pni
Fig. 6 :  a) $\beta  =0.25$ and b) $\beta=0.50\,;$
\pni
Fig. 7 :  a) $\beta =0.75$ and b) $\beta= 1.25\,;$
\pni
Fig. 8 :  a) $\beta = 1.50$ and b) $\beta=1.75\,.$

\vs\pni
III. For  the fundamental
solution of the {\it space-time-fractional diffusion} equation,
we exhibit the following plots:
\pni
Fig. 9\phantom{0} : $\alpha =0.50\,,\, \beta =0.50$
	with a) $\theta =0$ and b) $\theta=-0.50\,;$
\pni
Fig. 10 : $\alpha =1.50\,,\, \beta = 1.50$
     with a) $\theta =0$ and b) $\theta=-0.49\,;$
\pni
Fig. 11 : $\alpha =0.25\,,\, \beta =0.50$
  with a) $\theta =0$ and b) $\theta=-0.25\,;$
\pni
Fig. 12 : $\alpha =0.75\,,\, \beta =0.50$
	with a) $\theta =0$ and b) $\theta=-0.75\,;$
 \pni
Fig. 13 : $\alpha =1.50\,,\, \beta = 0.50$
     with a) $\theta =0$ and b) $\theta=-0.50\,;$
\pni
Fig. 14 : $\alpha =1.50\,,\, \beta =1.25$
  with a) $\theta =0$ and b) $\theta=-0.50\,.$
\pni


\begin{figure}
\centering
\includegraphics[width=0.40\textwidth]{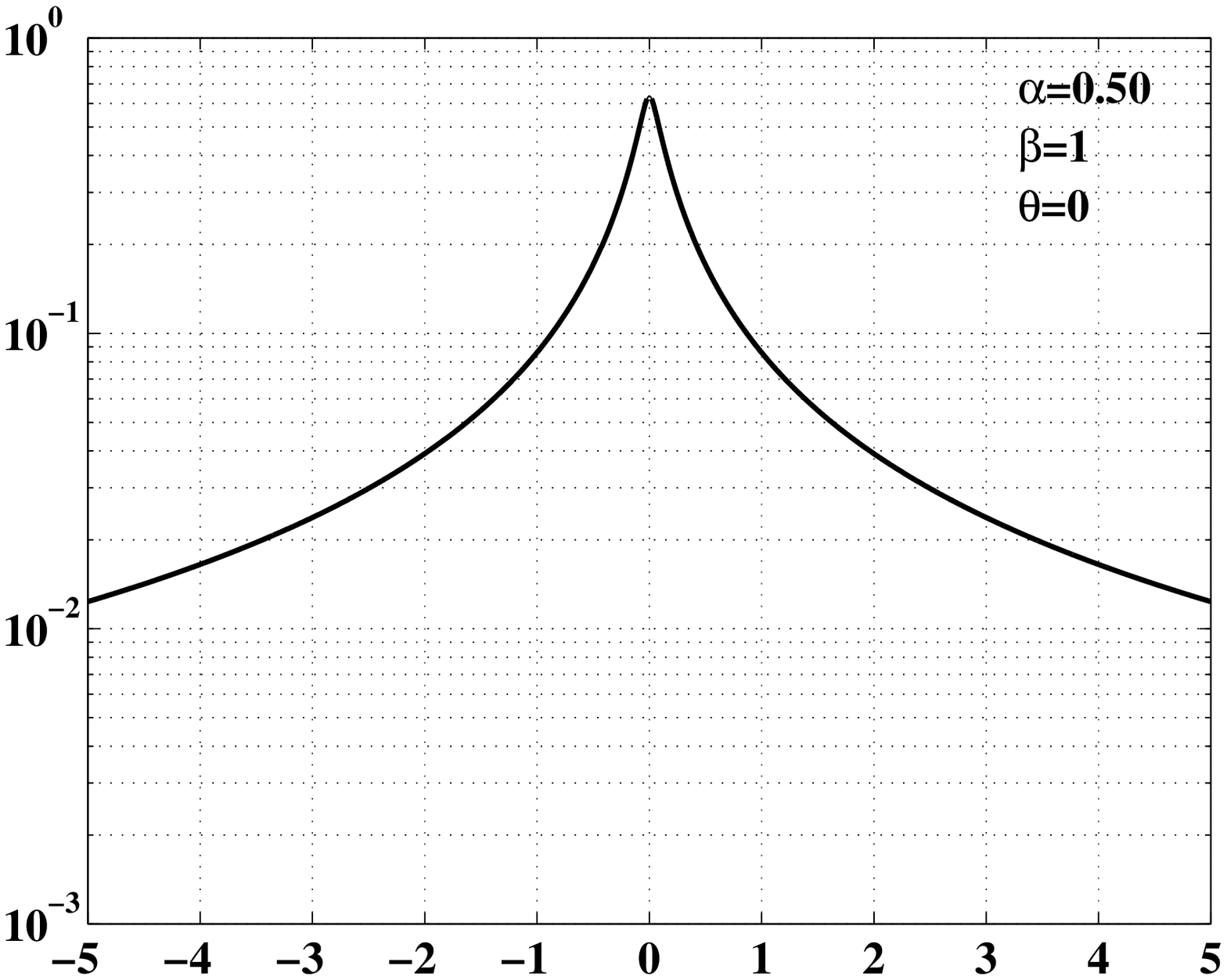}
\includegraphics[width=0.40\textwidth]{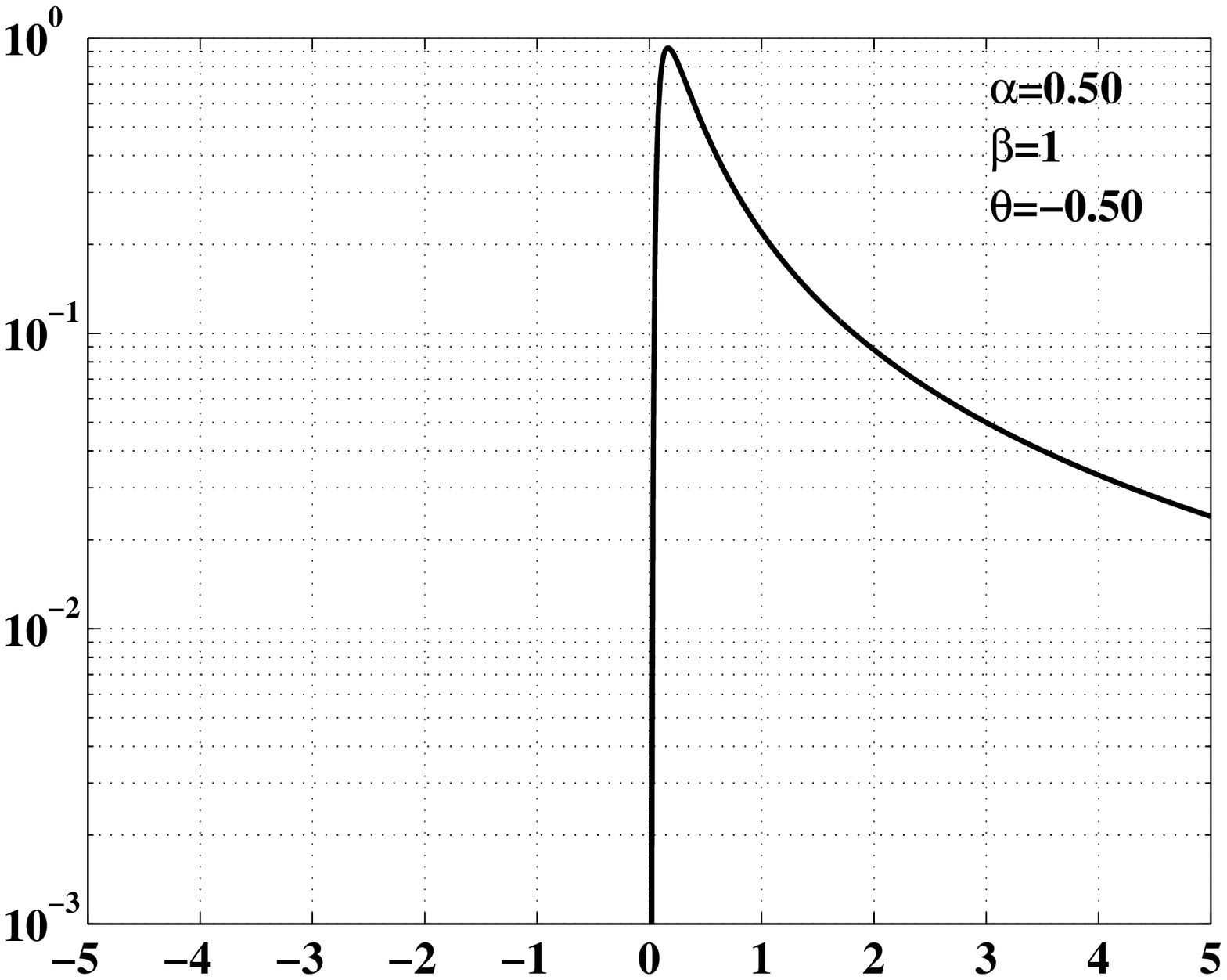}
\vskip 0.25truecm
\centerline{{\bf Fig. 3}}

\vskip 0.75truecm
\centering
\includegraphics[width=0.40\textwidth]{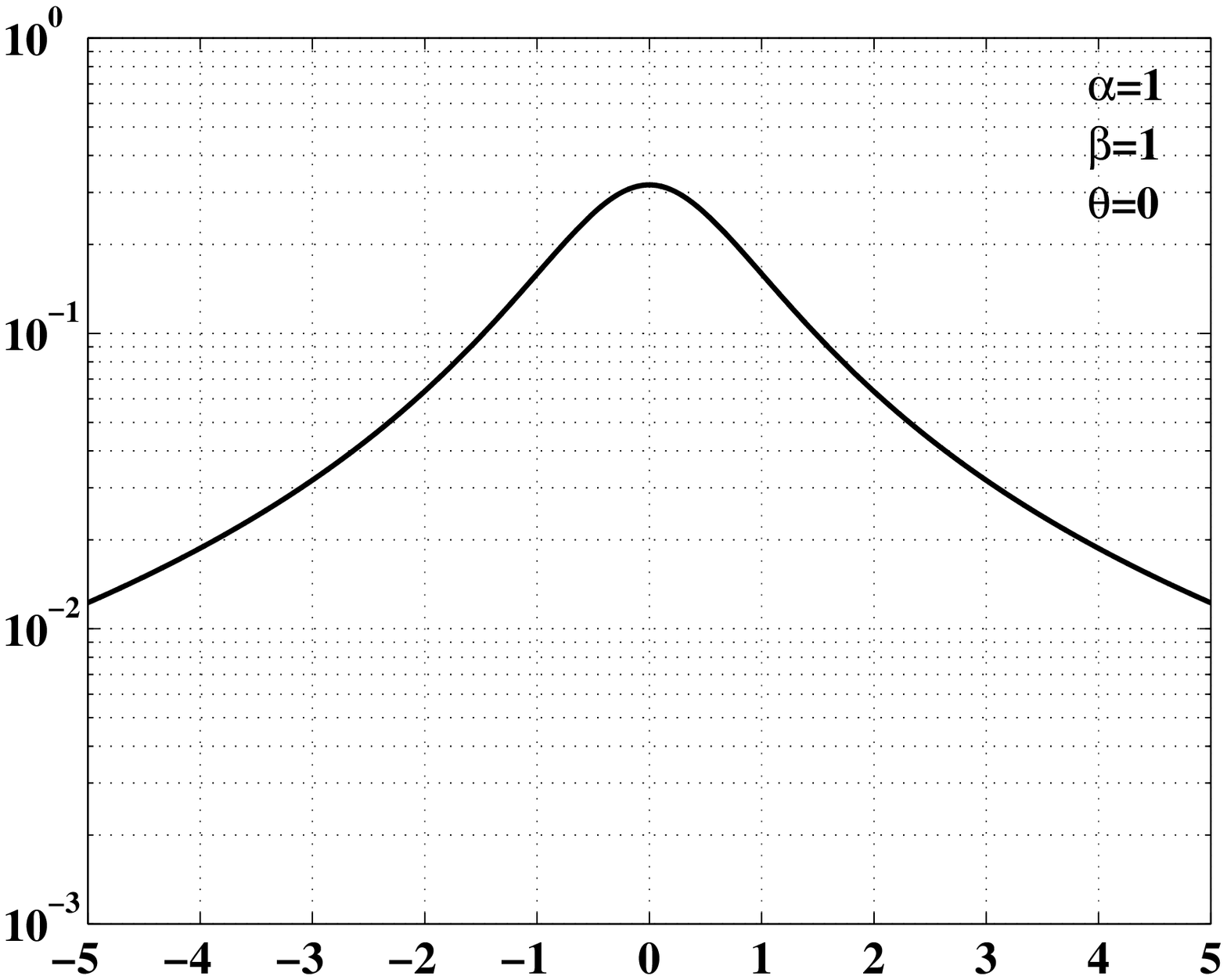}
\includegraphics[width=0.40\textwidth]{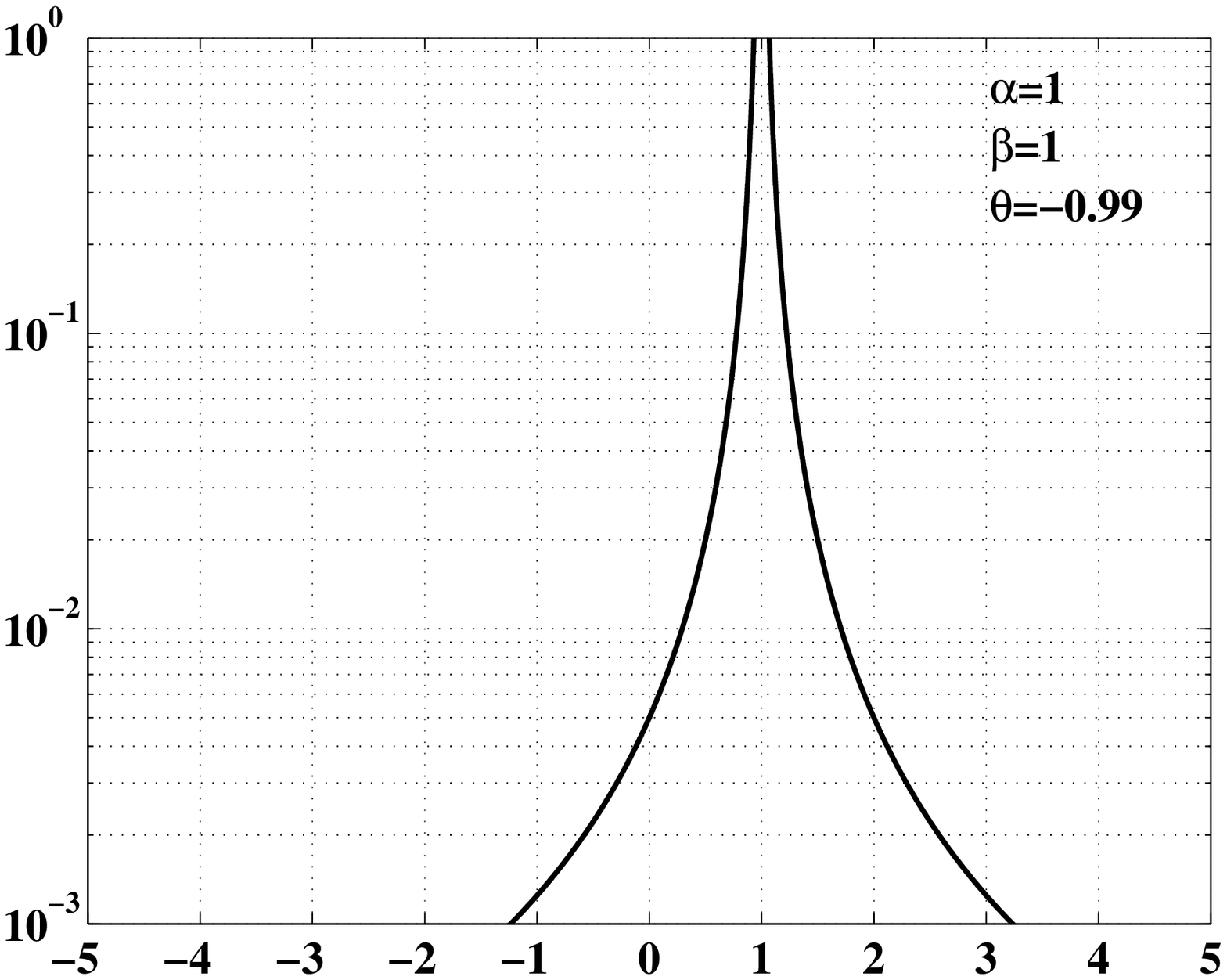}
\vskip 0.25truecm
\centerline{{\bf Fig. 4}}

\vskip 0.75truecm
\centering
\includegraphics[width=0.40\textwidth]{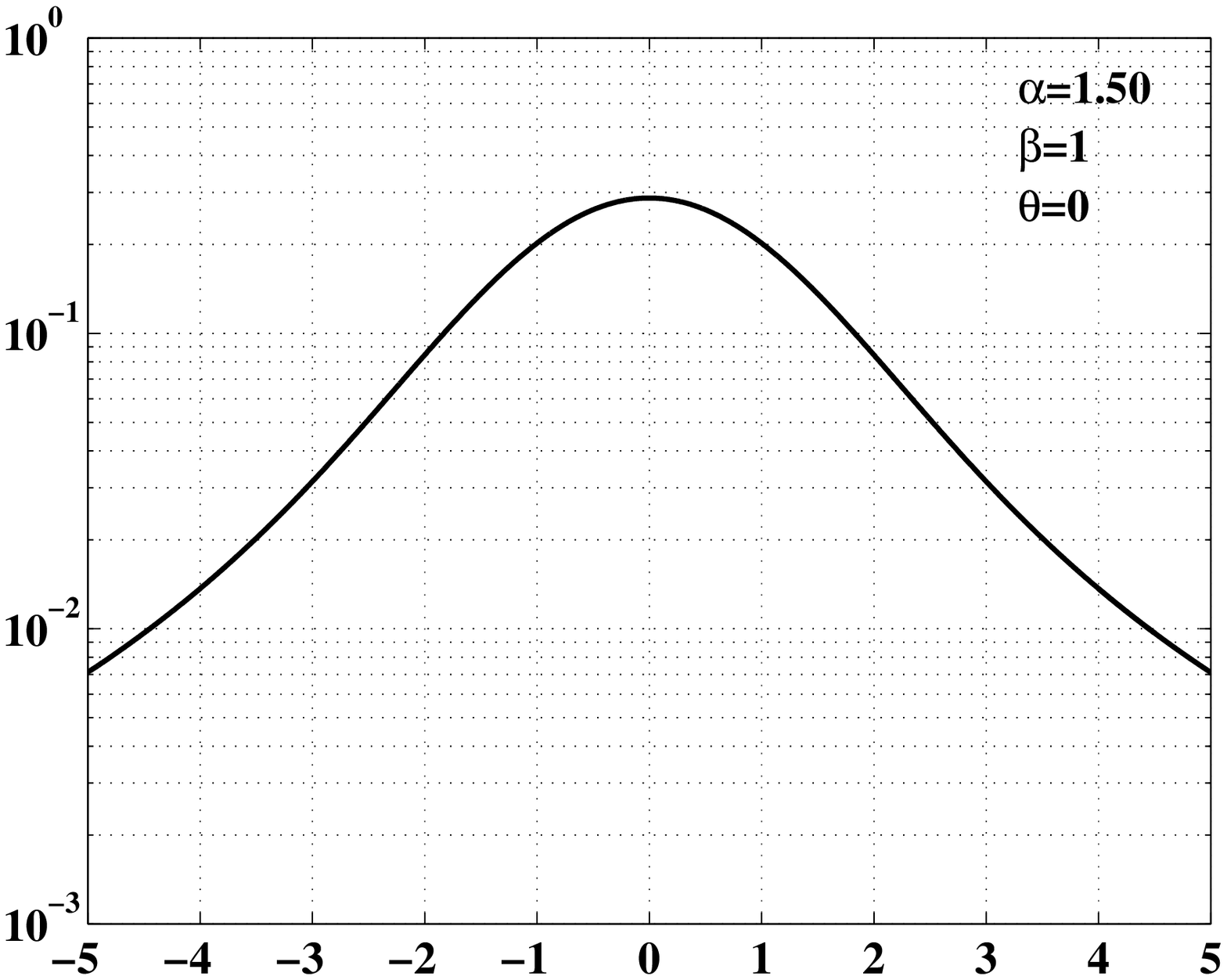}
\includegraphics[width=0.40\textwidth]{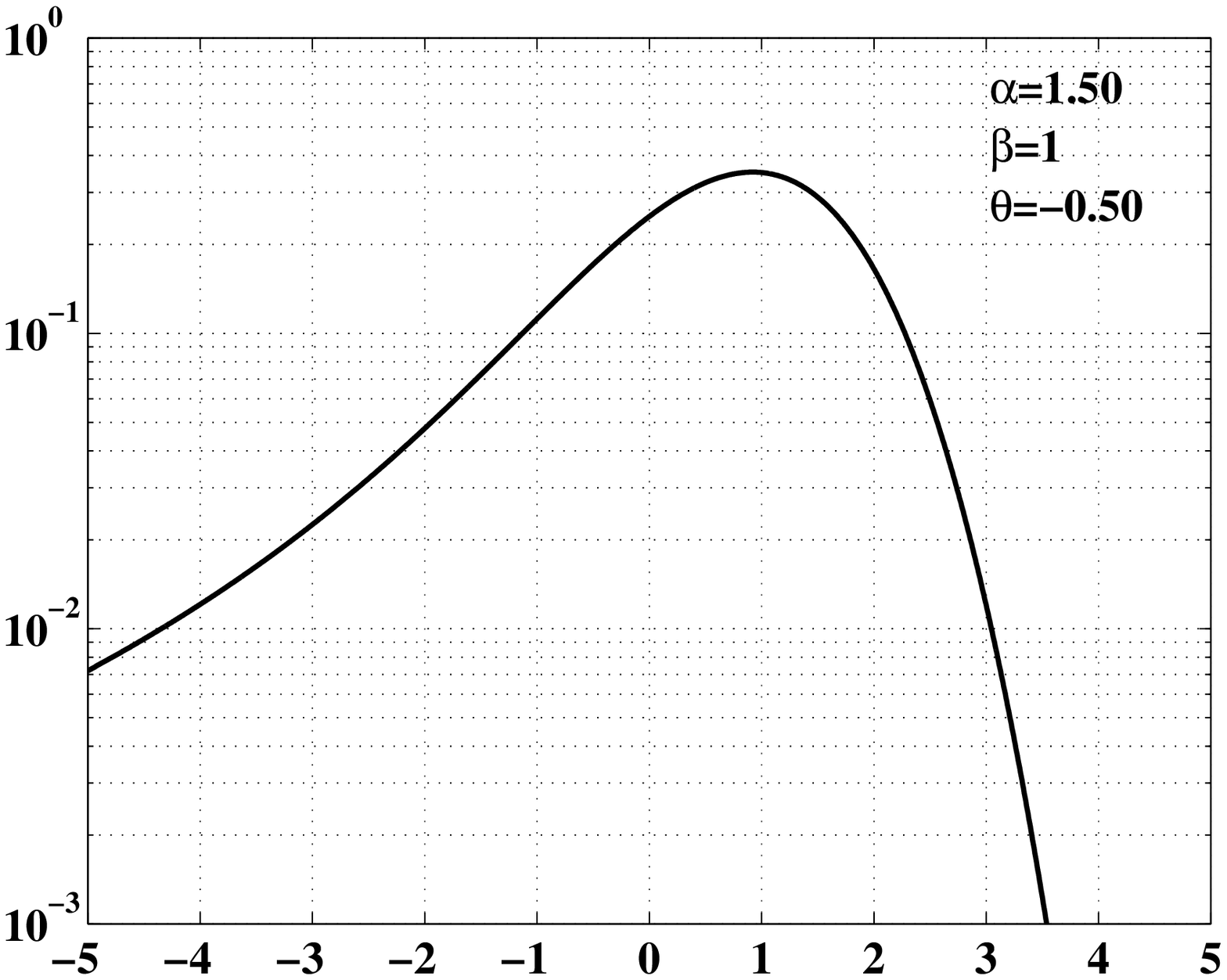}
\vskip 0.25truecm
\centerline{{\bf Fig. 5}}
\end{figure}



\begin{figure}
\centering
\includegraphics[width=0.40\textwidth]{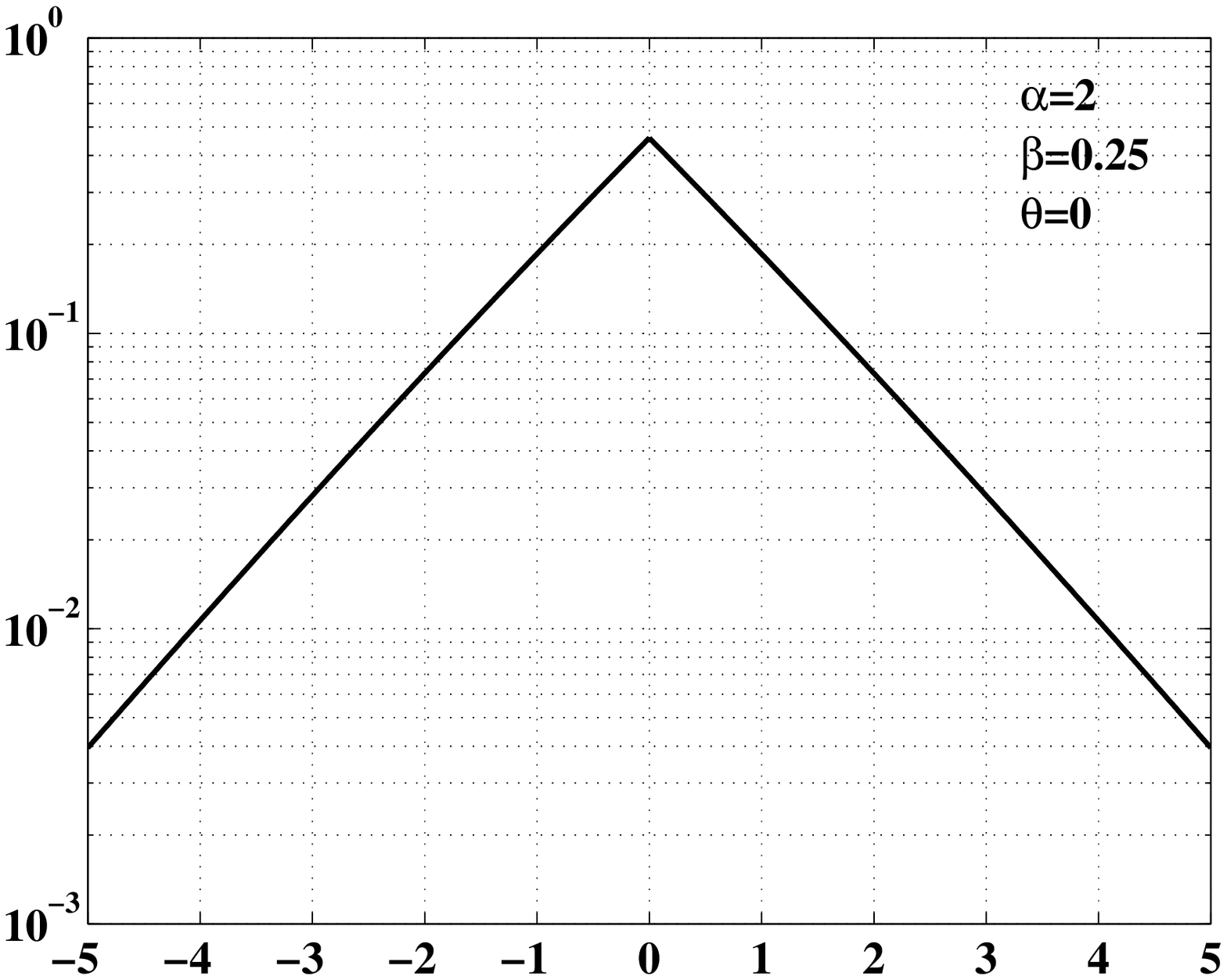}
\includegraphics[width=0.40\textwidth]{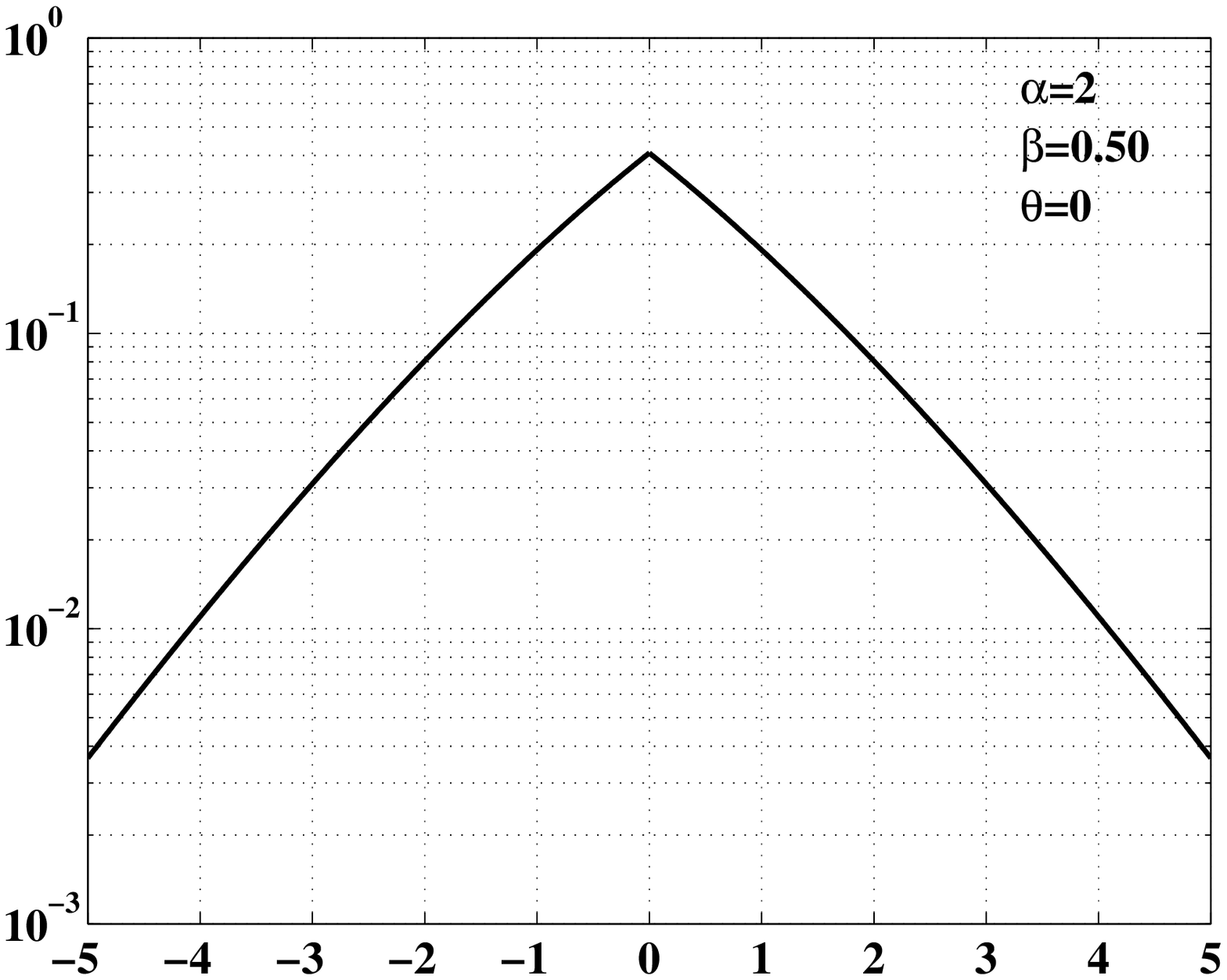}

\vskip 0.25truecm
\centerline{{\bf Fig. 6}}

\vskip 0.75truecm

\centering
\includegraphics[width=0.40\textwidth]{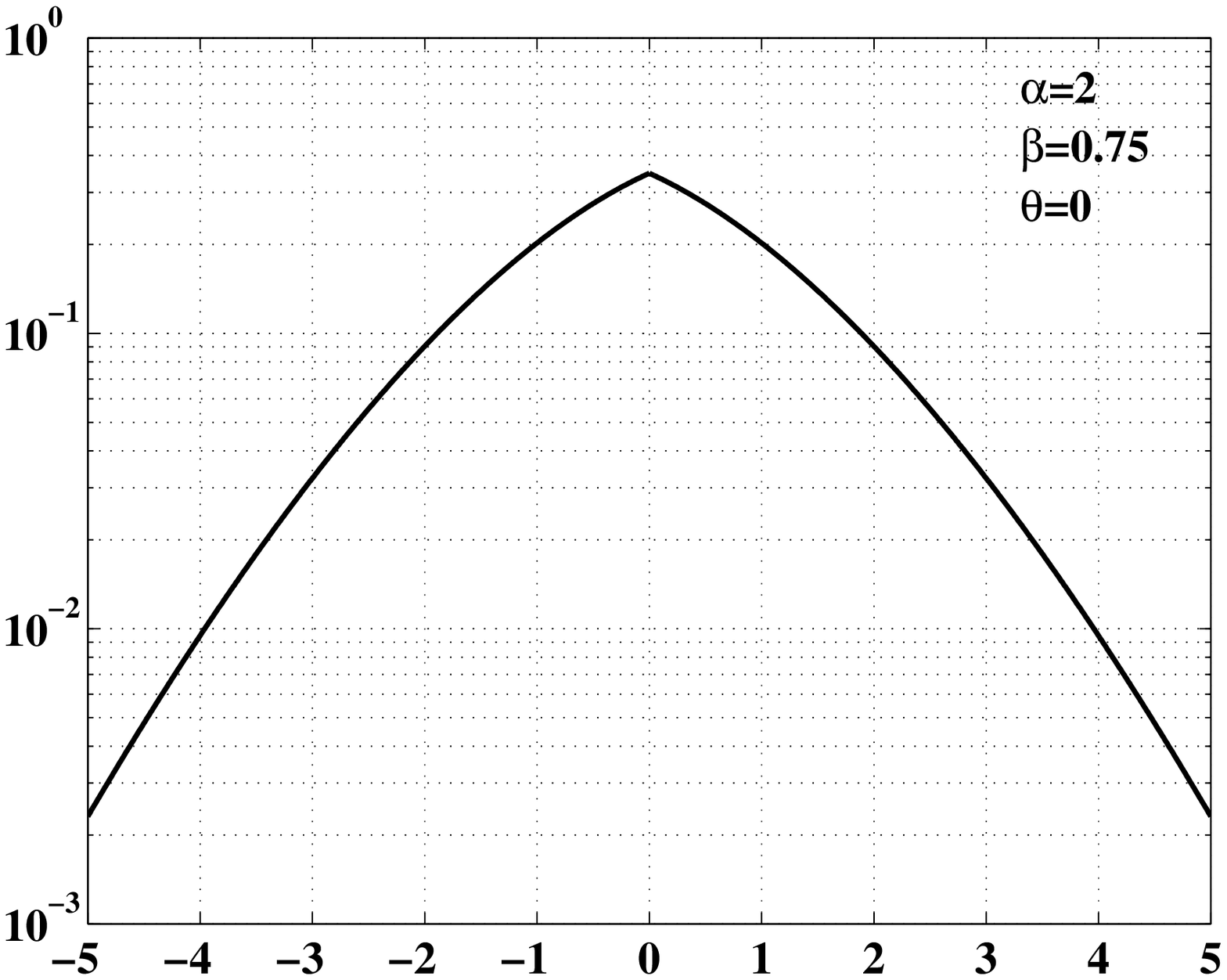}
\includegraphics[width=0.40\textwidth]{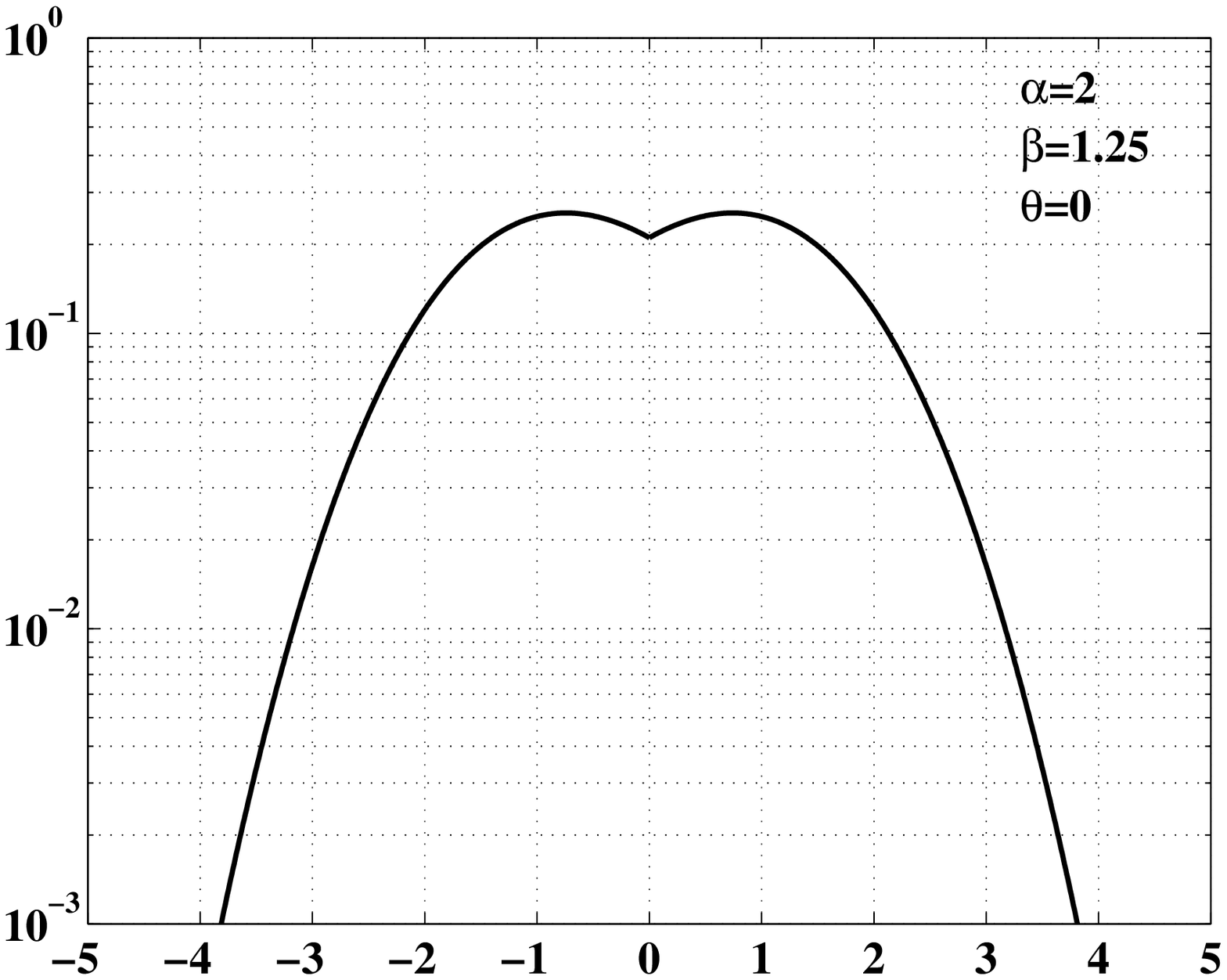}
\vskip 0.25truecm
\centerline{{\bf Fig. 7}}

\vskip 0.75truecm

\centering
\includegraphics[width=0.40\textwidth]{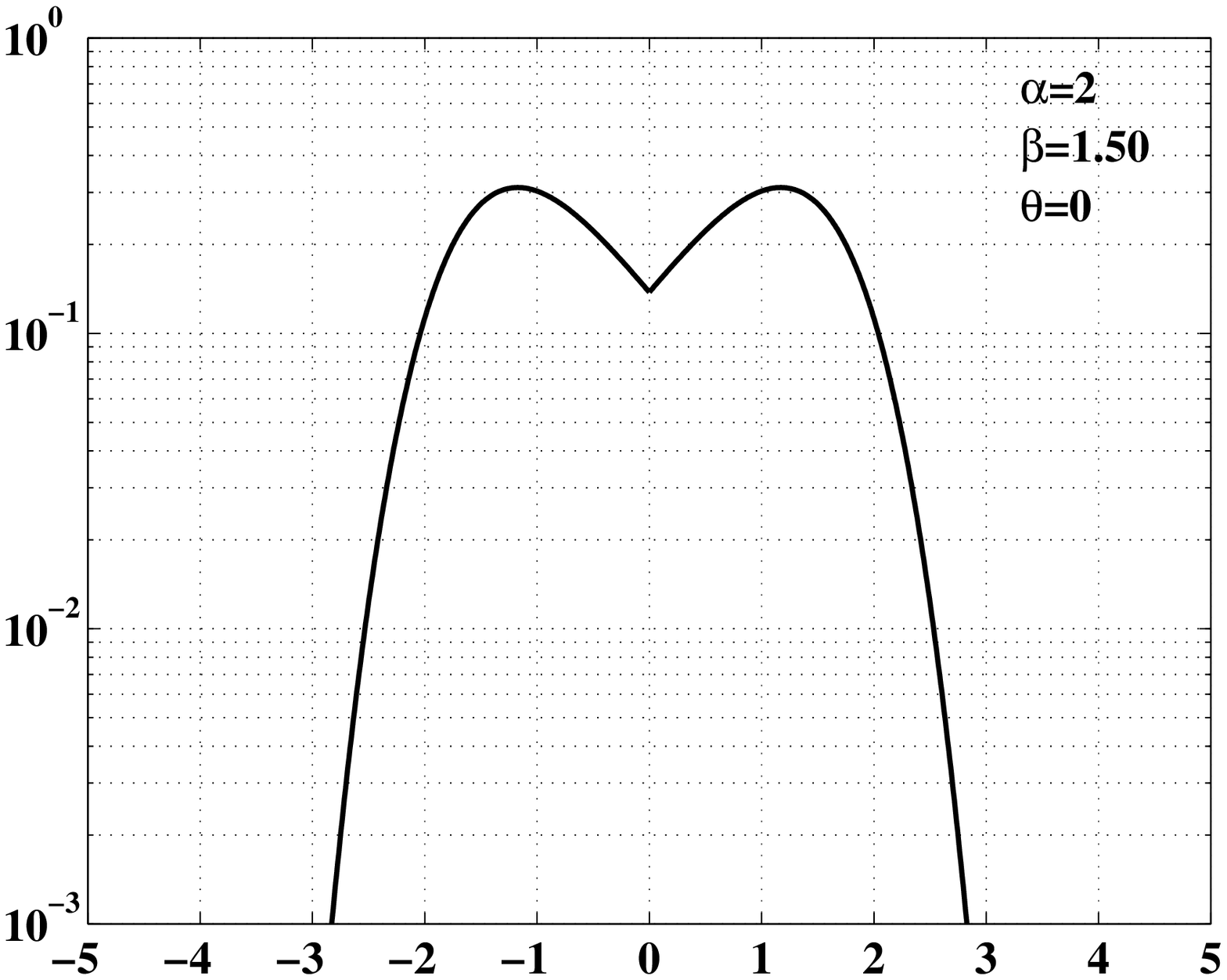}
\includegraphics[width=0.40\textwidth]{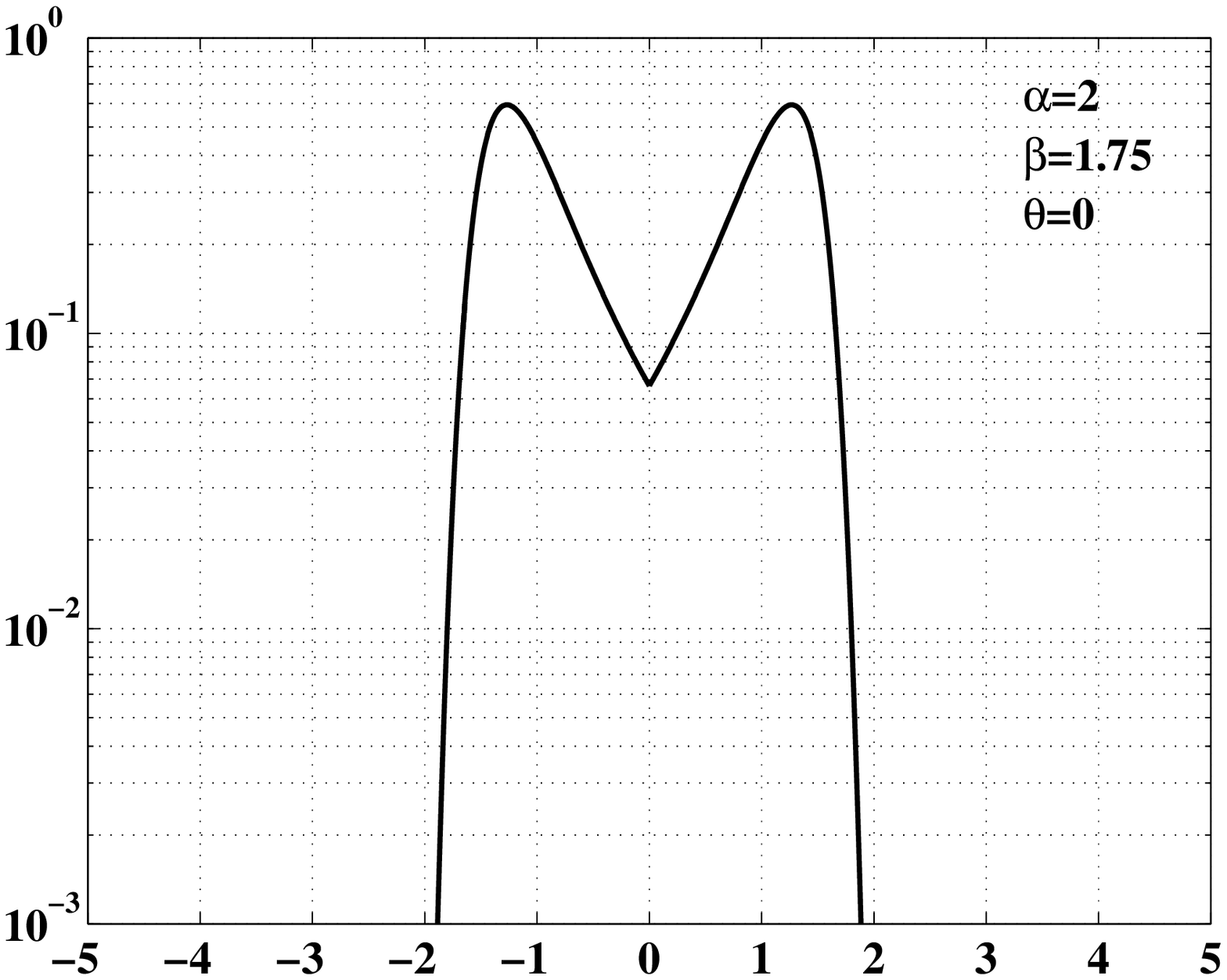}
\vskip 0.25truecm
\centerline{{\bf Fig. 8}}
\end{figure}

\begin{figure}
\centering
\includegraphics[width=0.40\textwidth]{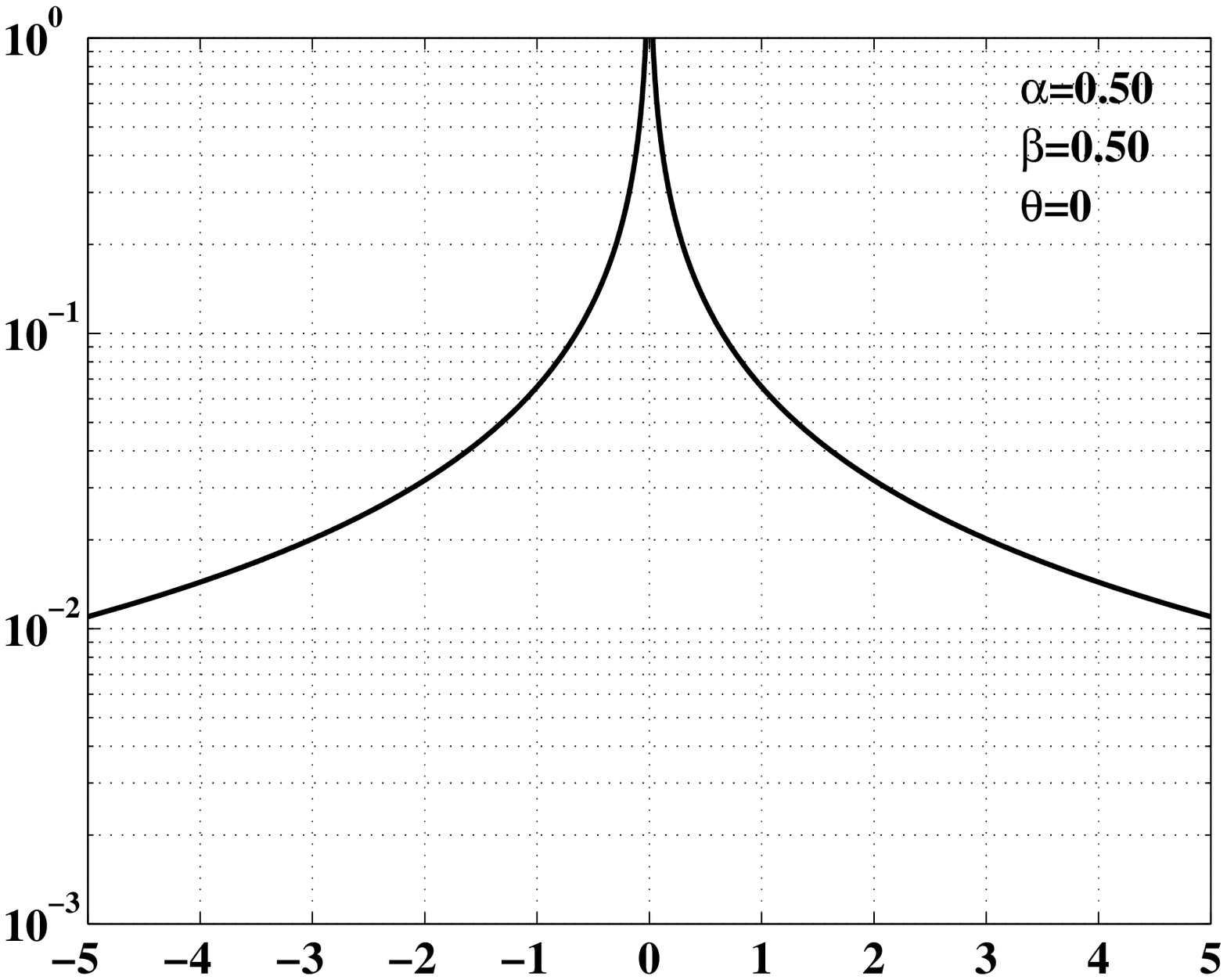}
\includegraphics[width=0.40\textwidth]{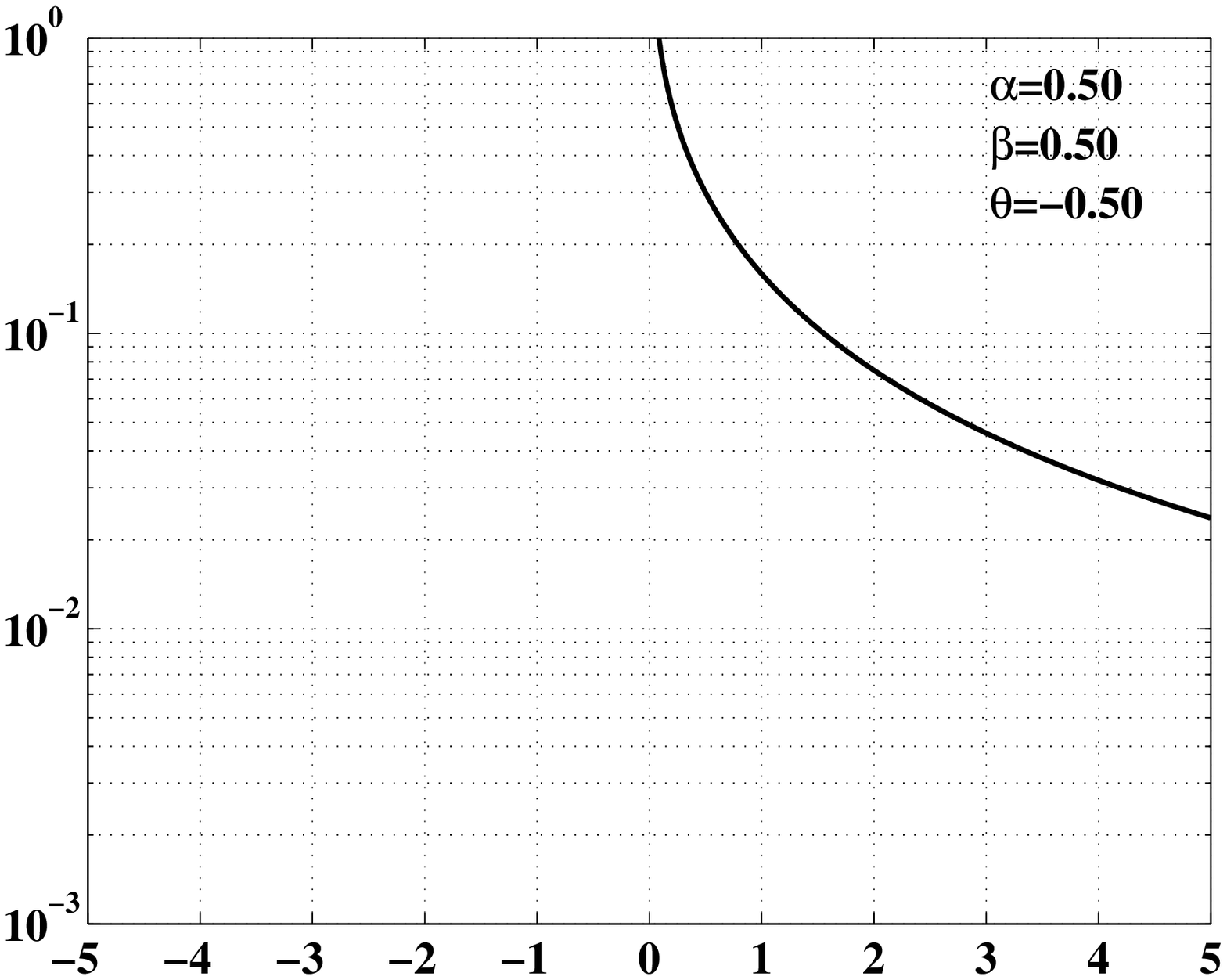}
\vskip 0.25truecm
\centerline{{\bf Fig. 9}}

\vskip 0.75truecm

\centering
\includegraphics[width=0.40\textwidth]{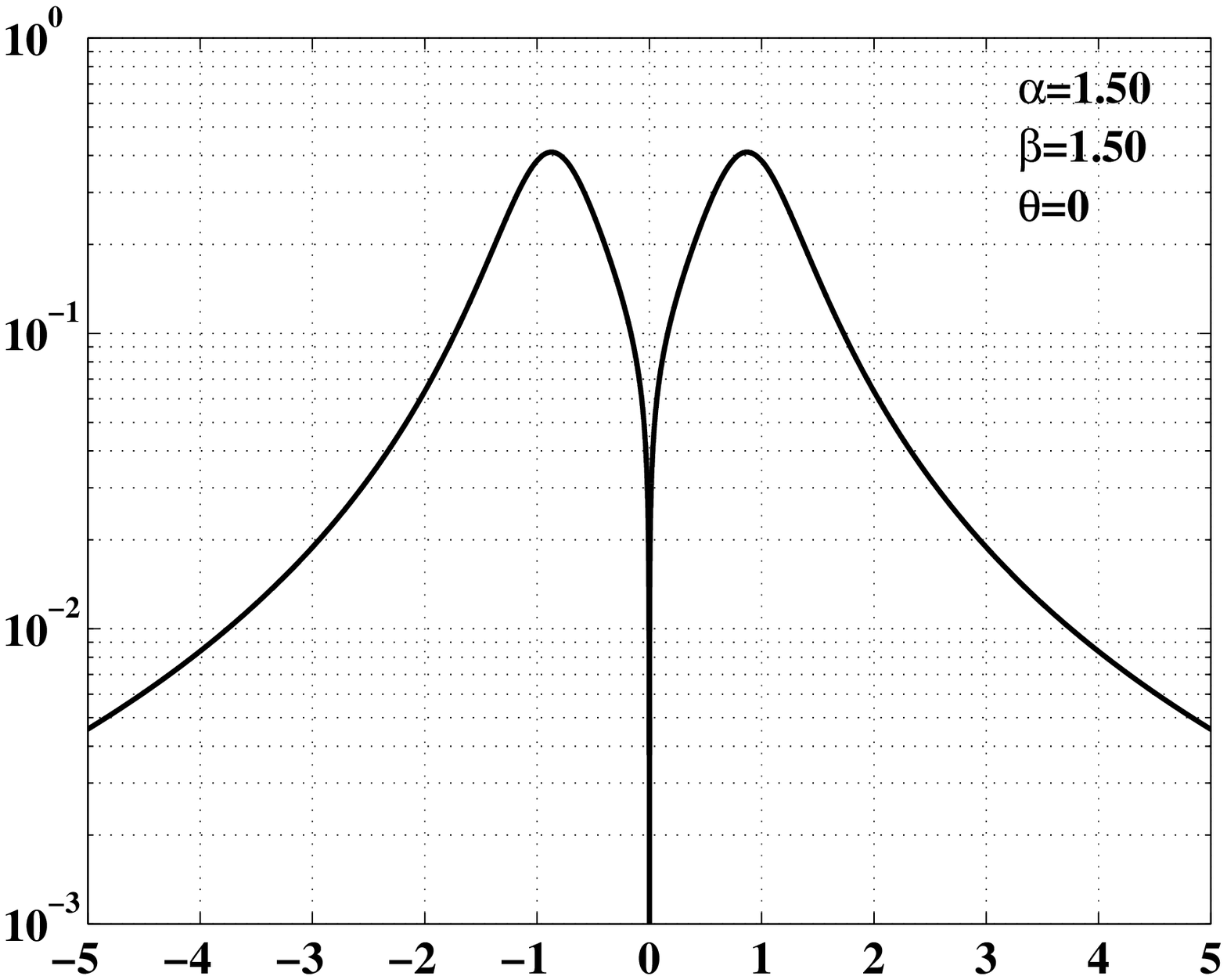}
\includegraphics[width=0.40\textwidth]{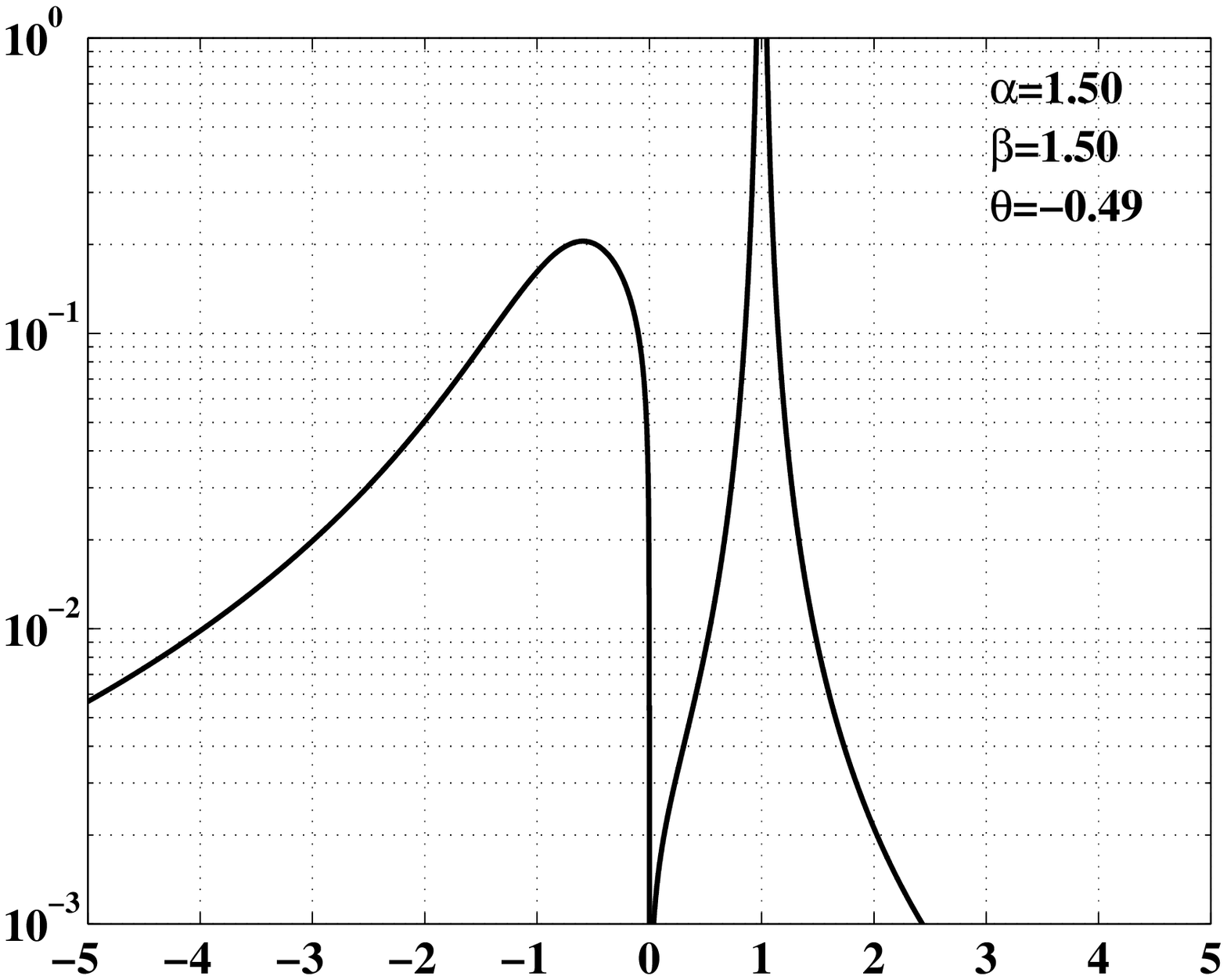}
\vskip 0.25truecm
\centerline{{\bf Fig. 10}}

\vskip 0.75truecm

\centering
\includegraphics[width=0.40\textwidth]{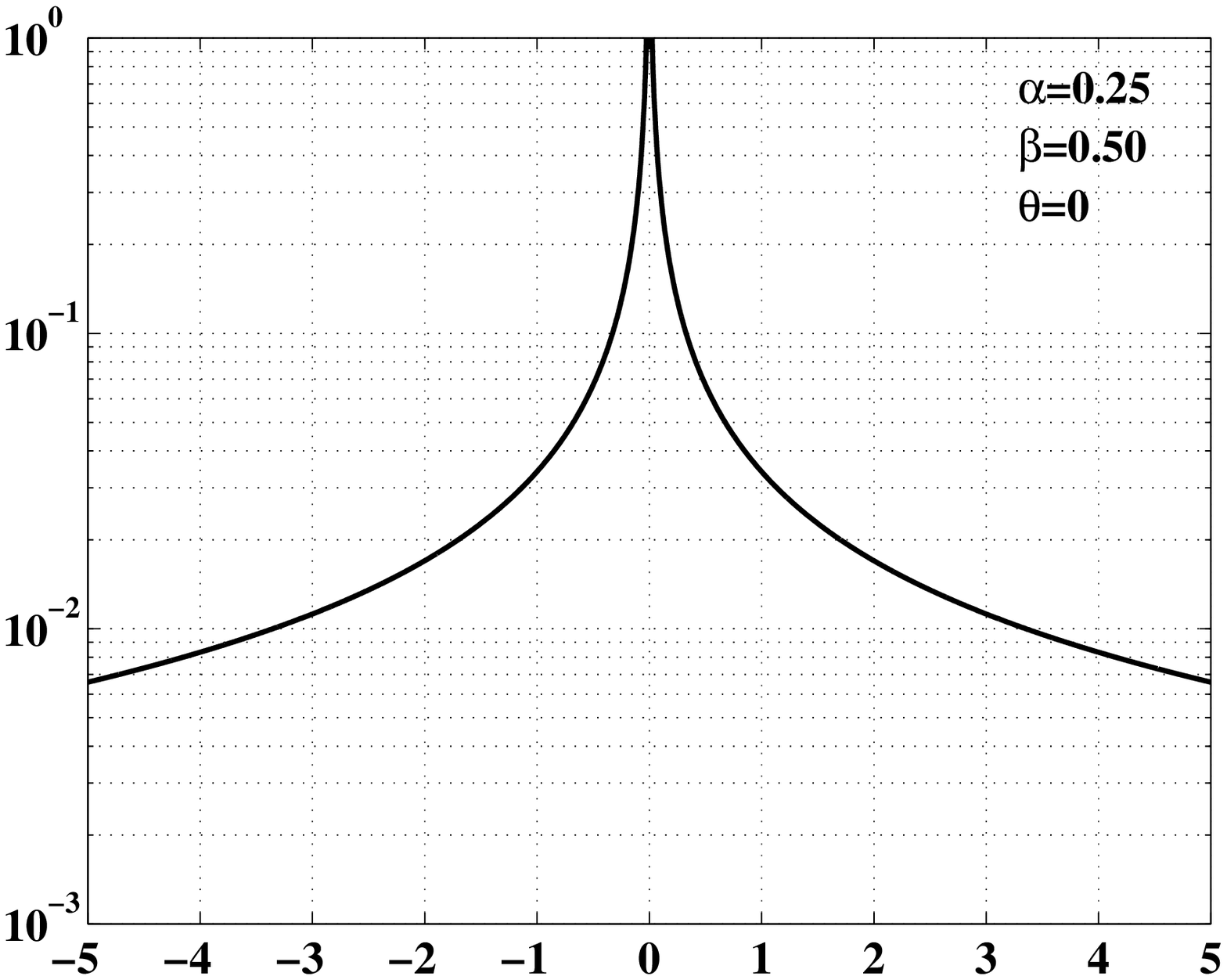}
\includegraphics[width=0.40\textwidth]{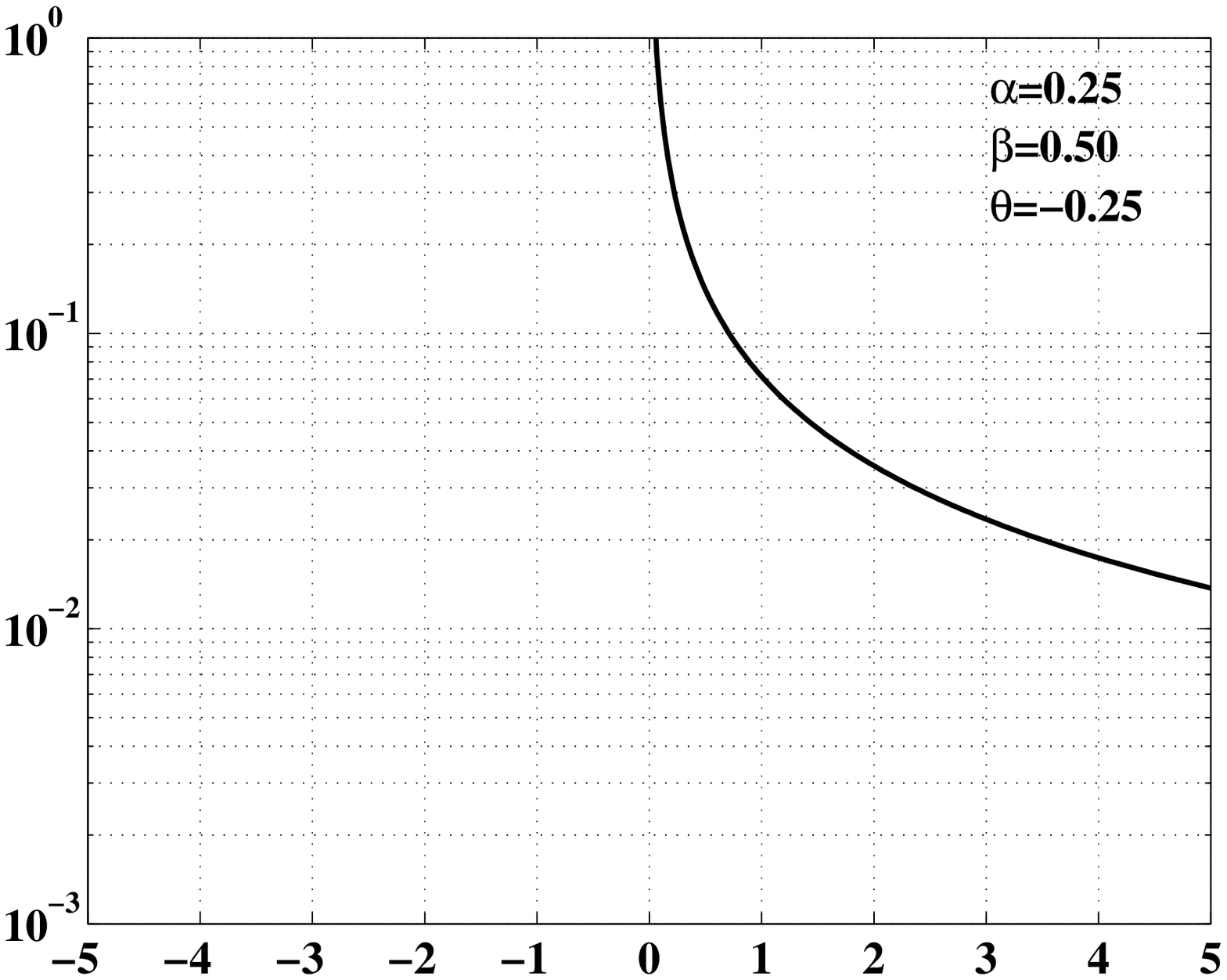}
\vskip 0.25truecm
\centerline{{\bf Fig. 11}}


\end{figure}
 


\begin{figure}
\centering
\includegraphics[width=0.40\textwidth]{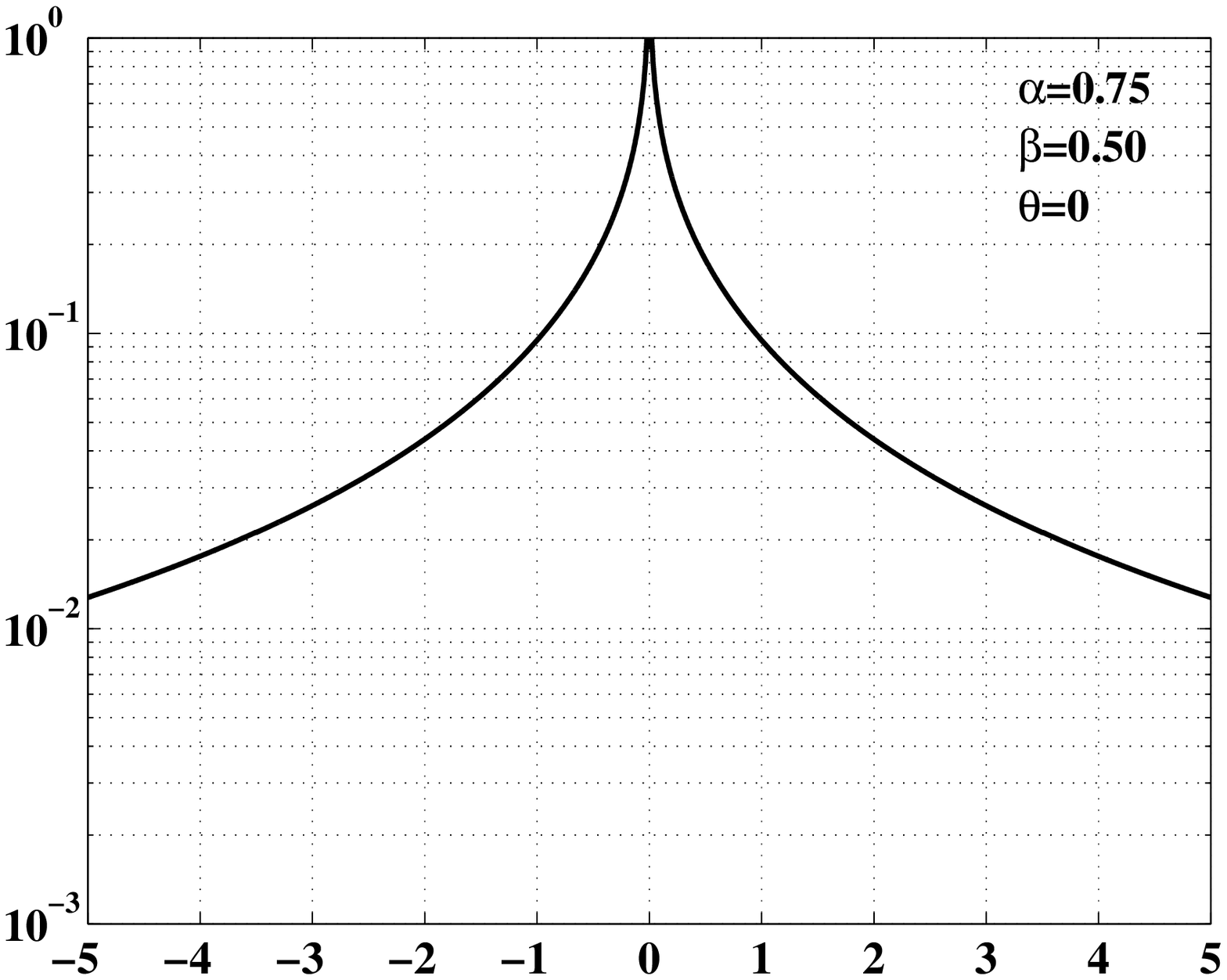}
\includegraphics[width=0.40\textwidth]{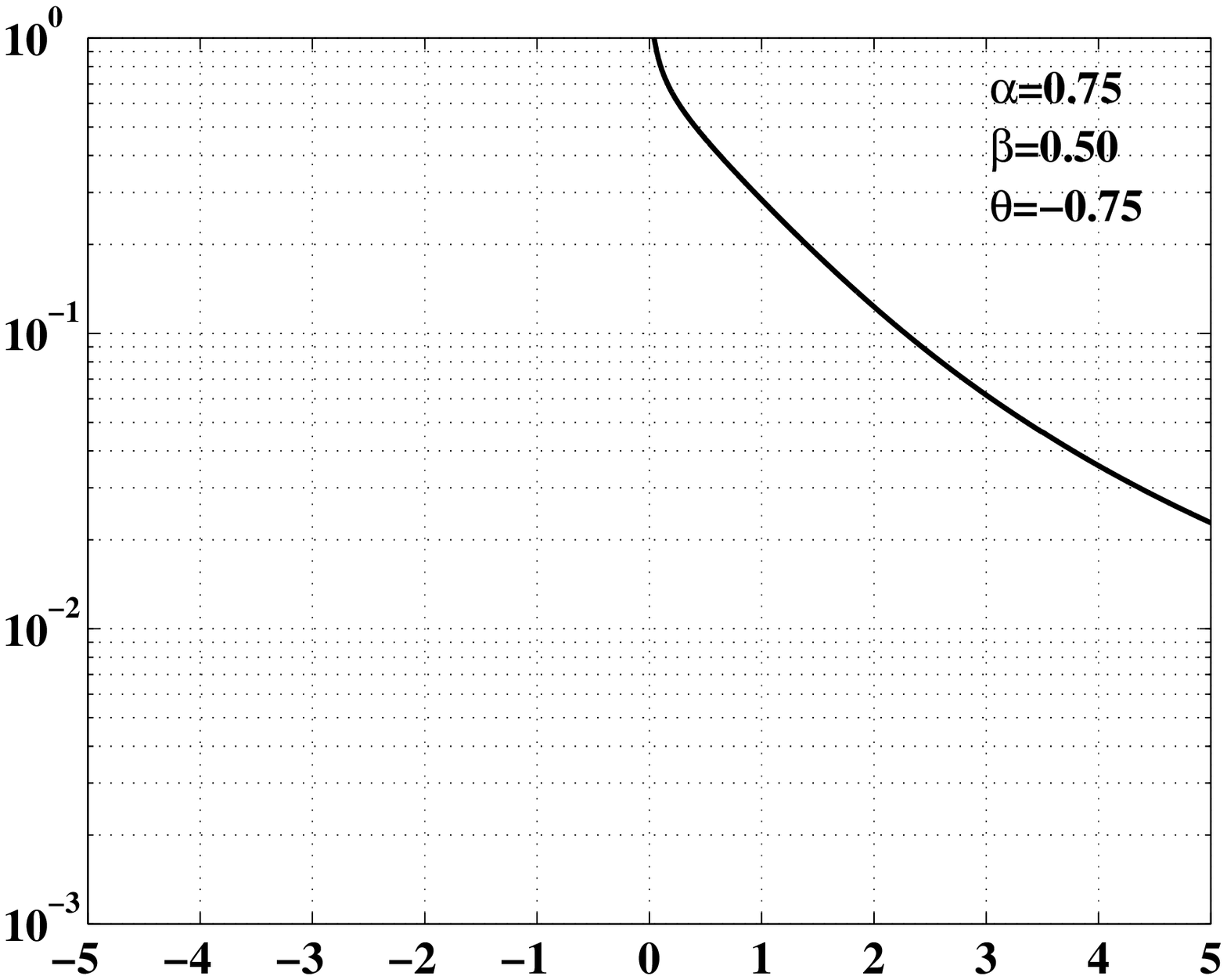}
\vskip 0.25truecm
\centerline{{\bf Fig. 12}}
\vskip 0.75truecm


\centering
\includegraphics[width=0.40\textwidth]{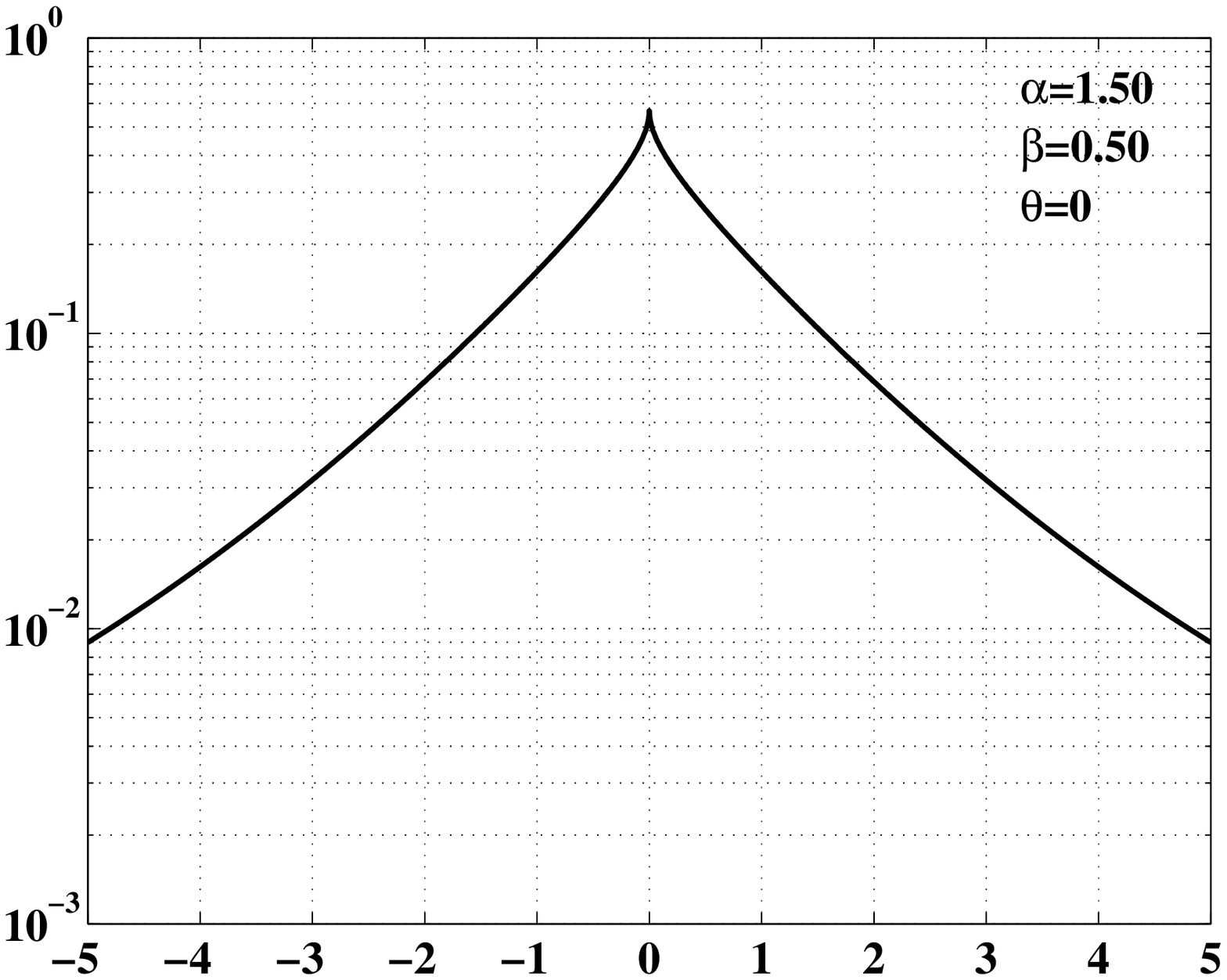}
\includegraphics[width=0.40\textwidth]{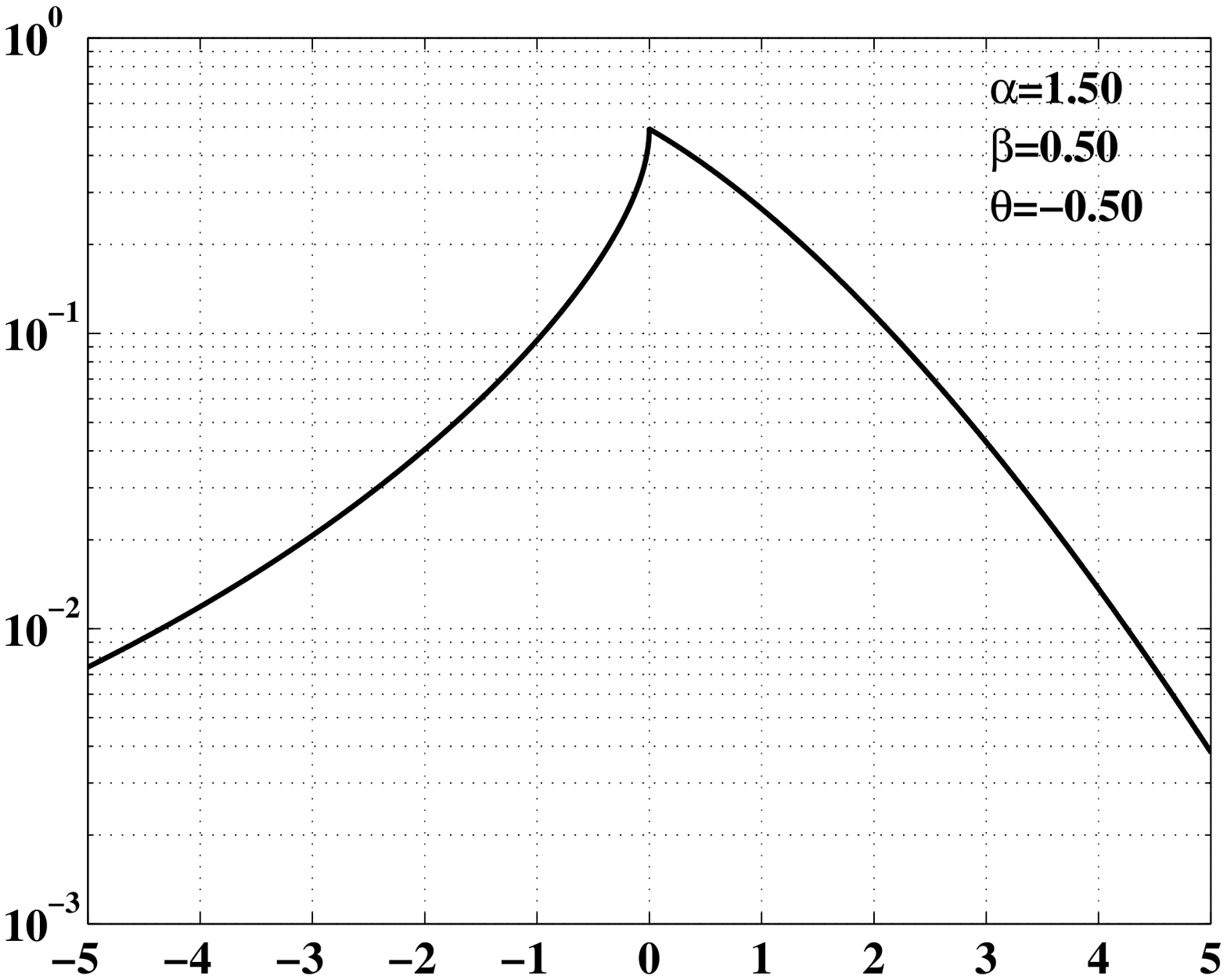}
\vskip 0.25truecm
\centerline{{\bf Fig. 13}}

\vskip 0.75truecm


\centering
\includegraphics[width=0.40\textwidth]{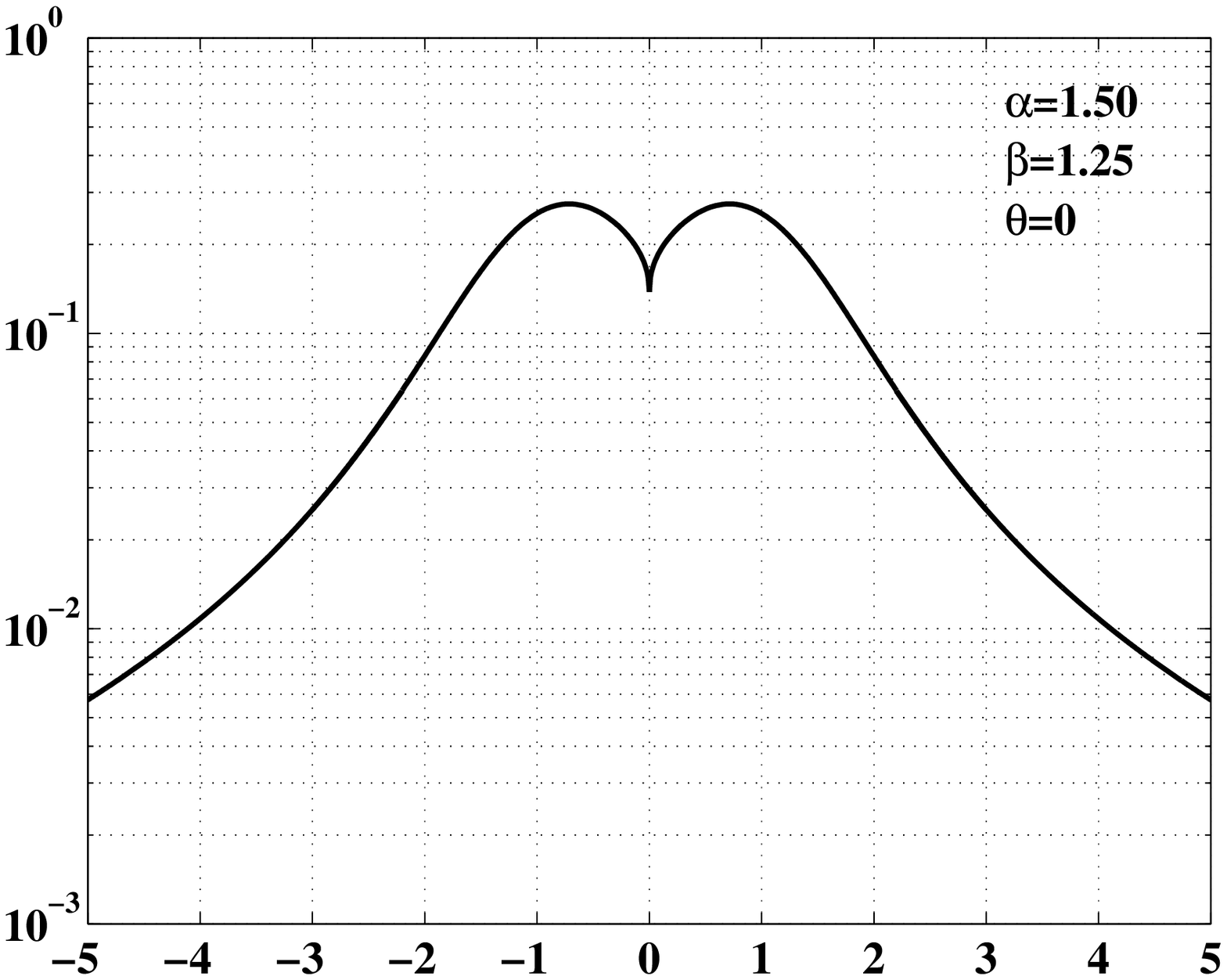}
\includegraphics[width=0.40\textwidth]{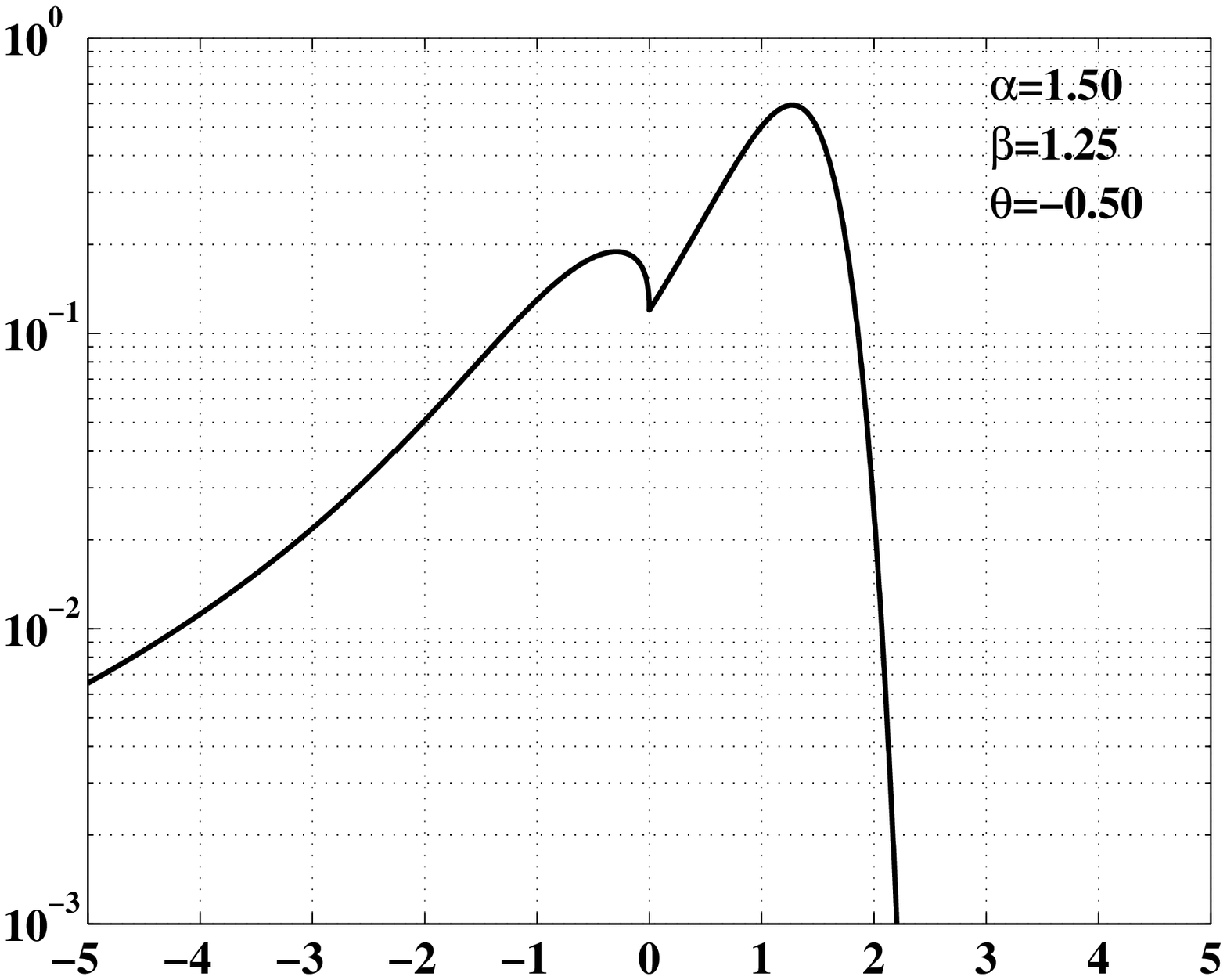}
\vskip 0.25truecm
\centerline{{\bf Fig. 14}}

\end{figure}

\newpage
\subsection*{Acknowledgements}
The authors are very grateful to Prof. Dr. R. Gorenflo
for the inspiring discussions and the helpful comments.



\end{document}